\documentclass[%
 reprint,
 amsmath,amssymb,
 aps,
]{revtex4-2}

\usepackage{graphicx}
\usepackage{dcolumn}
\usepackage{bm}
\usepackage{bbm}
\usepackage{siunitx}
\usepackage{mathtools}
\usepackage{braket}
\usepackage{chemformula}
\usepackage{tabls}
\usepackage[toc,page]{appendix}

\newcommand{\Rcal}{\mathcal{R}}
\newcommand{\Hcal}{{\hat{\mathcal{H}}}}

\newcommand{\Zcal}{\mathcal{Z}}

\newcommand{\ft}{\text{f}}

\newcommand{\bbf}{\mathbbm{f}}
\newcommand{\bbV}{\mathbb{V}}

\newcommand{\bA}{\boldsymbol{A}}

\newcommand{\bP}{\boldsymbol{P}}
\newcommand{\bR}{\boldsymbol{R}}

\newcommand{\bD}{\boldsymbol{D}}

\newcommand{\bG}{\boldsymbol{G}}
\newcommand{\bPi}{\boldsymbol{\Pi}}
\newcommand{\bu}{\boldsymbol{u}}

\newcommand{\bft}{\textbf{f}}

\newcommand{\bPhi}{\boldsymbol{\Phi}}

\newcommand{\bUps}{\boldsymbol{\Upsilon}}

\newcommand{\bRcal}{\boldsymbol{\Rcal}}

\newcommand{\bLambda}{\boldsymbol{\Lambda}}

\newcommand{\sss}[1]{\scriptscriptstyle{\text{#1}}}

\newcommand{\rschatrial}{\bRcal}

\newcommand{\Tr}[1]{\textup{Tr}\left[#1\right]}
\newcommand{\Avg}[2]{\left\langle #1\right\rangle_{#2}}

\newcommand{\Phithree}{\overset{\sss{(3)}}{\Phi}}
\newcommand{\bPhithree}{\overset{\sss{(3)}}{\bPhi}}

\newcommand{\Dthree}{\overset{\sss{(3)}}{D}}
\newcommand{\bDthree}{\overset{\sss{(3)}}{\bD}}

\newcommand{\Phifour}{\overset{\sss{(4)}}{\Phi}}
\newcommand{\bPhifour}{\overset{\sss{(4)}}{\bPhi}}
\newcommand{\Dfour}{\overset{\sss{(4)}}{D}}
\newcommand{\bDfour}{\overset{\sss{(4)}}{\bD}}

\newcommand{\Dtwo}{\overset{\sss{(2)}}{D}}
\newcommand{\bDtwo}{\overset{\sss{(2)}}{\bD}}

\renewcommand{\Im}{\mathrm{Im}}
\renewcommand{\Re}{\mathrm{Re}}

\renewcommand{\bf}{\boldsymbol{f}}

\newcommand{\rhoo}{{\rho}}
\newcommand{\hrhoo}{{{\hat \rho}}}

\newcommand{\Avgclassic}[1]{\Avg{#1}{\rho}}
\newcommand{\Avgclassict}[1]{\Avg{#1}{\rho(t)}}
\newcommand{\Avgquantum}[1]{\Avg{#1}{\hat\rho}}
\newcommand{\Avgquantumt}[1]{\Avg{#1}{\hat\rho(t)}}
\newcommand{\Avgclassiceq}[1]{\Avg{#1}{\rhoo}}
\newcommand{\Avgclassiceqz}[1]{\Avg{#1}{\rho^{(0)}}}
\newcommand{\Avgclassicpert}[1]{\Avg{#1}{\rho^{(1)}}}

\newcommand{\Avgquantumeq}[1]{\Avg{#1}{{\hat \rho}^{(0)}}}

\newcommand{\bTheta}{\boldsymbol{\Theta}}
\newcommand{\bC}{\boldsymbol{C}}
\newcommand{\bQ}{\boldsymbol{Q}}

\newcommand{\Gcal}{{\mathcal G}}
\newcommand{\bGcal}{\boldsymbol{\Gcal}}
\newcommand{\bU}{{\bm U}}
\newcommand{\br}{{\bm r}}

\newcommand{\Hbo}{{\hat H}}
\newcommand{\Htd}{{{\hat H}_{\text{td}}(t)}}

\newcommand{\norm}{{\mathcal N}}

\newcommand{\Vtot}{{V^{(\text{tot})}}}
\newcommand{\ftot}{f^{(\text{tot})}}
\newcommand{\bftot}{\boldsymbol{f}^{(\text{tot})}}
\newcommand{\Vext}{{V^{\text{ext}}}}

\newcommand{\Acal}{{\mathcal A}}
\newcommand{\Bcal}{{\mathcal B}}
\newcommand{\Ramant}{{\Xi}}
\newcommand{\bRamant}{\boldsymbol{\Ramant}}
\newcommand{\bchi}{\boldsymbol{\chi}}

\newcommand{\rhoone}{{{\hat\rho}^{(1)}}}
\newcommand{\Lsc}{ {{L}_\text{sc}}}
\newcommand{\Vsc}{ {\hat{ {V_{\text{sc}}}}^{(1)}}}
\newcommand{\comm}[2]{\left[#1, #2\right]}

\newcommand{\Scal}{{\mathcal{S}}}
\newcommand{\secname}{{Sec.}}
\newcommand{\Rc}{\Rcal}
\newcommand{\eqname}{{Eq.}}
\DeclareMathOperator{\tr}{Tr}
\newcommand{\bpm}{\begin{pmatrix}}
\newcommand{\epm}{\end{pmatrix}}

\begin{document}

\preprint{APS/123-QED}

\title{Time-Dependent Self Consistent Harmonic Approximation:\\Anharmonic nuclear quantum dynamics and time correlation functions}

\author{Lorenzo Monacelli}
 
\author{Francesco Mauri}%
 \affiliation{Physics Department, University of Rome ``Sapienza''}

\date{\today}

\begin{abstract}
Most material properties of great physical interest are directly related to nuclear dynamics, e.g. the ionic thermal conductivity, Raman/IR vibrational spectra, inelastic X-ray, and Neutron scattering. 
A theory able to compute from first principles these properties, accounting for the anharmonicity and quantum fluctuations in the nuclear energy landscape that can be implemented in systems with hundreds of atoms is missing.
Here, we derive an approximate theory for the quantum time evolution of lattice vibrations at finite temperature. This theory introduces the time dynamics in the Self-Consistent Harmonic Approximation (SCHA) and shares with the static case the same computational cost. It is nonempirical, as pure states evolve according to the Dirac least action principle and the dynamics of the thermal ensemble conserves both energy and entropy. The static SCHA is recovered as a stationary solution of the dynamical equations.
We apply perturbation theory around the static SCHA solution and derive an algorithm to compute efficiently quantum dynamical response functions. Thanks to this new algorithm, we have access to the response function of any general external time-dependent perturbation, enabling the simulation of phonon spectra without following any perturbative expansion of the nuclear potential or empirical methods. We benchmark the algorithm on the IR and Raman spectroscopy of high-pressure hydrogen phase III, with a simulation cell of 96 atoms. 
Our work also explores the nonlinear regime of the dynamical nuclear motion, providing a paradigm to simulate the interaction with intense or multiple probes, as in pump-probe spectroscopy, or chemical reactions involving light atoms, as the proton transfer in biomolecules.

\end{abstract}

\maketitle


\section{\label{sec:level1}
Introduction}

The computational power available for research dramatically increased in the last decades paving the way to the birth of a new field of science: material design. We can predict the physical properties of a material \emph{in silico}, anticipating the experimental data and enabling for an automatic search of target compounds before their synthesis.
An example of an impressive result is the prediction of record-breaking high-temperature superconductivity in hydrates, like hydrogen sulfide\cite{Li2014} and \ch{LaH10}\cite{Peng2017} that  came together with experimental result\cite{Drozdov2015,Somayazulu_2019,drozdov2019superconductivity}, thanks to the success of \emph{ab initio} crystal structure prediction\cite{CALYPSO,USPEX,AIRSS}.

Material science can also assist in interpreting experimental data, as it is possible, in principle, to anticipate the outcome of almost any experimental technique by calculating dynamical correlation functions. For example, IR and optical spectroscopy are obtained by simulating the dipole-dipole time correlation functions, Raman signal with the polarizability-polarizability dynamical correlation function, while Neutron and X-Ray scattering can be simulated calculating the dynamical structure factor.

However, first-principle calculations are still far from being perfect, and they often miss the required precision.
Correctly describing ionic vibrations in \emph{ab initio} simulation is challenging. Even if the energy of ionic vibration is orders of magnitude smaller than the typical energy involved in chemical bonds, the nuclear excited states are in the same range of energy as thermal excitation available at room temperature.
Therefore, ionic motion is responsible for almost all properties of materials that depend on temperature as thermal expansion and thermal conductivity.

The development of time-dependent Hartree-Fock (TD-HF) and time-dependent density functional theory\cite{Runge1984,TDDFT} (TD-DFT) paved the way to the simulation of dynamical quantum correlation functions for electrons. Still, a full quantum theory of nuclear motion that can be routinely applied in realistic systems (with hundreds of atoms in the simulation cell, treating \emph{ab initio} the electrons) is missing.

The usual approach to lattice dynamics is through perturbation theory. Here, the energy landscape expanded in the Taylor series around the static position (minimum of the total electronic energy of the system).
This approach fails in systems strongly anharmonic or close to the second-order phase transition, where harmonic phonons are unstable.


Ionic time correlation functions can be computed also in the presence of strong anharmonicity with \emph{ab initio} molecular dynamics (AIMD)\cite{car1985,Cupo2019}, but it neglects nuclear quantum dispersion. This is critical for molecular crystals or systems with light atoms, where the Debye temperature is comparable or above room temperature. In these cases, AIMD provides untrustable results.

The quantum equivalent of AIMD is path-integral molecular dynamics (PIMD), but, in its original imaginary time derivation, it is a static theory. PIMD has been extended in many ways to describe dynamical quantities, as an analytical continuation to real time\cite{Baym1961}, Fourier path-integral methods\cite{Chen1996,Kim1997}; but these methods are computationally heavy and possible only in prototypical cases, where the correlation function has few poles and if a force-field is available to clean stochastic noise\cite{Krilov1999}. To overcome the computational cost of the almost exact PIMD dynamical extension, approximate empirical methods have been developed, as centroid molecular dynamics\cite{Cao1994,Paesani2010}, Gaussian molecular dynamics\cite{Georgescu2010}, and ring-polymer molecular dynamics\cite{Poulsen2003,Hernandez1998,Sun1998,Bonella2010}. These methods have been successfully applied to study the IR spectrum of water employing force-fields, but, still, their application is computationally much more expensive than AIMD, limiting the range of applicability when coupled with an \emph{ab initio} description of the electronic degrees of freedom. Other semi-classical approaches have been developed to describe quantum dynamical properties, as Ehrenfest dynamics\cite{Li2005}, Gaussian wave-packet dynamics\cite{Heller1975}, and time-dependent self-consistent field\cite{Gerber1982}, however, the first two do not conserve energy during the wave-function propagation, and all of them are still far computationally more demanding than AIMD, preventing a systematic use in real cases.

Another method appreciated in the bio-chemistry community is the Nuclear Electronic-Orbital Method\cite{Pavoevi2020}. Here,  nuclei are treated with the same level as electrons. This approach allows for almost exact treatment of quantum fluctuations even beyond the Born-Oppenheimer approximation, however, their computational cost is extreme compared to the aforesaid methods, and it is suited only to simulate small molecules, or when the quantum mechanical degrees of freedom are a small fraction of the systems (as in QM/MM methods).

Relevant cases where all these techniques fail are materials with light atoms or close to structural instability, where ionic fluctuations are sizable. For example, the simulation of high-pressure hydrates requires a full quantum and anharmonic treatment of the nuclear motion. These materials are attracting a lot of attention due to the discovery of room-temperature superconductivity\cite{Snider2020}. On the other side, systems close to structural instabilities play a fundamental role in technological and industrial applications; among them, we have multiferroics, charge density waves in 2D layers, and ferroelectrics, where the huge anharmonic phonon scattering is exploited also to increase their thermoelectric efficiency. Moreover, quantum nuclear dynamics beyond the linear regime are fundamental for studying chemical reactions involving light atoms. A significant example is the proton transfer in biomolecules.

In this work, we formulate a new dynamical theory of quantum nuclear motion at finite temperature and derive an algorithm to calculate time-dependent correlation functions fully \emph{ab initio}.
This method has a computational cost of the same order as standard AIMD but fully accounts for the effect of quantum fluctuations non empirically. Our theory is a time-dependent (TD) extension of the Stochastic Self-Consistent Harmonic Approximation (SCHA), that already proved to be very efficient in describing equilibrium properties of many materials, as high-pressure hydrates\cite{Errea2014,Nature2016,Bianco2018,ErreaNature2020}, hydrogen\cite{Borinaga2016,Borinaga2017,MonacelliNatPhys2020}, charge density waves\cite{Bianco_2019,Bianco2020,Zhou2020} and thermoelectric materials\cite{Aseginolaza2019Phonon,Aseginolaza2019Strong,Ribeiro2018}.

In \secname~\ref{sec:full:dyn} we introduce the theoretical framework of this paper. We revise the general dynamical response to an external time-dependent perturbation and the relationship between the nuclear time-correlation functions to some common experimental techniques.
The Time-Dependent SCHA (TD-SCHA) is introduced in \secname~\ref{sec:tdscha}, where we derive the master equation for the density matrix by imposing that a Gaussian wave-packet must minimize the Dirac action. We also discuss generic features of the time-evolution as energy and entropy conservation, steady-state solutions, and the thermodynamic equilibrium (where the SCHA solution is recovered).
In \secname~\ref{sec:linear:response}, we linearize the TD-SCHA equations of motion around the equilibrium solution. We derive the explicit formula to compute the quantum dynamical response functions and derive a new Lanczos-based algorithm to compute them. 
In \secname~\ref{sec:applications}, we benchmark the method both in an illustrative 1D toy model and in the realistic case of high-pressure hydrogen phase III, where quantum dispersion and anharmonicity strongly affect the vibrational spectra. We prove, comparing the simulation of IR and Raman spectra to experiments, how the TD-SCHA can easily handle such a complex system with 96 atoms in the simulation cell computing forces between ions fully \emph{ab initio} within Density Functional Theory (DFT).

\section{The quantum nuclear evolution}
\label{sec:full:dyn}
Almost all experimental techniques used to characterize materials involve an exchange in energy between the probe and the sample below hundreds of \SI{}{\electronvolt}. At this energy scale, the only components of the material that interact with the probe are valence electrons and ions, which include the atomic nucleus plus the core electrons. These are the degrees of freedom we are interested in studying.
In the Born-Oppenheimer (BO) approximation, electrons are fast moving with respect to ions. For this reason, the electrons feel the ionic lattice frozen, and relax to their ground state for each ionic configuration. Thus, the full electron-ion wave-function $\ket\Psi$ can be factorized in a ionic and electronic part:
\begin{equation}
    \braket{\bR,\br|\Psi} = \braket{\bR|\psi_{\text{ion}}} \braket{\br|\psi_{\text{el}}[\bR]}
\end{equation}
where $\br$ are the electron positions, $\ket{\psi_{\text{ion}}}$ is the wave-function of ions and $\ket{\psi_{\text{el}}[\bR]}$ is the electronic ground state with the ions fixed in the $\bR$ position.

In the whole manuscript, we use bold fonts to indicate vectors or matrices; products between them are the standard rows-by-columns product. For quantum operators and wave-functions, we use the Dirac notation: we use a hat $\hat{\cdot}$ to distinguish quantum operators from real numbers. 

In this work, we focus on physical properties that depend only on ions. For this reason, we will drop the $_{\text{ion}}$ index from the wave-function, and $\ket{\psi}$ always refers to the nuclear wave-function $\ket{\psi_{\text{ion}}}$. Within BO approximation, the ionic wave-function obeys the Schroedinger equation with a BO Hamiltonian $\Hbo$:
\begin{equation}
    \Hbo = \sum_{a = 1}^{3N} \frac{\hat{p_a}^2}{2m_a} + V(\hat \bR)
\end{equation}
where $\hat p_a$ is the momentum operator of nucleus $a$, $m_a$ the mass of the $a$-th atom. 
To use a compact notation, each index indicates both atomic and Cartesian components, so, if not differently specified, it ranges from 1 to $3N$ ($N$ is the number of atoms). 
The BO potential $V(\bR)$ is the ionic energy landscape, obtained as the ground state energy of the electronic problem with ions fixed in positions $\bR$.
The BO energy landscape $V(\bR)$ is usually obtained \emph{ab initio} by solving the electronic problem with fixed nuclei within DFT or HF approximations. 

This work aims to describe how (nuclear) physical properties change in time when the system in equilibrium is perturbed by an external probe. This is the typical setup for any experiment: when $t < t_0$, the system is in equilibrium. At $t = t_0$, it starts interacting with an external time-dependent perturbation $\Vext(\bR, t)$. We are interested in the expectation value of a generic nuclear observable $\hat \Acal$ at time $t > t_0$.

At $T = \SI{0}{\kelvin}$, the nuclear wave-function in equilibrium satisfies the static Schroedinger equation:
\begin{equation}
    \Hbo \ket\psi = E_{\text{GS}} \ket\psi
\end{equation}

The presence of a time-dependent perturbation for $t \ge t_0$ introduces a time dependency in the nuclear wave-function $\ket{\psi(t)}$.
Thus, the quantum expectation value of the $\hat \Acal$ observable at time $t$ is:
\begin{equation}
    \Acal(t) =  \braket{\psi(t) | \hat \Acal | \psi(t)}
\end{equation}

The Schroedinger equation  governs the time-evolution of $\ket{\psi(t)}$:
\begin{equation}
i\hbar \frac{d}{dt}\ket{\psi(t)} = \Htd \ket{\psi(t)}.
\label{eq:schroedinger}
\end{equation}
where the $\Htd$ time-dependent Hamiltonian is:
\begin{equation}
    \Htd = \Hbo + \Vext(\hat \bR, t),\label{eq:H:t}
\end{equation}
and \eqname~\eqref{eq:schroedinger} is solved with the initial condition that the wave-function at $t = t_0$ is in the ground state:
\begin{equation}
    \ket{\psi(t_0)}= \ket \psi.
\end{equation}

At finite temperature ($T > \SI{0}{\kelvin}$), we just need to replace the wave-function with a density matrix $\hat \rho$ that can describe also mixture of states.
At equilibrium (when $t \le t_0$), the density matrix is ${\hat \rho}$:
\begin{equation}
    {\hat \rho} = \sum_{i} p_i \ket{\psi_i}\bra{\psi_i}.
\end{equation}
where the $\ket{\psi_i}$ are the eigenstates of the $\Hbo$ Hamiltonian with $E_i$ energy, and $p_i$ are the Boltzmann occupations ($k_b$ is the Boltzmann constant and $T$ the temperature):
\begin{equation}
    \Hbo \ket{\psi_i} = E_i \ket{\psi_i}
    \label{eq:schroedinger:static}
\end{equation}
\begin{equation}
    p_i = \frac{e^{-\beta E_i}}{Z} \qquad Z = \sum_i e^{-\beta E_i}\qquad
    \beta = \frac{1}{k_b T}
\end{equation}

After $t_0$, the system interacts with the external perturbation, and the density matrix depends on time $\hat \rho(t)$.
This time evolution is simply given by the evolution of each state in the mixture according to the Schroedinger equation:
\begin{equation}
    \hat \rho(t) = \sum_{i} p_i \ket{\psi_i(t)}\bra{\psi_i(t)}.
\end{equation}
Where $\ket{\psi_i(t)}$ are the same eigenstates of the BO Hamiltonian $\Hbo$ (\eqname~\ref{eq:schroedinger:static}) at $t = t_0$ evolving with the time dependent Hamiltonian $\Htd$.
In this work, we focus on isolated systems, therefore $p_i$ does not depend on time, and the density matrix satisfies the Liouville-Von Neumann equation:
\begin{equation}
    i\hbar \frac{d}{dt}\hat\rho (t) = \Htd \hat\rho(t) - \hat \rho(t) \Htd
    \label{eq:real:drho:dt}
\end{equation}

In this case, the average of the observable is:
\begin{equation}
     \Acal(t) = \Avgquantumt{\hat \Acal} = \Tr{\hat \rho(t) \hat \Acal}
    \label{eq:real:dA:dt}
\end{equation}

A case of particular interest is the linear regime (when the external perturbation does not change the status of the system, neither breaking chemical bonds, nor triggering a macroscopic rearrangement of atoms, nor heating the sample).
In this case the (small) external perturbation can be splitted in a time-independent coupling $\hat B= \Bcal(\hat \bR)$ between the external field and the ionic positions, and the time-envelope of the perturbation $\mathcal V(t)$:
\begin{equation}
    \Vext(\hat \bR, t) = V^{(1)}(\hat \bR, t) =  \Bcal(\hat \bR) \mathcal V(t)
    \label{eq:Vperturb}
\end{equation}
\begin{equation}
    \mathcal V(t) = 0 \qquad \text{for}\qquad t < t_0
\end{equation}
In this case, the time-dependence of an observable is given by the response function\cite{MahanBook}:
\begin{equation}
    \Acal(t) = \Acal(t_0) + \int_{-\infty}^\infty \chi_{\Acal\Bcal}(t - t') \mathcal{V}(t') \, dt'
    \label{eq:responset}
\end{equation}
The response function $\chi_{\Acal\Bcal}(t)$ is directly related to time-correlation function through the Kubo equation:
\begin{equation}
    \chi_{\Acal\Bcal}(t) = -\frac{i}{\hbar} \Avgquantumeq{[e^{\frac i \hbar \hat H t} \hat \Acal e^{-\frac i\hbar \hat H t}, \hat \Bcal]}\vartheta(t)
    \label{eq:kubo}
\end{equation}
Here, $e^{-\frac i\hbar \hat H t}$ is the time evolution operator in absence of the external perturbation and $\vartheta(t)$ is the Heaviside function.
In particular, the convolution of \eqname~\eqref{eq:responset} between $\chi_{\Acal\Bcal}$ and $\mathcal V$ becomes a simple product in the frequency domain:
\begin{equation}
    \Acal(\omega) - \Acal(t_0) = \chi_{\Acal\Bcal}(\omega) \mathcal V(\omega).
    \label{eq:chi:risp}
\end{equation}

The $\hat \Acal$ and $\hat \Bcal$ operators depend on the particular experiment we want to simulate. 

For example, to simulate IR spectra, we need to compute the dynamical nuclear correlation function $\chi_{MM}^{(ion)}(\omega)$, where $\hat \Acal = \hat \Bcal = M(\hat \bR)$, and $M(\hat\bR)$ is the dipole moment along the probe polarization of the system when nuclei are located in $\bR$. Analogously, the Raman signal is obtained measuring the energy exchanged between the incoming and outcoming radiation. This energy exchange occurs thanks to the polarizability tensor $\boldsymbol{\alpha}(\bR)$ of the system caused by the displacements in the ionic position $\bR$, induced by the probe. In this case, $\hat \Acal = \hat \Bcal = \alpha_{xy}(\hat \bR)$, where the $xy$ directions are the incoming and outcoming polarization of light\cite{Lazzeri_2003}.
In the same way, the dynamical structure factor and the thermal conductivity are dynamical nuclear correlation functions\cite{Simoncelli2019}.

Indeed, we can also simulate experiments beyond the linear regime, in which the system interacts with multiple probes, as in impulsive vibrational spectroscopy\cite{Monacelli_2017}. Here, the system is perturbed with two pulses: a first optical pulse at $t = t_0$ brings the system out-of-equilibrium, then, the probe pulse interacts with the sample at time $t_1 > t_0$ measuring its polarizability. 




\section{Time-Dependent Self-Consistent Harmonic Approximation}
\label{sec:tdscha}

The numerical simulation of the exact quantum dynamics requires the solution of \eqname~\eqref{eq:real:drho:dt} and \eqname~\eqref{eq:real:dA:dt}. This is computationally unfeasible for a system with more than few atoms, as the memory required to store the density matrix grows as $M^{(3N)^2}$, where $M$ is the dimension of the basis of the wave-function and $N$ the number of atoms in the simulation cell. The linear regime, also, requires the calculation of the time-correlation function in \eqname~\eqref{eq:kubo}, that involves the full diagonalization of the static interacting Hamiltonian $\Hbo$.

Here, we derive an approximate theory for the time evolution of the nuclear density matrix. Our theory allows simulating the time evolution of systems with hundreds of atoms with the BO energy landscape $V(\hat \bR)$ calculated \emph{ab initio} from the solution of the electronic problem.

In \secname~\ref{sec:scha} we revise the SCHA theory, the equilibrium solution of our dynamical equations.

\subsection{Static Self-Consistent Harmonic Approximation}
\label{sec:scha}

The SCHA is a mean-field theory developed to deal with interacting phonons and to compute thermodynamic properties of solids. 

Here, the exact interacting many-body ionic density matrix is replaced with a trial one, Gaussian in real space.
Starting from now through the rest of the paper, $\hrhoo$ is the SCHA equilibrium density matrix. As done in Hartree-Fock for electrons, in the SCHA, the $\hrhoo$ density matrix minimizes the free energy functional of a trial density matrix $\hat {\tilde \rho}$:
\begin{equation}
    F[\hrhoo] = \min_{\hat{\tilde\rho}} F[\hat{\tilde\rho}].
\end{equation}
\begin{equation}
    F[\hat {\tilde\rho}] = \Avg{\hat H}{\tilde{\hat\rho}} - T S[\hat{\tilde\rho}]
    \label{eq:free:energy},
\end{equation}
where $S[\hat {\tilde\rho}]$ is the entropy functional:
\begin{equation}
    S[\hat {\tilde\rho}] = -k_b\Tr{\hat {\tilde\rho} \ln \hat {\tilde\rho}}.
\end{equation}

The $\hat {\tilde\rho}$ is restricted to the most general Gaussian:
\begin{align}
    \braket{\bR|\hat{\tilde \rho}|\bR'} = \norm\exp&\bigg[ - \sum_{ab} \frac{\Theta_{ab}}{4}(R_a - \Rcal_a)(R_b - \Rcal_b) + \nonumber \\
    & -
    \sum_{ab} \frac{\Theta_{ab}}{4}(R_a' - \Rcal_a)(R_b' - \Rcal_b) + \nonumber \\
    &+ \sum_{ab} A_{ab}(R_a - \Rcal_a)(R_b' - \Rcal_b')\bigg]
    \label{eq:rho:scha}
\end{align}

The parameters that uniquely determine $\hat {\tilde \rho}$ are the vector $\bRcal$, the average ionic positions, and the real Hermitian matrices $\bTheta$ and $\bA$, the quantum and thermal fluctuations around the average positions, respectively. The $\norm$ factor is the normalization of the density matrix. The $\bTheta$ and $\bA$ matrices are not independent: they commute and are constrained so that $\hat{\tilde\rho}$ can be normalized. This condition is obtained if the real symmetric matrix $\bUps$ has only positive eigenvalues:
\begin{equation}
    \bUps = \bTheta - 2\bA
\end{equation}

Differently from the originally conceived SCHA\cite{Errea2014}, \eqname~\eqref{eq:rho:scha} is more general. Here, we optimize the free energy among all possible static Gaussian density matrices. 
\eqname~\eqref{eq:rho:scha} includes density matrices where each normal mode is thermally populated by a different auxiliary temperature, not necessarily the true one. However, as we prove in \appendixname~\ref{app:equilibrium}, in the minimum of the free energy, all the modes are always populated by the exact temperature, and the result coincides with ref.\cite{Errea2014}. The additional degrees of freedom on the off-diagonal elements of the density matrix are important to extend the theory out-of-equilibrium.

The SCHA is solved by substituting the expression of the trial density matrix $\hat {\tilde\rho}$ (\eqname~\ref{eq:rho:scha}) into the free energy functional (\eqname~\ref{eq:free:energy}), and minimizing with respect to $\bRcal$, $\bTheta$ and $\bA$.

The SCHA equilibrium density matrix $\hrhoo$
 satisfies the self-consistent equation:
\begin{equation}
    \hrhoo = \frac{\exp\left(-\beta \Hcal[\rhoo]\right)}{Z[\rhoo]},
\end{equation}
where $\Hcal[\rhoo]$ is a harmonic Hamiltonian that depends self-consistently on the nuclear equilibrium density $\rhoo(\bR)$:
\begin{equation}
    \rhoo(\bR) = \braket{\bR|\hrhoo|\bR}
\end{equation}

\begin{equation}
    \Hcal[\rhoo] = \sum_{i = 1}^{3N} \frac{{\hat p_i}^2}{2m_i} + \sum_{ij} \Avgclassiceq{\frac{\partial^2 V}{\partial R_i \partial R_j}} (\hat R_i - \Rc_i[\rhoo])(\hat R_j - \Rc_j[\rhoo]).
    \label{eq:H:scha}
\end{equation}
The functional $\Rc_i[\rhoo]$ is the average position of the $i$-th atom, and coincides with the $\Rc_i$ solution of the SCHA:
\begin{equation}
    \rschatrial[\rhoo] = \int  d\bR \;\rhoo(\bR) \bR,\label{def:Rci}
\end{equation}
and the averages are computed with the $\rho(\bR)$ probability density
\begin{equation}
    \Avgclassiceq{\frac{\partial ^2V}{\partial R_a \partial R_b}} = \int d\bR \rhoo(\bR) \frac{\partial ^2V(\bR)}{\partial R_a \partial R_b}.
\end{equation}
When $\hat{\tilde\rho}$ is the equilibrium solution $\hrhoo$, we have:
\begin{equation}
    A_{ab} = \sqrt{m_am_b}\sum_{\mu} \frac{2\omega_\mu n_\mu (n_\mu + 1)}{\hbar (2n_\mu + 1)} e_\mu^a e_\mu^b
    \label{eq:static:A}
\end{equation}
\begin{equation}
    \Upsilon_{ab}= \sqrt{m_a m_b}\sum_{\mu} \frac{2\omega_\mu}{\hbar(2n_\mu + 1)} e_\mu^a e_\mu^b
    \label{eq:static:Y}
\end{equation}
Where $\boldsymbol{e_\mu}$ and $\omega_\mu$ are the normal modes and frequencies of the self-consistent harmonic Hamiltonian $\Hcal[\rho]$, and $n_\mu$ is the Bose-Einstein occupation number:
\begin{equation}
    n_\mu = \frac{1}{ e^{\beta\hbar\omega_\mu} - 1}
    \label{eq:static:bose}
\end{equation}


A particular case is the $T = \SI{0}{\kelvin}$ limit, when the equilibrium SCHA density matrix is a pure state. Here, the SCHA equilibrium pure state $\ket\psi$
is the ground state of the self-consistent Hamiltonian $\Hcal[\rho]$:
\begin{equation}
    \Hcal[\rho] \ket \psi = E_{GS} \ket\psi
\end{equation}
and the density $\rho(\bR)$ is:
\begin{equation}
    \rho(\bR) = \braket{\bR | \psi}\braket{\psi|\bR} =  \left|\braket{\bR|\psi}\right|^2
\end{equation}
The pure state is a Gaussian wave-packet:
\begin{equation}
    \braket{\bR|\psi} = \norm^\frac 12 \exp\left[ - \frac 14 \sum_{ab}\Upsilon_{ab}(R_a - \Rcal_a)(R_b - \Rcal_b)\right]
    \label{eq:psischa:t0}
\end{equation}

\subsection{Dynamics of a pure quantum state}

The SCHA introduced in \secname~\ref{sec:scha} is a static theory: it cannot describe dynamical properties, like phonons observed experimentally.

In this section, we derive a new theory, the Time-Dependent Self-Consistent Harmonic Approximation (TD-SCHA), to describe correctly, without any empirical approximation, the response of the system to any (small or not) external time-dependent probe that interacts with ions.

We start from a pure state, i.e. the equilibrium solution at $T = \SI{0}{\kelvin}$. At time $t = t_0$, we switch on a perturbation $\Vext(\bR, t)$ and the overall Hamiltonian becomes $\Htd(t)$ defined in \eqname~\eqref{eq:H:t}.

We constrain the wave-packet to the most general time-dependent Gaussian:
\begin{align}
    &\braket{\bR|\psi(t)}  = \norm^{\frac 12}(t) \exp\bigg(
    i \sum_a Q_a(t)[R_a - \Rc_a(t)] + \nonumber \\ 
    & - \sum_{ab}\left[\frac{\Theta_{ab}(t)}{4} - iC_{ab}(t)\right][R_a - \Rc_a(t)][R_b - \Rc_b(t)]\bigg).
    \label{eq:psi:t0}
\end{align}

In \eqname~\eqref{eq:psi:t0} we have two new parameters with respect to the static solution of \eqname~\eqref{eq:psischa:t0}: $\bQ$ and $\bC$. They add a complex phase to our wave-packet, and represent the momentum of the $\bR$ and $\bTheta$ variables. 
The time-dependency of the $\bQ$, $\bRcal$, $\bTheta$, and $\bC$ parameters is found by minimizing the Dirac action along the time-evolution path:
\begin{equation}
    A = \frac{1}{t - t_0} \int_{t_0}^{t}\braket{\psi(t') | \hat H_{\text {td}}(t') - i\hbar \frac{d}{dt'} |\psi(t')} \, dt'
    \label{eq:action}
\end{equation}

The full equations of motion are derived in \appendixname~\ref{app:least:action}.

It can be proved (as we show in \appendixname~\ref{app:least:action}) that the same equations of motion are obtained if \eqname~\eqref{eq:psi:t0} is evolved by a self-consistent Schroedinger equation:
\begin{equation}
    i\hbar \frac{d}{dt} \ket {\psi(t)} = \Hcal[\rho(t)] \ket {\psi(t)}
    \label{eq:tdscha:t0}
\end{equation}

The self-consistent Hamiltonian that defines the time-evolution in the TD-SCHA is:
\begin{align}
    \Hcal[&\rho(t)] = \sum_{a = 1}^N \frac{{\hat p_a}^2}{2m_a} + \sum_a \Avgclassict{\frac{\partial \Vtot}{\partial R_a }} (\hat R_a - \Rc_a[\rho(t)]) +  \nonumber\\
    & + \sum_{ab}\Avgclassict{\frac{\partial^2 \Vtot}{\partial R_a \partial R_b}}(\hat R_a - \Rc_a[\rho(t)])(\hat R_b - \Rc_b[\rho(t)]),\label{eq:h:tdscha}
\end{align}
where $\Vtot(\bR, t)$ is the total potential: the BO energy landscape $V(\bR)$ plus time-dependent external potential $\Vext(\bR, t)$
\begin{equation}
    \Vtot(\bR, t) = V(\bR) + \Vext(\bR, t).
\end{equation}
The $\Hcal[\rho(t)]$ Hamiltonian depends on the $\rho(\bR, t)$ probability distribution of finding the ions in the $\bR$ configurations at time $t$:
\begin{equation}
    \rho(\bR, t) =\left| \braket{\bR |\psi(t)}\right|^2
    \label{eq:rho:t}
\end{equation}

The time-dependent self-consistent Hamiltonian
(\eqname~\ref{eq:h:tdscha}) has one extra linear term in $\hat \bR$ compared to the static one (\eqname~\ref{eq:H:scha}). This extra linear term is zero when the self-consistency of the equilibrium SCHA is achieved, as the average of the derivative of the BO potential (forces) on the equilibrium SCHA distribution is a necessary condition for the SCHA self-consistency\cite{Errea2014}.

The nuclear self-consistent Schroedinger equation (\eqname~\ref{eq:tdscha:t0}) has the same shape of other mean-field theories for electrons, as TD-HF or TD-DFT.
It is worth noticing that \eqname~\eqref{eq:tdscha:t0} minimizes the action only if the wave-function is a Gaussian wave-packet. Notably, as we show in the next section, if $\ket\psi$ is Gaussian, \eqname~\eqref{eq:tdscha:t0} is a closed equation: a Gaussian wave-packet evolving in a general self-consistent harmonic Hamiltonian keeps its Gaussian form. 

The Dirac least-action principle (\eqname~\ref{eq:action}) and the self-consistent Schroedinger equation (\eqname~\ref{eq:tdscha:t0}) are equivalent as they lead to the same dynamics.

\subsection{Dynamics of a mixture of states}
\label{sec:gaussian}
\eqname~\eqref{eq:tdscha:t0} describes the dynamics just of pure states. The equilibrium solution of the SCHA is a pure state only if $T = \SI{0}{\kelvin}$.

We can derive the nuclear time-dependent evolution of a mixture of states by replacing the time-dependent Schroedinger equation (\eqname~\ref{eq:tdscha:t0}) with the Liouville-von Neumann equation (\eqname~\ref{eq:real:drho:dt}), as usually done in TD-DFT\cite{Li1985} and TD-HF:
\begin{equation}
    i\hbar \frac{d}{dt} \hat \rho(t) = \Hcal [\rho(t)]\hat \rho(t) - \hat \rho(t) \Hcal[\rho(t)],
    \label{eq:tdscha:ft}
\end{equation}
\begin{equation}
    \rho(\bR, t) = \braket{\bR | \hat\rho(t) | \bR},
\end{equation}
where $\Hcal[\rho(t)]$ is given by \eqname~\eqref{eq:h:tdscha}.

Thanks to \eqname~\eqref{eq:tdscha:ft}, we can describe the dynamics also of mixtures of states, starting from the equilibrium SCHA solution at any temperature. Moreover, as \eqname~\eqref{eq:tdscha:t0}, this is a closed equation for a Gaussian wave-packet (as we show in this section). We also prove that \eqname~\eqref{eq:tdscha:ft} correctly conserves both the energy and the entropy, as expected from the correct evolution of an isolated quantum system.

As we did for the pure state (\eqname~\ref{eq:psi:t0}), we can represent explicitly the Gaussian density matrix:
\begin{align}
    \bra{\bR'}\hat\rho&(t)\ket{\bR}  = \norm(t) \exp\bigg( 
     i \sum_a Q_a(t)(R_a' - R_a) \nonumber+ \\ 
     &- \sum_{ab} \left[ \frac{\Theta_{ab}(t)}{4} - i C_{ab}(t)\right] [R_a - \Rc_a(t)][R_b - \Rc_b(t)] + \nonumber\\
     &-
    \sum_{ab} \left[ \frac{\Theta_{ab}(t)}{4} + iC_{ab}(t)\right] [R'_a - \Rc_a(t)][R'_b - \Rc_b(t)]\nonumber \\ 
    & + \sum_{ab} A_{ab}(t) [R_a - \Rc_a(t)][R_b' - \Rc_b(t)] \bigg).\label{eq:density:matrix}
\end{align}
In addition to the parameters already introduced for the pure state (\eqname~\ref{eq:psi:t0}), the time-dependent density matrix (\eqname~\ref{eq:density:matrix}) has one more parameter: the  $\bA(t)$ complex Hermitian matrix.
$\Rc_i$ is the average position of the $i$-th atom and coincides with the definition of \eqname~\eqref{def:Rci}. The variable $Q_i$ is a linear phase modulation; multiplied by $\hbar$, it represents the momentum of the $i$-th atom. In a flat potential (where $\Hcal$ only contains the kinetic operator) the $i$-th atom average position drifts with constant velocity $v_i = \frac{\hbar Q_i(t)}{m_i}$. 
The $\bm Q$ and $\rschatrial$ variables are similar to those of a classical molecular dynamics. The $\bTheta$, $\bA$ matrices describe the quantum and thermal fluctuations. In particular, by looking at the diagonal elements of the density operator (the density distribution, \eqname~\ref{eq:rho:t}), we get the covariance matrix of quantum-thermal fluctuations:
\begin{equation}
    \frac{\rho(\bR, t)}{\norm(t)} =\exp\left\{ -\sum_{ab}\frac 12\Upsilon_{ab}(t)[R_a - \Rc_a(t)][R_b - \Rc_b(t)]\right\}
\end{equation}
\begin{equation}
     \Upsilon_{ab}(t) = \Theta_{ab}(t) - 2\Re A_{ab}(t),
     \label{eq:upsilon}
\end{equation}
where $\Re{}$ and $\Im{}$ identify the real and imaginary part. 
$\bUps$ is the inverse the covariance matrix of the Gaussian distribution:
\begin{equation}
    \left(\bUps^{-1}\right)_{ab}(t) = \Avgclassict{[R_a - \Rcal_a(t)][R_b - \Rcal_b(t)]}.
\end{equation}
We give intuitive picture on the physical meaning of the parameters. The $\bTheta$ matrix encodes pure quantum fluctuations and $\Re \bA$ the thermal ones: if $\bA = 0$, \eqname~\eqref{eq:density:matrix} is a pure state and we recover \eqname~\eqref{eq:psi:t0}.
$\bC$ is a quadratic phase and represents the \emph{chirp} along the quantum fluctuations. Its role is very similar to the chirp in signal propagation, and represent a gradient in the speed of particles in different positions in the wave-packet, as discussed in ref.\cite{Monacelli_2017}.
On the other side, the $\Im \bA$ plays the role of the momentum for the thermal fluctuations. It is nonzero only when there are more than 1 degree of freedom ($\Im \bA$ is anti symmetric) and if the system is not in a pure quantum state.

The $\norm(t)$ is the density matrix normalization:
\begin{equation}
    \norm(t) = \sqrt{\frac{\det{\bUps(t)}}{(2\pi)^{3N}}}
    \label{eq:norm}
\end{equation}

\eqname~\eqref{eq:density:matrix} can be substituted in \eqname~\eqref{eq:tdscha:ft} to get the dynamical equations for the parameters. 
For a convenient choice of the notation, it is better to express the parameters rescaled by the masses. We indicate with a $\tilde{\cdot}$ the rescaled matrices and vectors as:
\begin{subequations}
\begin{equation}
\tilde C_{ab}(t) = \frac{C_{ab}(t)}{\sqrt{m_a m_b}}, \qquad
\tilde A(t)= \frac{ A_{ab}(t)}{\sqrt{m_am_b}},
\end{equation}
\begin{equation}
\tilde \Upsilon_{ab}(t) = \frac{\Upsilon_{ab}(t)}{\sqrt{m_a m_b}}, \qquad
\tilde \Theta(t)= \frac{ \Theta_{ab}(t)}{\sqrt{m_am_b}},
\end{equation}
\begin{equation}
\tilde \Rc_a(t) = \sqrt m_a \Rc_a(t)\qquad 
\tilde Q_a(t) = \frac{Q_a(t)}{\sqrt m_a}
\end{equation}
\label{eq:mass:rescaled}
\end{subequations}

The final equations of motion are:
\begin{subequations}
\begin{equation}
\frac{d\tilde\bRcal}{dt} = \hbar \tilde\bQ \qquad
\frac{dQ_a}{dt} = \frac{\Avgclassict {\ftot_a}}{\hbar}\label{eq:newton}
\end{equation}
\begin{equation}
\frac{d\tilde \bUps}{dt} = \hbar\left[\tilde \bUps(2 \tilde \bC  + \Im \tilde \bA) + (2\tilde \bC - \Im\tilde \bA)\tilde \bUps\right] 
\label{eq:ups:dyn}
\end{equation}
\begin{equation}
\frac{d \Re \tilde \bA}{dt} =\frac{\hbar}{2}\left(
4\tilde \bC \Re\tilde \bA + 4\Re\tilde \bA \tilde \bC - \tilde \bTheta \Im \tilde \bA + \Im\tilde \bA \tilde \bTheta\right)
\end{equation}
\begin{equation}
    \frac{d\Im \tilde \bA}{dt} = 
    \frac{\hbar}{2} \left(
    4\tilde \bC \Im\tilde \bA + 4\Im\tilde \bA \tilde \bC + 
    \tilde\bTheta \Re \tilde \bA - 
    \Re\tilde \bA \tilde\bTheta\right)
\label{eq:dia:dt}
\end{equation}
\begin{equation}
    \frac{d\tilde \bC}{dt} = \frac{1}{2\hbar} \Avgclassict{\frac{\partial^2 \Vtot}{\partial \tilde \bR\partial\tilde \bR}} + \frac{\hbar}{2}\left[
    4 \tilde \bC^2 - \frac 1 4 \tilde \bTheta^2 + \Re (\tilde \bA \tilde \bA^\dagger)\right].
    \label{eq:C:dyn}
\end{equation}
\label{eq:dynamics}
\end{subequations}
Here, we dropped the explicit time-dependency of these variables for brevity: they represent the $\hat\rho(t)$ density matrix and not the equilibrium one.

The products of matrices is the standard rows-by-columns. The symbol ${\cdot}^\dagger$ after a matrix indicates the Hermitian conjugate, and $\Avgclassict{\ftot_a}$ is the average of the total force (BO force plus the time-dependent external potential) acting on the $a$-th atom:
\begin{equation}
    \ftot_a(\bR, t) = -\frac{\partial \Vtot(\bR, t)}{\partial R_a},
\end{equation}
while the derivative with respect the $\tilde\bR$ variable indicates the mass rescale:
\begin{equation}
    \frac{\partial^2 \Vtot}{\partial \tilde R_a \partial \tilde R_b} = \frac{1}{\sqrt{m_am_b}} \frac{\partial^2 \Vtot}{\partial R_a \partial R_b}.
\end{equation}

\eqname~\eqref{eq:newton} are the semi-classical equation of motion: they resemble the Newton dynamics, but the force is averaged on the ionic probability distribution.
The dynamics preserves $\bUps$, $\Re \bA$ and $\bC$ symmetric and $\Im \bA$ antisymmetric. The details of the derivation of \eqname~\eqref{eq:dynamics} is reported in appendix~\ref{app:full:dynamical}. These equations reduce to the evolution of the pure state that minimizes the Dirac action, if we set $\bA = 0$ (see \appendixname~\ref{app:least:action}).

Notably, substituting the Gaussian wave-packet \eqname~\eqref{eq:density:matrix} into TD-SCHA equation of motion \eqname~\eqref{eq:tdscha:ft}, we get an extra condition on the time-dependency of the $\norm(t)$ parameter.
\begin{equation}
    \frac{d\norm(t)}{dt} = 2\hbar \norm(t) \Tr{\tilde \bC(t)}\label{eq:norm:evo}
\end{equation}

This condition is automatically satisfied by \eqname~\eqref{eq:dynamics} if we substitute \eqname~\eqref{eq:norm} into \eqname~\eqref{eq:norm:evo}. Since \eqname~\eqref{eq:norm:evo} does not depend explicitly on $\hat \bR$ operators and it is automatically satisfied by the equation of motions, the Gaussian wave-packet evolution is closed with the TD-SCHA equation (\eqname~\ref{eq:tdscha:ft}). In fact, the application of the Liouville operator on a Gaussian density matrix:
\begin{equation}
\Hcal[\rho(t)]\hat \rho(t) - \hat \rho(t) \Hcal[\rho(t)]
\label{eq:liouville:op}
\end{equation}
with $\Hcal[\rho(t)]$ Harmonic, generates a polynomial of the same order than the time derivative for the density matrix $i\hbar \frac{d}{dt} \hat\rho(t)$.
This means that, if the density matrix is Gaussian at $t = t_0$ (equilibrium), it remains Gaussian for the whole time-evolution, as \eqname~\eqref{eq:liouville:op} does not provide any term shifting $i\hbar \frac{d}{dt} \hat\rho(t)$ from a Gaussian.

Very interestingly, the full dynamics is just obtained by standard rows-by-columns product of small matrices (they are $3N\times 3N$, with $N$ the number of atoms in the simulation cell). The only two quantities depending from the physical system (the real BO Hamiltonian $\hat H$) are
\begin{equation}
    \Avgclassict{\bftot} \qquad
    \Avgclassict{\frac{\partial^2 \Vtot}{\partial \tilde \bR \partial \tilde\bR}}.
\end{equation}
The calculation of these averages is also needed for a static SCHA calculation and can be computed stochastically as described in \cite{Errea2014, Bianco2017}.

In particular, an efficient method to compute the average of the second derivatives of the BO potential is obtained exploiting the methodology introduced in ref.\cite{Bianco2017}, that takes advantage of integration by parts:
\begin{equation}
    \Avgclassict{\frac{\partial^2 \Vtot}{\partial  R_a\partial  R_b}} = \sum_{p} \Upsilon_{ap} \Avgclassict{(R_p - \Rcal_p)f_a}
\end{equation}
In this way, only \emph{ab initio} forces are required.
Therefore, the implementation of the TD-SCHA equations has the same overall computational cost as a static calculation.

Different kinds of Gaussian wave-packet dynamics are discussed in literature\cite{Huber1987,Huber1988,Pal2016,Haegeman2011,Guaita2019,Hackl2020}. For example, refs.\cite{Haegeman2011,Guaita2019, Hackl2020} project the real dynamics into the manifold of Gaussian states, while refs.\cite{Huber1987,Huber1988,Pal2016} exploit the Wentzel-Kramers-Brillouin (WKB) method to derive semiclassical equations expanding the Schroedinger equation linearly around $\hbar = 0$. 

We can prove that the TD-SCHA equation of motion satisfy both energy and entropy conservation.
The total energy is computed as the average of the time-dependent Hamiltonian on the time-dependent density matrix:
\begin{equation}
    E(t) = \Avgquantumt{\Htd} = \tr\left[ \hat \rho(t)  \Htd\right].
\end{equation}
In \appendixname~\ref{app:energy:cons}, we prove that
\begin{equation}
    \frac{dE}{dt} = \Avgclassict{\frac{d\Vext(\bR, t)}{dt}}.
\end{equation}
This states the energy conservation in absence of an external time-dependent perturbation (that can transfer energy to the system).

Similarly, it is possible to prove that also entropy is conserved during the dynamics. This is consequence of the reversibility of the quantum dynamical equations in a closed quantum system.
This is a general feature of any Hamiltonian dynamics, and the TD-SCHA makes no exception, even if the Hamiltonian depends self-consistently from the density matrix. The entropy defined on the many-body density matrix is:
\begin{equation}
    S[\hat\rho(t)] = - k_b \tr \left[\hat\rho(t)\log\hat\rho(t)\right]
\end{equation}
and in \appendixname~\ref{app:entropy:cons} we show how:
\begin{equation}
    \frac{dS}{dt} = 0.
\end{equation}

The entropy conservation derives from the unitary time evolution: we are evolving a closed quantum system and there is no de-coherence in the dynamics. In other words, the dynamics is reversible, as if we change the initial sign of $\Im \bA$, $\bC$, and $\bQ$, the evolution proceeds backward in time.

\subsection{Steady-states and equilibrium}
\label{sec:steady:state}
Thanks to \eqname~\eqref{eq:tdscha:ft}, the steady state solution of the dynamical equations occurs when the density matrix $\hat\rho$ commutes with the self-consistent Hamiltonian $\Hcal[\rho]$. This means that there is a basis that simultaneously diagonalizes both the Hamiltonian $\Hcal[\rho]$ and the density matrix.
This condition can also be inferred from the equation of motions \eqname~\eqref{eq:dynamics}, imposing that the time-derivatives are zero.
In this case we have:
\begin{equation}
    C_{ab} = 0 \qquad 
    \Im A_{ab} = 0 \qquad 
    Q_a = 0\label{eq:stationary}
\end{equation}

This is quite intuitive, as discussed in \secname~\ref{sec:gaussian}, these variables are related to the instantaneous average momentum and chirp.
The steady state solution obtained is not the equilibrium SSCHA result (see \appendixname~\ref{app:steady:state}). 
In particular, the steady-state solutions are equal to the product of equilibrium noninteracting quantum Harmonic oscillators, where each normal mode has a thermal occupation number $n_\mu$ with a temperature that depends on the mode. This is not, indeed, the equilibrium solution of the SCHA, which requires all normal modes populated by the same temperature.
However, not all steady-states are equilibrium solutions: if we prepare the system in a mixture of state with eigenstates of the Hamiltonian, the exact time-dependent density matrix is stationary even if the occupation probabilities are not the Boltzmann factors.  



The system reaches equilibrium if we introduce an interaction with an external bath (or a dephasing mechanism). In an isolated system, the equilibrium solution is the one that maximizes the entropy among all possible steady states at fixed energy.
Maximizing the entropy fixing the energy is equivalent in minimizing the Helmholtz free energy. This is the starting point for the static SCHA. In \appendixname~\ref{app:equilibrium} we prove that this condition correspond imposing an uniform $\beta$ on each mode.

Therefore, we recover the SCHA as the stationary solution of the TD-SCHA that maximizes the entropy. It is worth noting that the TD-SCHA will not spontaneously evolve into the SCHA solution, unless the equations are modified to account for the coupling with a reservoir that provides a mechanism for quantum decoherence\cite{Breuer2007}, allowing the entropy of the subsystem to increase. The extension of TD-SCHA to describe the dynamics of an open quantum system is beyond the scope of the current work.

\section{Linear response theory}
\label{sec:linear:response}
Almost all experimental data are collected by probing the response of the system to a time-dependent external perturbation. This perturbation could be either electromagnetic radiation (static electric fields, IR, optical light, X-ray) or particles like electrons and neutrons.
If the perturbation does not provide enough energy to heat the system, we are in the 
linear regime\cite{Kohn1957,PribramJones2016}.
Typical experiments that involve interactions with ionic degrees of freedom are Raman and IR spectroscopy, neutron, and X-Ray scattering.
In this section, we present the dynamical linear response of the TD-SCHA equations on top of the static SCHA solution. This enables the computation of the response function for any experiment probing the nuclear motion fully \emph{ab initio} and considering both quantum/thermal fluctuations and anharmonicity beyond perturbation theory.

Since the probe heating of the sample (Joule effect) is a higher-order process, the linear response does not depend on the coupling with the thermal bath\cite{Kohn1957,PribramJones2016}. Therefore, even if the TD-SCHA introduced in this work describes closed quantum systems, the results we derive in this section are general and apply also to systems coupled with a bath.

We start from the thermodynamic equilibrium (the SCHA solution) and then we add ``small'' time-dependent external potential $V^{(1)}(\bR, t)$ that acts on the nuclei for $t \ge t_0$.
The density matrix $\hat\rho(t)$ is the equilibrium solution $\hat\rho^{(0)}$ plus a small perturbation $\rhoone(t)$:

\begin{equation}
\hat \rho(t) = \hat \rho^{(0)} + \rhoone(t)
\qquad
    \Vtot(\bR, t) = V(\bR) + V^{(1)}(\bR, t),
\end{equation}
We indicate with ${}^{(0)}$ the equilibrium quantities and ${}^{(1)}$ a small perturbation around the equilibrium SCHA solution.
We perform a linear expansion of the TD-SCHA equation around equilibrium (\eqname~\ref{eq:tdscha:ft}):
\begin{equation}
    i\hbar \frac{d}{dt} \rhoone(t) = \Lsc \rhoone(t) + \comm{ \Vsc(t)}{ \hat\rho^{(0)}}
    \label{eq:linear:start}
\end{equation}
where $\Lsc$ is the super-operator that describes the unperturbed (anharmonic) evolution according to the self-consistent Hamiltonian, and $\Vsc(t)$ is the interaction with the external potential $V^{(1)}(\bR, t)$: 
\begin{equation}
    \Lsc\rhoone(t) = \comm{\Hcal[\rho^{(0)}]}{ \rhoone(t)} + \comm{\Hcal[\rho^{(1)}(t)]}{ \hat \rho^{(0)}} \label{eq:Lsc}
\end{equation}
\begin{align}
    \Vsc(t)& = \frac 12 \sum_{ab}(\hat R_a - \Rcal_a^{(0)}) \Avgclassiceqz{\frac{d^2 V^{(1)}(t)}{dR_a dR_b}}(\hat R_b - \Rcal_b^{(0)}) + \nonumber \\
    & + \sum_a \Avgclassiceqz{\frac{dV^{(1)}(t)}{dR_a}} (\hat R_a - \Rcal_a^{(0)})\label{eq:harmonic:external},
\end{align}
the $\Hcal[\rho^{(0)}]$ and $\Hcal[\rho^{(1)}]$ are given by \eqname~\eqref{eq:h:tdscha} and we used the square brackets to indicate the commutator:
$$
\comm{\hat A}{\hat B} = \hat A \hat B - \hat B\hat A
$$

The $\Lsc$ is the free propagator for the interacting nuclei. The first commutator in \eqname~\eqref{eq:Lsc} describes the evolution of $\rhoone$ with the self-consistent harmonic Hamiltonian $\Hcal[\rho^{(0)}]$ computed with the equilibrium nuclear density. If we evolve the system only according to this term, the resulting dynamics are the same as a harmonic oscillator with frequencies and polarization vectors re-normalized by anharmonicity.  The second commutator, instead, accounts for how the self-consistent Hamiltonian changes with the density. This term gives phonons finite lifetimes. We discuss this more in detail in \secname~\ref{sec:ansatz} by computing the phonon Green function.

\eqname~\eqref{eq:linear:start} is very similar to the linear response in other self-consistent theories, as for electrons TD-DFT\cite{Rocca2008}. The main difference is that here we are dealing with phonons (that are bosons) and the fact that the external perturbation $V^{(1)}(\bR, t)$ does not act directly on the time evolution, but affects the equations as an external time-dependent harmonic potential $\Vsc(t)$ (\eqname~\ref{eq:harmonic:external}). This modification of the external potential has no impact if the perturbation has a linear or quadratic coupling with nuclear displacements. However, it can excite only up to two phonons simultaneously (it contains at most a quadratic dependency on the $\hat \bR$ operator), meaning that the theory does not account for the excitation of three or more phonons by the external perturbation. We deepen this discussion in \secname~\ref{sec:two:phonons}.

To solve the linear response theory we just need to pass in Fourier space, and we get:
\begin{equation}
    \rhoone(\omega) = (\hbar \omega - \Lsc)^{-1} \comm{\Vsc(\omega)}{\hat\rho^{(0)}}.
    \label{eq:linear:evolution}
\end{equation}

The $(\hbar\omega - \Lsc)^{-1}$ term in \eqname~\eqref{eq:linear:evolution} is the Green function, and describe the free evolution of the system. The poles of this function are the ionic excitation energies, i.e. the physical anharmonic phonon frequencies. These are different from the phonons obtained from the equilibrium self-consistent harmonic Hamiltonian $\Hcal[\rho^{(0)}]$.

\eqname~\eqref{eq:linear:evolution} is a very compact expression in the Hilber space. However, for a practical calculation of the linear response, is better to work in the restricted space of Gaussians for $\rhoone(\omega)$. In this way, instead of working in a infinite dimension Hilbert space of $N$ particles, we have a finite linear space of dimension of about $3N\times 3N$. This means that $\rhoone(t)$ is uniquely defined by the parameters of the time dependent density matrix (\eqname~\ref{eq:rho:t}):
\begin{subequations}
\begin{equation}
    \tilde \Rc_i(t) = {\tilde \Rc}_i^{(0)} + \tilde \Rc_i^{(1)}(t) \qquad
    \tilde Q_i(t) = \tilde Q_i^{(1)}(t)
\end{equation}
\begin{equation}
    \tilde \Upsilon_{ab}(t) = {\tilde \Upsilon}_{ab}^{(0)} + \tilde \Upsilon_{ab}^{(1)}(t) \qquad 
    \tilde C_{ab}(t) =  \tilde C_{ab}^{(1)}(t)
\end{equation}
\begin{equation}
 \tilde A_{ab}(t) = {\tilde  A}_{ab}^{(0)} + \tilde A_{ab}^{(1)}(t)
\end{equation}
\label{eq:perturbed:variables}
\end{subequations}  
\begin{equation}
    \rhoone(\omega) \coloneqq \bpm \tilde\bUps^{(1)}(\omega) \\ \tilde\bA^{(1)} (\omega) \\
    \tilde\bC^{(1)}(\omega)\\ 
    \tilde\bQ^{(1)}(\omega) \\
    \tilde\bRcal^{(1)}(\omega)\epm\label{eq:full:rhoone}
\end{equation}

Since \eqname~\eqref{eq:linear:evolution} is a linear equation in the density matrix, it corresponds to an analogous linear system for the vector of \eqname~\eqref{eq:full:rhoone}.
In this case, since we restrict to perturbations and responses depending only on the atomic positions, we can get rid of the momentum variables $\bQ^{(1)},\bC^{(1)}$, and $\Im \bA^{(1)}$ deriving in time \eqname~\eqref{eq:linear:start} and transform the linear system of first order differential equations in a smaller system of second-order (the details of the calculation are reported in \appendixname~\ref{app:linear:response}). 

\begin{equation}
    \bpm {\tilde\bUps}^{(1)}(\omega) \\
    \Re{\tilde\bA}^{(1)}(\omega) \\ 
    {\tilde\bRcal}^{(1)}(\omega)\epm = - (\omega^2 + {\mathcal L})^{-1}
    \bpm \bf_{\Upsilon}^{(1)} \\
    \bf_{\Re A}^{(1)} \\
    \bf_{\Rcal}^{(1)} \epm
    \label{eq:perturb:cart}.
\end{equation}

Here, $\omega^2$ comes from the second derivative in time. The $\mathcal{L}$ kernel and the $\bf$ vector represent the free (anharmonic) evolution and the coupling of the phonons with the bare perturbation, respectively, in the space of the parameter of the Gaussian.
The explicit expression of the $\mathcal L$ in the polarization basis of the equilibrium solution is reported in Appendix~\ref{app:full:system}.

The $\bf_\Upsilon^{(1)}$, $\bf_{\Re A}^{(1)}$, and $\bf_\Rc^{(1)}$ represent how the bare perturbation $V^{(1)}(\bR, t)$ enters in the equation of the motion. It is comes from $\comm{\Vsc(\omega)}{ \hat\rho^{(0)}}$ in \eqname~\eqref{eq:linear:evolution}. In the polarization basis (i.e. the eigenmodes of the equilibrium self-consistent harmonic Hamiltonian), they are:
\begin{subequations}
\begin{equation}
    {f_\Upsilon^{(1)}}_{\mu\nu} = \frac{1}{\hbar}\left(\frac{2\omega_\mu}{2n_\mu + 1} + \frac{2\omega_\nu}{2n_\nu + 1}\right) \Avgclassiceqz{\frac{\partial^2 V^{(1)}}{\partial \tilde R_\mu \partial \tilde R_\nu}},
\end{equation}
\begin{align}
    {f_{\Re A}^{(1)}}_{\mu\nu} = \frac{1}{\hbar}&\bigg(\frac{2(n_\mu + 1)n_\mu\omega_\mu}{2n_\mu + 1} + \nonumber\\
    & + \frac{2(n_\nu +1)n_\nu\omega_\nu}{2n_\nu + 1}\bigg) \Avgclassiceqz{\frac{\partial^2 V^{(1)}}{\partial \tilde R_\mu \partial \tilde R_\nu}}
\end{align}
\begin{equation}
    {f^{(1)}_\Rc}_\mu = - \Avgclassiceqz{\frac{\partial V^{(1)}}{\partial \tilde R_\mu}}
\end{equation}
\label{eq:perturbations}
\end{subequations}
where $\omega_\mu$ and $n_\mu$ refers to the frequencies and populations of the equilibrium SCHA self-consistent harmonic Hamiltonian ($\hat\Hcal[\rho]$), and the derivative in $\tilde R_\mu$ refers to a collective atomic displacement directed along the polarization mode associated to the $\omega_\mu$ eigenvalue.
We can introduce the Green function $\bG(\omega)$ as:
\begin{equation}
\bm G(\omega) = -\left(\omega^2 + \mathcal{L}\right)^{-1},
\label{eq:full:green:function}
\end{equation}
we can compute the general linear response of the system to any external perturbation on nuclei $V^{(1)}(\bR, \omega)$
\begin{equation}
\bpm {\tilde\bUps}^{(1)}(\omega) \\
\Re{\tilde\bA}^{(1)}(\omega) \\
{\tilde \bRcal}^{(1)}(\omega) \epm = 
    \bm G(\omega) \bpm 
    \bf^{(1)}_\Upsilon(\omega) \\ 
    \bf^{(1)}_{\Re A}(\omega) \\ 
    \bf^{(1)}_\Rcal (\omega)\epm.
    \label{eq:response}
\end{equation}

\subsection{The general response function}
\label{sec:response}
In this section, we derive the general expression of the response function $\chi_{\Acal\Bcal}(\omega)$ within the TD-SCHA for any couple of ionic time-independent observables $\hat \Acal$ and $\hat \Bcal$.
Here, we assume that $\hat \Acal$ and $\hat\Bcal$ depends only on ionic positions.

The $\chi_{\Acal\Bcal}(\omega)$ describes how an external perturbation, interacting with the ions through $\hat \Bcal$, affects the average of $\hat \Acal$, as described in \secname~\ref{sec:full:dyn}.
\begin{equation}
    V^{(1)}(\hat \bR, t) =  \Bcal(\hat \bR) \mathcal V(t)
\end{equation}
\begin{equation}
    \Avg{\Acal(\bR)}{\rho^{(1)}(\omega)} = \chi_{\Acal \Bcal}(\omega) \mathcal V(\omega)
    \label{eq:chi:resp:1}
\end{equation}

Since we are in linear regime, we expand the average of $\hat \Acal$ at first order around the equilibrium solution:
\begin{align}
    \Avg{\Acal(\bR)}{\rho^{(1)}(\omega)} &=
    \sum_{\mu\nu} \frac{\partial \Avg{ \Acal(\bR)}{\rho(\omega)}}{\partial {\tilde\Upsilon}_{\mu\nu}} {\tilde\Upsilon}^{(1)}_{\mu\nu} (\omega)+ \nonumber \\
    &
    + \sum_{\mu\nu} \frac{\partial \Avg{ \Acal(\bR)}{ \rho(\omega)}}{\partial \Re {\tilde A}_{\mu\nu}} \Re{\tilde A}^{(1)}_{\mu\nu}(\omega) +  \nonumber \\
    & +
    \sum_\mu \frac{\partial \Avg{\Acal(\bR)}{ \rho(\omega)}}{\partial {\tilde\Rc}_{\mu}} {\tilde\Rc}^{(1)}_{\mu}(\omega) \label{eq:obs}
\end{align}
where the derivatives in \eqname~\eqref{eq:obs} are evaluated with the equilibirum SCHA density matrix. Also here, with greek letters indices ($\mu,\nu$) we indicate the basis of polarization vectors of the self-consistent harmonic Hamiltonian in equilibrium.
In particular, since the $\hat\Acal$ observable depends only on the ionic positions $\hat\bR$, only the ${\tilde\bUps}^{(1)}$ and ${\tilde \bRcal}^{(1)}$ affects its average ($\Re\bA$ vanishes in the diagonal elements of the density matrix in the basis of the ionic positions, see \eqname~\ref{eq:rho:t}).

By exploiting the integration by parts as illustrated in ref.\cite{Bianco2017}, we get:
\begin{subequations}
\begin{equation}
    \frac{\partial \Avg{\Acal(\bR)}{ \rho(\omega)}}{\partial {\tilde \Upsilon}_{\mu\nu}}  = -\hbar^2 \frac{(2n_\mu + 1)(2n_\nu + 1)}{8\omega_\mu \omega_\nu} \Avgclassiceq{\frac{\partial^2 \Acal}{\partial\tilde R_\mu \partial\tilde R_\nu}}
\end{equation}
\begin{equation}
    \frac{\partial \Avg{\Acal(\bR)}{ \rho(\omega)}}{\partial \Re{\tilde A}_{\mu\nu}}  = 0
\end{equation}
\begin{equation}
    \frac{\partial \Avg{ \Acal(\bR)}{ \rho(\omega)}}{\partial {\tilde\Rc}_{\mu}}  = \Avgclassiceq{\frac{\partial\Acal}{\partial\tilde R_\mu}}
\end{equation}
\end{subequations}
In the same way, we employ \eqname~\eqref{eq:perturbations}
to express how the perturbation $\hat \Bcal$ affects the density matrix dynamics:
\begin{subequations}
\begin{equation}
    {f_\Upsilon^{(1)}}_{\mu\nu}(\omega) = \frac{1}{\hbar}\left(\frac{2\omega_\mu}{2n_\mu + 1} + \frac{2\omega_\nu}{2n_\nu + 1}\right) \Avgclassiceq{\frac{\partial^2 \Bcal}{\partial \tilde R_\mu \partial \tilde R_\nu}} \mathcal V(\omega),
\end{equation}
\begin{align}
    {f_{\Re A}^{(1)}}_{\mu\nu}(\omega) = \frac{1}{\hbar}&\bigg(\frac{2(n_\mu + 1)n_\mu\omega_\mu}{2n_\mu + 1} + \nonumber\\
    & + \frac{2(n_\nu +1)n_\nu\omega_\nu}{2n_\nu + 1}\bigg) \Avgclassiceq{\frac{\partial^2 \Bcal}{\partial \tilde R_\mu \partial \tilde R_\nu}} \mathcal V(\omega)
\end{align}
\begin{equation}
    {f^{(1)}_\Rc}_\mu(\omega) = - \Avgclassiceq{\frac{\partial \Bcal}{\partial \tilde R_\mu}} \mathcal V(\omega)
\end{equation}
\label{eq:pert:ft}
\end{subequations}
Since $\bf^{(1)}$ is proportional to $\mathcal V$, it is convenient to define $\bft^{(1)}$ as:
\begin{equation}
    \bf_x^{(1)}(\omega) = \bft_x^{(1)} \mathcal V(\omega) 
    \qquad 
    x = \tilde\Upsilon,\Re\tilde A, \tilde\Rcal
\end{equation}

We can express the response function with a standard linear-algebra matrix-vector multiplications. We introduce the response vector $\bm p$ and the perturbation vector $\bm q$ as:
\begin{equation}
    \boldsymbol{p} = 
    \bpm
    \frac{\partial \Avg{\hat \Acal}{\hat \rho(\omega)}}{\partial {\tilde\bUps}} &
    \frac{\partial \Avg{\hat \Acal}{\hat \rho(\omega)}}{\partial \Re {\tilde \bA}} &
    \frac{\partial \Avg{\hat \Acal}{\hat \rho(\omega)}}{\partial {\tilde\bRcal}}\epm
    \label{eq:p1},
\end{equation}
\begin{equation}
    \boldsymbol{q} = \bpm {\bft_\Upsilon^{(1)}} \\ 
    {\bft_{\Re A}^{(1)}} \\
    {\bft^{(1)}_\Rc} \epm 
    \label{eq:q1}.
\end{equation}

In this way, the average of $\hat \Acal$  (\eqname~\ref{eq:obs}) is a simple scalar product: 
\begin{equation}
    \Avg{\hat \Acal}{\rho^{(1)}(\omega)} = \boldsymbol{p}\cdot \bpm {\tilde\bUps}^{(1)}(\omega) \\
    \Re{\tilde\bA}^{(1)}(\omega) \\
    \tilde\bRcal^{(1)}(\omega)\epm
    \label{eq:avgApert}
\end{equation}

The last vector in \eqname~\eqref{eq:avgApert} is the result of the linear response system (\eqname~\ref{eq:response}), and it is the product between the Green function $\bm G(\omega)$ to the $\bm q$ vector:
\begin{equation}
    \bpm {\tilde\bUps}^{(1)}(\omega) \\
    \Re{\tilde\bA}^{(1)}(\omega) \\
    \tilde\bRcal^{(1)}(\omega)\epm = 
    \bG(\omega) \boldsymbol{q} \mathcal V(\omega).
    \label{eq:pert11}
\end{equation}
Combining \eqname~\eqref{eq:avgApert} with \eqname~\eqref{eq:pert11}, we get the expression of the response function in the TD-SCHA linear response formalism:
\begin{equation}
    \Avg{\hat \Acal}{\rho^{(1)}(\omega)} = 
    \boldsymbol{p} \bG(\omega) \boldsymbol{q}\mathcal V(\omega)
    \label{eq:response:final:1}
\end{equation}
Comparing \eqname~\eqref{eq:response:final:1} with the definition of the response function $\chi_{\Acal\Bcal}$ (\eqname~\ref{eq:chi:resp:1}) we get:
\begin{equation}
    \chi_{\Acal\Bcal}(\omega) = \boldsymbol{p} \bG(\omega) \boldsymbol{q},
    \label{eq:response:final}
\end{equation}
where $\bG(\omega)$ is defined in \eqname~\eqref{eq:full:green:function}, $\boldsymbol{p}$ in \eqname~\eqref{eq:p1} and $\boldsymbol{q}$ in \eqname~\eqref{eq:q1}.

\section{Interacting one-phonon Green function}

The one-phonon Green function describes lattice excitations inside the material, and its trace is the so called ``spectral function''. Phonon spectral functions are probed by X-ray and neutron scattering and are related to theory of thermal transport and superconductivity. The poles of the spectral function are the energies of the physical phonons in the system, and the imaginary part is their lifetime.

The one-phonon Green function is the dynamical response function of the system to two atomic displacements (re-scaled by the masses of the atom): where $\hat \Acal = \sqrt m_a \hat R_a$ and $\hat \Bcal = \sqrt m_b\hat R_b$.

\begin{equation}
    \Avg{\sqrt m_a \hat R_a}{\rho^{(1)}(\omega)} = \tilde \Rcal^{(1)}_a(\omega)
\end{equation}
Thus, the response $\bm p$ vector (\eqname~\ref{eq:p1}) is:
\begin{equation}
    \bm p = \bpm 0 & 0 & \boldsymbol{\delta}_a \epm 
\end{equation}
where we indicate with $\boldsymbol{\delta}_a$ the vector with 1 in the $a$-th atom/Cartesian index and zeros elsewhere.

On the other side, the perturbation vector $\boldsymbol{q}$ can be obtained substituting $\hat\Bcal$ into \eqname~\eqref{eq:pert:ft}:

\begin{equation}
    \bft_{\Upsilon}^{(1)} = 0\qquad
    \bft_{\Re A}^{(1)} = 0
\end{equation}
\begin{equation}
{\ft_{\Rcal}^{(1)}}_c(\omega) = -\delta_{cb}.
\end{equation}
Then, from \eqname~\eqref{eq:q1}, we get:
\begin{equation}
\boldsymbol{q} = -\bpm 0 \\ 0 \\ \boldsymbol{\delta}_b\epm
\end{equation}

From which we get the interacting one-phonon Green function, as the last block of the Green function:
\begin{equation}
    \chi_{\sqrt m_a R_a, \sqrt m_b R_b}(\omega) = \mathcal G_{ab}(\omega) =  - \bpm 0 & 0 & \boldsymbol{\delta}_a \epm {\bm G}(\omega) \bpm 0 \\ 0 \\ \boldsymbol{\delta}_b \epm
    \label{eq:one:phonon:full}
\end{equation}

The full expressions for the Green function is quite complex. 
For simplicity, we can look what happens for a system with negligible odd anharmonicity. If we take the expression of $\mathcal{L}$ from \appendixname~\ref{app:linear:response} and \ref{app:full:system}, we see that, if we neglect odd contribution of anharmonicity in the atomic displacements, the $\bRcal^{(1)}$ block of the $\mathcal L$ matrix is isolated and diagonal in the polarization space (\eqname~\ref{eq:Z2mu}):

\begin{equation}
    { {\mathcal G}^{(0)}}_{\mu\nu}(\omega) = \delta_{\mu\nu}(\omega^2 -  \omega_\mu^2)^{-1} = \frac{\delta_{\mu\nu}}{\omega^2 - \omega_\mu^2}.
    \label{eq:green:one:nonint}
\end{equation}

This is the Green function we would obtain considering only the dynamics accoring to the first commutator of $\Lsc$ in \eqname~\eqref{eq:Lsc}, that is the propagator of the self-consistent harmonic Hamiltonian with the density fixed at equilibrium. This already accounts for anharmonicity, as 
the $\omega_\mu$ are frequencies of the self-consistent harmonic Hamiltonian in equilibrium.
Thanks to the self-consistency,  the $\omega_\mu$ are affected by the anharmonicity (even and odd) and, thus, they are temperature dependent. 
Therefore, the static SCHA auxiliary frequencies $\omega_\mu$ coincides with the poles of the dynamical Green function only if odd anharmonicity is negligible. However, this is not true in the most general case, and to find the poles of the Green function, we must invert \eqname~\eqref{eq:full:green:function}.

Notably, this is a general misunderstanding in empirical mean-field methods to deal with strong ionic anharmonicity, as TDEP\cite{TDEP}, ALAMODE\cite{ALAMODE} and the static SCHA\cite{Errea2014}. The self-consistent frequencies $\omega_\mu$ extracted from these methods are not the phonon frequencies probed by dynamical experiments.
The reason is that the $\omega_\mu$s extracted from this methods do not coincide with the poles of the interacting one-phonon Green function.

\subsection{TD-SCHA self-energy}
\label{sec:ansatz}

Ref.\cite{Bianco2017} introduced the response function of equilibrium SCHA for static perturbations. They introduced a (static) self-energy $\bPi(0)$ that when added to the SCHA equilibrium dynamical matrix, generates the free energy Hessian for the atomic position.
Based on analogy with field theory, they proposed an \emph{ansatz} for the dynamical Green function by the analytical continuation of it at finite frequency:
\begin{equation}
    \bGcal^{-1}(\omega) = \left[\bGcal^{(0)}(\omega)\right]^{-1} - \bPi(\omega),\label{eq:self:energy:def}
\end{equation}
with $\bGcal^{(0)}(\omega)$ given in \eqname~\eqref{eq:green:one:nonint}, and
\begin{equation}
    \bPi(\omega) = \bDthree \left(-\frac 12 \bchi(\omega)\right)\left[ \mathbbm{1} -  \bDfour \left(-\frac 12 \bchi(\omega)\right)\right]^{-1} \bDthree
    \label{eq:self:energy:ansatz}
\end{equation}
\begin{equation}
    \Dtwo_{\mu\nu} = \Avgclassiceqz{\frac{\partial^2 V}{\partial \tilde R_\mu \partial \tilde R_\nu}} = \omega_\mu^2 \delta_{\mu\nu}
\end{equation}
\begin{equation}
    \Dthree_{\mu\nu\eta} = 
    \Avgclassiceqz{\frac{\partial^3 V}{\partial \tilde R_\mu \partial \tilde R_\nu \partial \tilde R_\eta}},
    \label{eq:D3}
\end{equation}
\begin{equation}
    \Dfour_{\mu\nu\eta\lambda} = 
    \Avgclassiceqz{\frac{\partial^4 V}{\partial \tilde R_\mu \partial \tilde R_\nu \partial \tilde R_\eta \partial \tilde R_\lambda}},
    \label{eq:D4}
\end{equation}
and $\bchi(\omega)$ the standard two phonon propagation:
\begin{align}
    \chi_{\mu\nu\mu\nu}(\omega) =  \frac{1}{2\omega_\mu\omega_\nu}&\bigg[\frac{(\omega_\mu + \omega_\nu)(n_\mu + n_\nu + 1)}{(\omega_\mu + \omega_\nu)^2 - \omega^2} + \nonumber \\
    &-\frac{(\omega_\mu - \omega_\nu)(n_\mu - n_\nu)}{(\omega_\mu - \omega_\nu)^2 - \omega^2}\bigg].
    \label{eq:chi:twophonons}
\end{align}

Ref.\cite{Bianco2017} proved the correctness of this \emph{ansatz} in the limit of small anharmonicity (perturbation theory) and for the static case $\omega = 0$.

Within TD-SCHA, we can compute the dynamical self energy $\bPi(\omega)$ by substituting the TD-SCHA one-phonon Green function \eqname~\eqref{eq:one:phonon:full} into the equation that defines the self-energy (\eqname~\ref{eq:self:energy:def}).
We report the details of this calculation in \appendixname~\ref{app:ansatz:proof}, where we prove that the expression of the self-energy $\bPi(\omega)$ given in \eqname~\eqref{eq:self:energy:ansatz} is exact: \eqname~\eqref{eq:self:energy:ansatz} is valid at any order, including strongly anharmonic and high-frequency regimes. 

\begin{figure}
    \centering
    \includegraphics[width=\columnwidth]{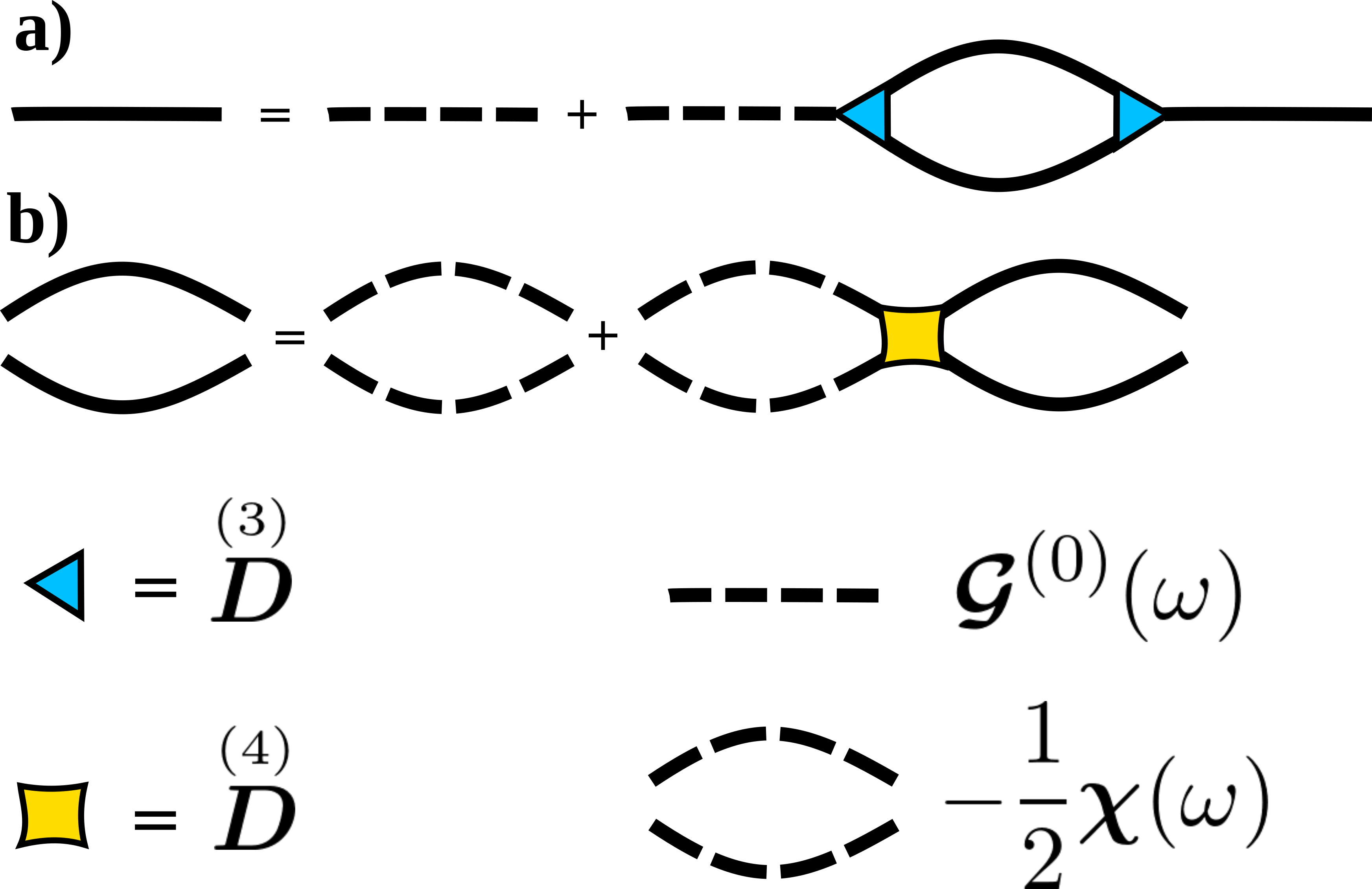}
    \caption{Diagrammatic representation of the TD-SCHA self-energy. Panel \textbf{a}: The Dyson equation for the one Green function. \textbf{b}: The Dyson equation for the two-phonons Green function. The proof of the equivalence of this diagrammatic expression for the self-energy in \eqname~\eqref{eq:self:energy:ansatz} is reported in ref.\cite{Bianco2017}.}
    \label{fig:selfenergy}
\end{figure}

In \figurename~\ref{fig:selfenergy} we report diagrammatic representation of the self-energy of \eqname~\eqref{eq:self:energy:ansatz}. \figurename~\ref{fig:selfenergy}(\textbf{a}) represents the diagrammatic expression for the one-phonon Green function of \eqname~\eqref{eq:self:energy:def}. The static phonons of the self-consistent harmonic Hamiltonian at equilibrium $\Hcal[\rho^{(0)}]$ (identified by dashed lines) are corrected with a bubble diagram that contains two three phonon vertices (blu triangles, the $\bDthree$) and an interacting two-phonon propagator.
Also the interacting two phonon propagator is expressed by a Dyson equation (\figurename~\ref{fig:selfenergy}\textbf{b}). It depends on the four-phonon scattering vertex (yellow square, the $\bDfour$). Notably, here the three and four phonon scattering vertices are already dressed by anharmonicity, as they are obtained averaging the derivative of the BO potential in the nuclear equilibrium distribution (\eqname~\ref{eq:D3} and \eqname~\ref{eq:D4}).

From the diagrammatic representation of the TD-SCHA Green function, we get some useful insight. The dressed bubble diagram of the self-energy introduces new poles in the response function in sum and differences of phonon frequencies (\eqname~\ref{eq:chi:twophonons}). This accounts for excitations coming from two phonon processes, including overtones. Moreover, there are no diagrams where a phonon decays in more than two phonons.  The result is the absence of overtones at three or more times the fundamental frequency, and it is related to the Gaussian constrain on the density matrix.
Since the computational cost to invert the Dyson equation for the two-phonon propagator (\figurename~\ref{fig:selfenergy}\textbf{b}) is extremely high, one usually replaces the interacting two-phonon propagator with the non-interacting one. This is called the bubble approximation and corresponds to neglecting fourth-phonon scattering processes in the diagrammatic expression ($\bDfour = 0$).
Indeed, TD-SCHA accounts for four-phonons scattering processes in the self-energy beyond the bubble approximation, which is the standard way to get finite lifetimes for phonons in other theories like\cite{TDEP}.


Contemporary to this work, another independent study\cite{lihm2020gaussian} proved the dynamical \emph{ansatz} of the SCHA, underlying the relevance of the topic.
They derived the linear response equations from the time-dependent variational principle with Gaussian wave-packets, similar to our Dirac least action principle we employed only for pure states. The difference between the derivation here presented and the one of ref.\cite{lihm2020gaussian} is for mixtures of states;  we derived the dynamics following the self-consistent Schroedinger equation for mixtures of states (\eqname~\ref{eq:tdscha:ft}), while they employed a variational principle at finite temperature. Both derivations are nonempirical, and lead to the same result for the one-phonon self-energy at any temperature.

\subsection{IR and Raman response}

\label{sec:ir:raman}
In this section, we derive the explicit expression of the IR and Raman response. To compute the general response function of the two observables $\hat\Acal$ and $\hat\Bcal$ in the TD-SCHA formalism, we follow the procedure introduced in \secname~\ref{sec:response}: we get the $\bm p$ and $\bm q$ vectors of \eqname~\eqref{eq:p1} and \eqref{eq:q1} from the observables $\hat \Acal$ and $\hat \Bcal$, and compute the response $\chi_{\Acal\Bcal}(\omega)$ with \eqname~\eqref{eq:response:final}.

Let us analyze the IR response first. This is related to the ionic dipole-dipole correlation function:
\begin{equation}
    \hat\Acal = M_\alpha(\hat \bR)\qquad 
    \hat \Bcal = M_{\alpha'}(\hat \bR)
\end{equation}
where $M_\alpha(\hat \bR)$ is the $\alpha$ Cartesian component of the net dipole moment when the ions are displaced in the $\bR$ position. 

To compute the IR response we replace $\Acal = M_\alpha$ and $\Bcal = M_{\alpha'}$ inside the expressions of vectors $\bm p$ (\eqname~\ref{eq:p1}) and $\bm q$ (\eqname~\ref{eq:q1}).

So the quantities we need to compute to obtain $\bm p$ and $\bm q$ are the averages of the dipole derivatives:

\begin{equation}
    \Avgclassiceqz{\frac{\partial M_\alpha(\bR)}{\partial R_a}} = \Avgclassiceqz{\Zcal_{\alpha a}(\bR)}
    \label{eq:eff:av}
\end{equation}
where $\Zcal(\bR)$ is the effective charge of the system when ions are displaced along $\bR$, and, by exploiting integration by parts as introduced in ref.\cite{Bianco2017},
\begin{equation}
\Avgclassiceqz{\frac{\partial^2 M_\alpha(\bR)}{\partial R_a\partial R_b}} = \sum_c \Upsilon_{ac} \Avgclassiceqz{(R_c - \Rcal_c)\Zcal_{\alpha b}(\bR)}
\label{eq:deff:av}.
\end{equation}

Both \eqname~\eqref{eq:eff:av} and \eqname~\eqref{eq:deff:av} can be computed by averaging the effective charges in a random ensemble extracted according to the equilibrium distribution $\rho(\bR)$. Since most standard \emph{ab initio} codes calculate effective charges, the full IR response is accessible fully \emph{ab initio}.

In general, the IR response involves the full interacting Green function.
A particular case is when effective charges do not depend on the ionic position.
In this case, the calculation of IR response is much easier, as \eqname~\eqref{eq:eff:av} becomes:
\begin{equation}
 \Avgclassiceqz{\Zcal_{\alpha a}(\bR)} = \Zcal_{\alpha a}(\bRcal),
\end{equation} 
while \eqname~\eqref{eq:deff:av}:
\begin{equation}
    \sum_c \Upsilon_{ac} \Avgclassiceqz{(R_c - \Rcal_c)\Zcal_{\alpha b}(\bR)} = 0
\end{equation}

In this simple case, only the $\bRcal$ block of $\bm p$ and $\bm q$ is different from zero, thus we can link the IR response to the interacting one-phonon Green function:
\begin{equation}
    \chi_{M_\alpha M_{\alpha'}}(\omega) =\sum_{ab}\frac{\Zcal_{\alpha a} \Zcal_{\alpha'  b}}{\sqrt{m_a m_b}} \Gcal_{ab}(\omega)
    \label{eq:ir:one}
\end{equation}

Notably, this equation, which is the standard approximation for IR signal, is only valid for effective charges that do not depend on the ionic displacement. TD-SCHA allows computing the IR response even in the general scenario.

The same procedure holds also for the Raman response. Here, the $\Acal$ and $\Bcal$ observables are the polarizability $\alpha_{\alpha\beta}(\hat \bR)$ of the sample along $\alpha$ and $\beta$ direction (the incoming and outcoming polarization of the light) when ions are displaced in the $\bR$ position.
\begin{equation}
    \Acal = \alpha_{\alpha\beta}(\hat \bR)\qquad 
    \Bcal = \alpha_{\alpha'\beta'}(\hat \bR)
\end{equation}

The procedure of deriving $\bm p$ and $\bm q$ is the same as for the IR. If we define the Raman tensor $\boldsymbol{\Ramant}$ as:
\begin{equation}
    \Ramant_{\alpha\beta a}(\bR) = \frac{\partial \alpha_{\alpha\beta}(\bR)}{\partial R_a},
\end{equation}
then we have:

\begin{equation}
    \Avgclassiceqz{\frac{\partial \alpha_{\alpha\beta}(\bR)}{\partial R_a}} = \Avgclassiceqz{\Ramant_{\alpha \beta a}(\bR)}
    \label{eq:raman:av}
\end{equation}

\begin{equation}
\Avgclassiceqz{\frac{\partial^2 \alpha_{\alpha\beta}(\bR)}{\partial R_a\partial R_b}} = \sum_c \Upsilon_{ac} \Avgclassiceqz{(R_c - \Rcal_c)\Ramant_{\alpha\beta b}(\bR)}
\label{eq:draman:av}.
\end{equation}
Also in this case, the full response function can be obtained by simply calculating the raman tensor $\bRamant$ in a randomly distributed ensemble of ionic configurations according to $\rho(\bR)$. The Raman tensor can be calculated \emph{ab initio} efficiently with the method introduced by ref.\cite{Lazzeri_2003}. If the Raman tensor depends on nuclear positions, \eqname~\eqref{eq:draman:av} is different from zero, and we have a contribution to the response function from the complete interacting Green function. However, if we neglect the dependence of $\bRamant$ from the nuclear position, we can express the Raman response only from the one-phonon Green function, as we did for the IR:
\begin{equation}
    \chi_{\alpha_{\alpha\beta}\alpha_{\alpha'\beta'}}(\omega) = \sum_{ab} \frac{\Ramant_{\alpha\beta a} \Ramant_{\alpha'\beta' b}}{\sqrt{m_a m_b}} \Gcal_{ab}(\omega).
    \label{eq:raman:one}
\end{equation}

We remark that the calculation of the Raman tensor and the effective charges for the displaced ionic configuration is an input for the calculation of the Raman spectrum within the TD-SCHA. One needs to employ a specific theoretical framework accounting for electrons, as linear response DFT. A certain degree of nonadiabatic effects can be included in the Raman spectrum calculating the Raman tensors at the frequency of the incoming radiation within finite-differences of the Bethe-Saltpeter equation\cite{Gillet2013,Miranda2017}.

In the next section, we give a schematic overview of the processes neglected by considering the response only due to the interacting one-phonon Green function.

\subsection{One-two and two-phonons Green functions}
\label{sec:two:phonons}

The one-phonon Green function does not provide the response to any general external perturbation $V^{(1)}(\bR, t)$, but only to those that depend linearly on the ionic displacements $\bR$, as we show for the specific cases of IR and Raman (\secname~\ref{sec:ir:raman}). If the perturbation or the observable we probe are nonlinear in $\bR$, then the vectors $\bm p$ and $\bm q$ that determine the response function $\chi_{\Acal\Bcal}(\omega)$ (\eqname~\ref{eq:response:final}) have a non zero contribution also in the ${\tilde\bUps}^{(1)}$ and $\Re{\tilde\bA}^{(1)}$ sector (\eqname~\ref{eq:p1} and~\ref{eq:q1}).

This contribution is important if ionic fluctuations are sizable (e.g. in presence of light atoms or close to a second-order phase-transition), or if the linear term in $\bR - \bRcal$ of $\Acal(\hat \bR)$ and $\Bcal(\hat \bR)$ is zero by symmetry. In these conditions, quadratic terms in the coupling between the probe and ionic displacement may become important. 
For example, they are fundamental to explain IR/Raman spectra of ice VII and X\cite{Putrino2002} and liquid water\cite{Silvestrelli1997}. In these works, the authors employed molecular dynamics to go beyond the one-phonon Green function (neglecting quantum fluctuations). Another class of materials with Raman spectra arising from a nonlinear coupling between the probe and the ionic displacement are high-symmetry structures like diamond\cite{Windl1993}. Here, the authors calculated the Raman signal within the harmonic approximation. A full treatment with both anharmonicity and quantum fluctuations with a nonlinear probe is missing in the literature.

\begin{figure}
\centering
\includegraphics[width=\columnwidth]{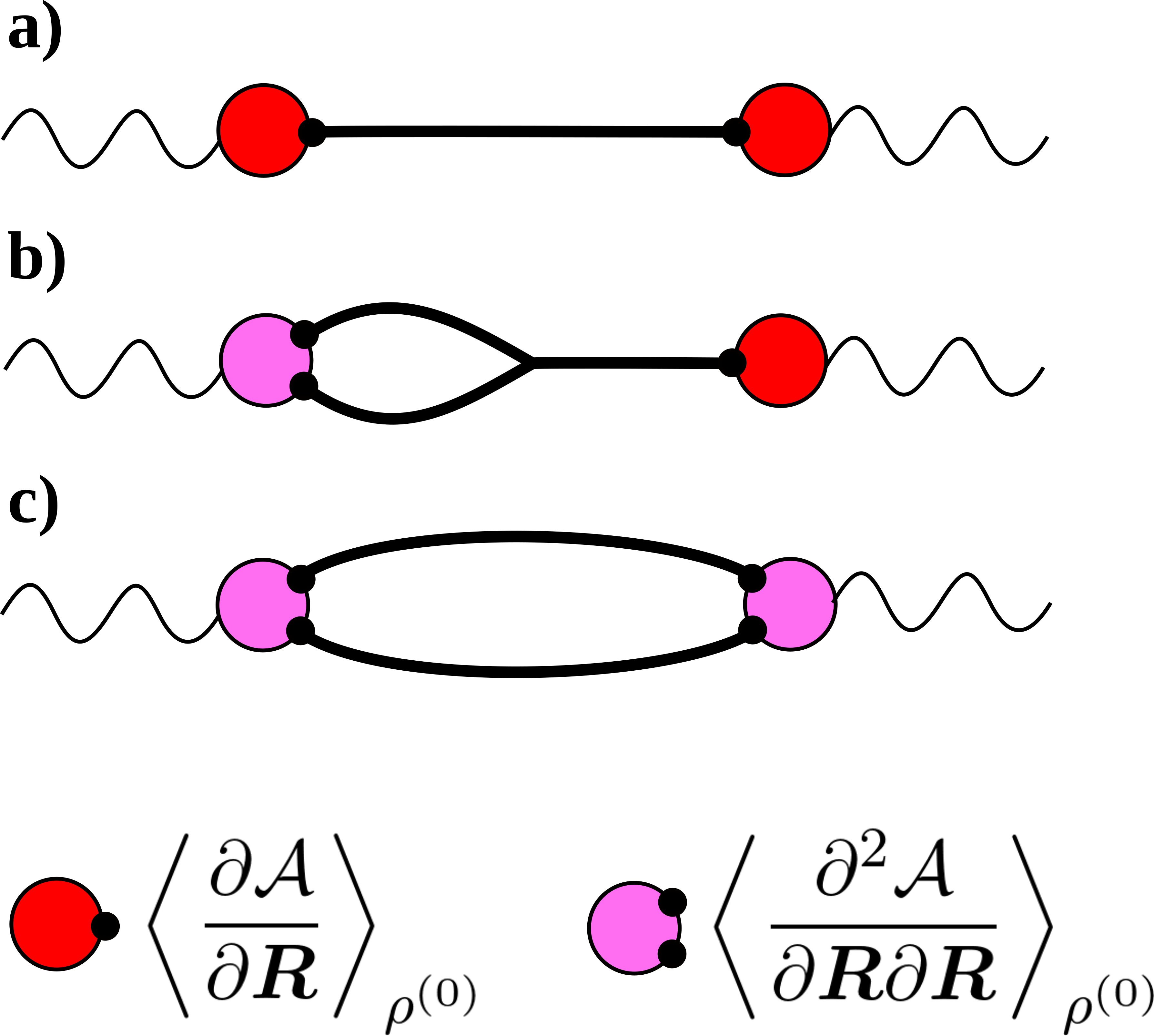}
\caption{Schematic representation of all processes captured by the TD-SCHA response function $\chi_{\Acal\Acal}(\omega)$.
The circles represent the interaction between the perturbation and the ionic displacements, the wave-like line indicates the probe, while the solid bold lines indicates the phonon interacting Green function. We have two possible interaction, the linear one give rise to a single phonon, the quadratic one excites two-phonons. Panel \textbf{a} represent a simple process where the probe excites only one phonon. In this case, the response function only depends on the one-phonon Green function as for the Raman in \eqname~\eqref{eq:raman:one}) or IR in \eqname~\eqref{eq:ir:one}). If the perturbation has a quadratic dependency on the ionic positions, we have contribution to the total response function also from diagrams \textbf{b} and \textbf{c}. In particular, the process in \textbf{b} cannot occur in purely harmonic systems, it requires a non diagonal term between one-phonon and two-phonons Green functions (involves phonons scattering).}
\label{fig:diagram}
\end{figure}

In \figurename~\ref{fig:diagram} we report a diagrammatic representation of the response function when $\hat\Acal = \hat\Bcal$.
When the probe interacts with the sample, it can either excite a single phonon (the linear dependence of $\Acal$) or two phonons with the quadratic dependency of $\Acal$ on ionic positions. 
Then, the excited phonons evolve interacting through anharmonicity, and the dynamics may end with a different phonon with respect to the original one. Since the evolution is anharmonic, it is also possible that a phonon splits in two phonons, or vice-versa, as illustrated in \figurename~\ref{fig:diagram}(\textbf{b}). 
Therefore, the total contribution to the response function is given by the anharmonic one-phonon Green function (\figurename~\ref{fig:diagram}\textbf{b}), the mixed one-two phonon Green function (\figurename~\ref{fig:diagram}\textbf{b}), and the two-phonon Green function (\figurename~\ref{fig:diagram}\textbf{c}).
Harmonic systems do not have a contribution from mixed Green functions, as phonons cannot decay or scatter during the free propagation.

The expression of the IR (\eqname~\ref{eq:ir:one}) and Raman (\eqname~\ref{eq:raman:one}) in the approximation of linear coupling with the probe is equivalent to neglect the diagrams \figurename~\ref{fig:diagram}(\textbf{b,c}).

Interestingly, the two-phonon propagation gives a non-zero response also in purly harmonic crystals. For example, the harmonic two-phonon IR signal is:
\begin{equation}
    \chi^{(2ph)}_{M_\alpha M_{\alpha'}}(\omega) = 
    \sum_{abcd\mu\nu} \frac{e_\mu^a e_\nu^b e_\mu^c e_\nu^d}{\sqrt{m_am_bm_cm_d}} \frac{d\Zcal_{\alpha a}}{dR_b}\chi_{\mu\nu\mu\nu}(\omega)\frac{d\Zcal_{\alpha' c}}{dR_d}
\end{equation}
where $\chi_{\mu\nu\mu\nu}(\omega)$ is the harmonic two-phonon Green function.
The interacting two-phonon Green function can be computed within the SSCHA formalism as the response function between quadratic displacements:
\begin{equation}
    \hat \Acal = \sqrt{m_am_b}  (\hat R_a - \Rcal_a)(\hat R_b - \Rcal_b)
\end{equation}
\begin{equation}
    \hat \Bcal =\sqrt{m_am_b}  (\hat R_c - \Rcal_c)(\hat R_d - \Rcal_d)
\end{equation}

Since, in second-quantization formalism, each position operator $\hat \bR$ is proportional to the creation-annihilation of one phonon, $\hat \Acal$ and $\hat\Bcal$ contain creations and annihilations of two phonons; this is the reason why it is called the two-phonon Green function.

We compute the $\bm p$ and $\bm q$ vectors (\eqname~\ref{eq:p1} and~\ref{eq:q1}) from $\hat \Acal$ and $\hat \Bcal$ and insert them into \eqname~\eqref{eq:response:final} to get the interacting two-phonon Green functions. 

Indeed, if one consider a perfectly harmonic oscillator, we show in \appendixname~\ref{app:twophonons} that the response function coincides with \eqname~\eqref{eq:chi:twophonons}. This is the well-known two-phonon propagator for harmonic systems, and it coincides with the one obtained with the standard many-body approach.

Notably, since usually the non-linear interaction between the probe and the ionic position is small, the diagram \figurename~\ref{fig:diagram}(\textbf{b}) provides a lower order signal than the two-phonon one (\textbf{c}).

Indeed, all the processes represented in \figurename~\ref{fig:diagram} are automatically included in the response function of the TD-SCHA introduced in \secname~\ref{sec:response}.
In principle, also higher-order processes exist, where the probe interacts with more than two phonons. However, the Gaussian constrain on the TD-SCHA density matrix does not allow these excitations; they are accounted within the mean-field approach, affecting the average of the derivatives of $\Acal$ and $\Bcal$ observables on the ionic probability distribution, providing a temperature dependency for the vertexes in \figurename~\ref{fig:diagram}. This is the result of replacing the perturbation $V^{(1)}(\bR, t)$ into $\Vsc(t)$ in the linear response theory (\eqname~\ref{eq:linear:evolution}).

\section{Lanczos algorithm to compute the response function}
\label{sec:lanczos}
While a numerical implementation to calculate the dynamical one-phonon Green function has been presented\cite{Bianco2017}, its computational cost diverges quickly for systems with more than $N=10$ atoms, as the self-energy of \eqname~\eqref{eq:self:energy:ansatz} requires, for any frequency to probe, the inversion of the matrix
\begin{equation}
\left[ \mathbbm{1} - \bDfour \left(-\frac 12 \bchi(\omega)\right)\right]^{-1}.
\label{eq:big:matrix}
\end{equation}
This matrix has a dimension of $(3N)^2\times (3N)^2$; its numerical inversion is a heavy computational task. For example, just to store in memory the matrix of \eqname~\eqref{eq:big:matrix} in a system with 100 atoms with 64-bit floating-point precision, more than 60 Gb are needed. 
The inversion of \eqname~\eqref{eq:big:matrix} requires a LU decomposition of a matrix of size $n = 9N^2$. The LU decomposition scales a $n^3$; the overall scaling for computing the inversion is $N^6$, which quickly diverges for realistic systems with more than 10 atoms. For this reason, the application of the full dynamical one-phonon Green function has been performed in realistic systems always under the assumption that $\bDfour = 0$.

Moreover, as discussed in \secname~\ref{sec:ansatz}, the one-phonon Green function does not provide the most general response to the experimental probe, but it is limited to probes interacting linearly with atomic displacements.

In this section, we derive a very efficient algorithm that allows computing the elements of the response function to any general external perturbation, that is computationally achievable in systems of hundreds of atoms in the full anharmonic regime (even with $\bDfour \neq 0$). 

We recall the expression of the response function from \eqname~\eqref{eq:response:final}:
\begin{equation}
    \chi_{\Acal\Bcal}(\omega) = -\boldsymbol{p} (\mathcal L + \omega^2)^{-1} \boldsymbol{q}
\end{equation}
where $\boldsymbol{p}$ and $\boldsymbol{q}$ are defined in \eqname~\eqref{eq:p1} and \eqname~\eqref{eq:q1} from $\hat\Acal$ and $\hat\Bcal$.

The algorithm we discuss in this section computes in one shot \eqname~\eqref{eq:response:final} for any value of $\omega$.
This is an easy task if we use a basis where $\mathcal L$ is tridiagonal (as we show later). Therefore, the purpose of the algorithm is to find the basis on which $\mathcal L$ is tridiagonal.

The bi-conjugate Lanczos algorithm does exactly this job: it builds with an iterative procedure the basis-changing $Q$ matrix (non-singular) so that:
\begin{equation}
    Q^{-1}\mathcal L Q = {\mathcal T},
\end{equation}
where $\mathcal T$ is tridiagonal:
\begin{equation}
    {\mathcal T} = \bpm \alpha_1 & \gamma_1 &  & \cdots & 0 \\ 
    \beta_1 & \alpha_2 & \ddots & & \vdots \\ 
    & \ddots & \ddots & \ddots & \\
    \vdots & & \ddots & \ddots & \gamma_{n-1} \\ 
    0 & \cdots & & \beta_{n-1} & \alpha_n \epm
\end{equation}

The basis (non orthonormal) in which $\mathcal L$ is tridiagonal is represented by the columns of the $Q$ matrix:
\begin{equation}
    Q = \bpm   {\bm q_1} & \cdots & {\bm  q_n}\epm\label{eq:Q}
\end{equation}
The rows of the $Q^{-1}$ matrix define the conjugate vectors:
\begin{equation}
    Q^{-1} = \bpm {\bm p_1} \\ 
    \vdots \\  {\bm p_n}\epm.\label{eq:P}
\end{equation}
The coefficients of the $\mathcal T$ matrix (the $\alpha_k$, $\beta_k$, $\gamma_k$), the $\bm q_k$ and $\bm p_k$ vectors are found with the iterative bi-conjugate Lanczos algorithm\cite{numericalrecipies}:
\begin{subequations}
\begin{equation}
    \alpha_k = \bm{p_k} \cdot {\mathcal L} \bm{ q_k}
\end{equation}
\begin{equation}
    \beta_k  \bm {q_{k+1}} = \bm {r_k} = (\mathcal L - \alpha_k) \bm {q_k} - \gamma_{k-1} \bm {q_{k-1}}
    \label{eq:beta:k}
\end{equation}
\begin{equation}
    \gamma_k \bm {p_{k+1}} = \bm s_k = ({\mathcal L}^\dagger - \alpha_k)  \bm{ p_k} - \beta_{k-1}\bm { p_k}
    \label{eq:gamma:k}
\end{equation}
\begin{equation}
    \beta_k = | \bm {r_k}|
\end{equation}
\begin{equation}
    \gamma_k = \frac{\bm{ s_k} \cdot \bm{ r_k}}{\beta_k}
\end{equation}
\end{subequations}
This recursion formally ends either when either $\bm q_k$ or $\bm p_k$ are linear combinations of the previous vectors or if $\bm p_k \cdot \bm q_k = 0$. Unless the system is perfectly harmonic, this condition is usually never reached in practical runs, and the algorithm is truncated after a maximum number of steps $N_{\text{max}}$.

To facilitate the calculation of the Green function, we initialize the algorithm in the following way:
\begin{equation}
    \bm{ q_1} =   \frac{\boldsymbol{q}}{|\boldsymbol{q}|}
\end{equation}
\begin{equation}
    \bm{p_1} = \bm{p}\frac{|\boldsymbol{q}|}{\bm{p} \cdot \bm{q}}
\end{equation}

The response function can be rewritten as:
\begin{equation}
    \chi_{\Acal\Bcal}(\omega) =- ({\bm p} \cdot {\bm q})\; \bm {p_1} \cdot (\mathcal L + \omega^2)^{-1} \bm {q_1}
\end{equation}
\begin{equation}
    \chi_{\Acal\Bcal}(\omega) =-  ({\bm p} \cdot {\bm q})\; \bm{ p_1} Q Q^{-1} \cdot (\mathcal L + \omega^2)^{-1}  Q Q^{-1}\bm{ q_1}
\end{equation}
Thanks to the definitions in \eqname~\eqref{eq:Q} and~\eqref{eq:P} we have 
\begin{equation}
    Q^{-1} \bm{q_1} = \bpm 1 \\ 0 \\ \vdots \\ 0 \epm 
\end{equation}
\begin{equation}
    \bm {p_1} Q = \bpm 1 & 0 & \cdots & 0 \epm 
\end{equation}
Thus, the response function is the first element of the inverse matrix in the tridiagonal basis:
\begin{equation}
    \chi_{\Acal\Bcal}(\omega) =- ({\bm p} \cdot {\bm q}) \left[\left(\mathcal T + \omega^2\right)^{-1}\right]_{11}
\end{equation}

We can compute the first element of the inverse of a tridiagonal matrix as a continued fraction:
\begin{equation}
    \chi_{\Acal\Bcal}(\omega) = \cfrac{-({\bm p} \cdot {\bm q})}{\alpha_1 + \omega^2 - \cfrac{\gamma_1\beta_1}{\alpha_2 + \omega^2 - \cfrac{\gamma_2\beta_2}{\ddots}}}
    \label{eq:green:lanczos}
\end{equation}

Indeed, the application of this method is limited to response functions where the perturbation and the response are not orthogonal, i.e. the product $\bm{p}\cdot\bm{q} \neq 0$. 
The condition $\bm{p}\cdot\bm{q} = 0$ is met, for example, by off-diagonal elements of the interacting one-phonon Green function. Luckily, the spectral function depends only on the trace of the interacting one-phonon Green function, so only diagonal elements are required (for which $\bm p\cdot\bm q = 1$).

When the recursion is truncated without reaching the stopping condition, we can extend the $\alpha$,$\beta$ and $\gamma$ above $N_{\text{steps}}$ by assuming them as constant:
$$
k \ge N_{\text{steps}}
$$
\vspace{-0.85cm}
\begin{equation}
    \alpha_k = \alpha_{N_{\text{max}}} \qquad\beta_k = \beta_{N_{\text{max}}} \qquad
    \gamma_k = \gamma_{N_{\text{max}}}.
\end{equation}

Under this hypothesis, the last part of the continued fraction satisfy the self-consistent equation
\begin{equation}
    g(\omega)= \frac{1}{\alpha_{N_{\text{max}}} + \omega^2 - \gamma_{N_{\text{max}}}\beta_{N_{\text{max}}} g(\omega)},
\end{equation}
that is solved:
\begin{equation}
    g(\omega) = \frac{\alpha_{N_{\text{max}}} + \omega^2}{2\gamma_{N_{\text{max}}}\beta_{N_{\text{max}}}} - \frac{\sqrt{(\alpha_{N_{\text{max}}} + \omega^2)^2 - 4\gamma_{N_{\text{max}}}\beta_{N_{\text{max}}}}}{2\gamma_{N_{\text{max}}}\beta_{N_{\text{max}}}}.
    \label{eq:terminator}
\end{equation}

This introduces an imaginary term to the Green function when the argument inside the square root becomes negative (branch cut).
This termination of the recursion is similar to the proposed one for the turbo Lanczos algorithm in TD-DFT\cite{Rocca2008}.  Additionally, to introduce a finite lifetime, we also add a Lorentzian smearing $\eta$ by replacing $\omega\rightarrow \omega + i\eta$; the convergence of the spectrum is achieved as $N_\text{steps} \rightarrow \infty$ and $\eta\rightarrow 0$.
With this method, we compute the response function at any frequency, with a single tri-diagonalization procedure: the application of $\mathcal L$ to $\bm p_i$ and $\bm q_i$ in \eqname~\eqref{eq:beta:k} and~\eqref{eq:gamma:k} is the expensive part of the algorithm.
The $\alpha_i,\beta_i,\gamma_i$ coefficients depend on the $\Acal$ and $\Bcal$ observables (on the vectors $\bm p$ and $\bm q$ used to start the Lanczos algorithm) but not on the frequency $\omega$.
The application of $\mathcal L$ to a vector does not require any matrix inversion; it is efficient and parallelizable. 
In particular, even if $\mathcal L$ is a dense matrix of $[3N(1 + 6N)]^2 \sim N^4$ elements ($3N\times 3N$ the $\bRcal$ block and $9N^2\times 9N^2$ both the $\bUps$ and $\Re\bA$ blocks), the application of $\mathcal L$ to a vector scales as $N^3$, much better than the inversion of \eqname~\eqref{eq:big:matrix} ($N^6$).
In particular, the $\mathcal L$ matrix can be divided into the harmonic and anharmonic contribution:
\begin{equation}
    {\mathcal L} = {\mathcal L}^{\text{har}} + {\mathcal L}^{\text{anh}},
\end{equation}
where ${\mathcal L}^{\text{har}}$ is the propagation according to the self-consistent Hamitlonian in equilibrium, while the anharmonic part comes from how the self-consistent Hamiltonian changes under the changes of $\rhoone$ during the evolution. The harmonic contribution is diagonal in the polarization space (its application to a vector scales as $N^2$). In \appendixname~\ref{app:lanczos}, we show that the anharmonic term can be applied to a vector as:
\begin{equation}
     {\mathcal L}^{\text{anh}}\bpm {\tilde \bUps}^{(1)} \\ {\tilde\Re\bA}^{(1)} \\ {\tilde\bRcal}^{(1)} \epm = \bpm \tilde\bUps \Avgclassicpert{\frac{d^2\mathbb{V}}{d\tilde\bR d\tilde\bR}} + \Avgclassicpert{\frac{d^2\mathbb{V}}{d\tilde\bR d\tilde\bR}} \tilde\bUps \\ 
     \Re \tilde\bA \Avgclassicpert{\frac{d^2\mathbb{V}}{d\tilde\bR d\tilde\bR}} + \Avgclassicpert{\frac{d^2\mathbb{V}}{d\tilde\bR d\tilde\bR}} \Re\tilde\bA \\ 
     - \Avgclassicpert{\frac{d{\mathbb V}}{d\tilde\bR}}\epm,\label{eq:apply:L}
\end{equation}
where the average $\Avgclassicpert{\cdot}$ is computed on the perturbed ensemble defined by the vector that multiplies ${\mathcal L}^{\text {anh}}$, and $\mathbb{V}$ is the difference between the BO energy landscape and the potential energy of the equilibrium SSCHA auxiliary Hamiltonian:
\begin{equation}
    \mathbb{V}(\bR) = V(\bR) - \frac 12 \sum_{ab}(R_a - \Rcal_a)\Avgclassiceqz{\frac{d^2V}{dR_a dR_b}}(R_b - \Rcal_b).
\end{equation}

Notably, \eqname~\eqref{eq:apply:L} scales as the matrix product between $N\times N$ matrices, i.e. the overall cost of applying $\mathcal{L}$ scales with $N^2$ (in the polarization basis the $\tilde\bUps$ and $\Re\tilde\bA$ matrices are diagonal).
To calculate the full response function, we need to apply the $\mathcal L$ matrix for each iteration. 
Thus the overall computational cost scales as $N^2 \cdot N_{\text{steps}}$, where $N_{\text{steps}}$ is the number of iterations of the Lanczos algorithm.
The computational most expensive part comes from the average
\begin{equation}
    \Avgclassicpert{\frac{d^2\mathbb{V}}{d\tilde \bR d\tilde \bR}}\label{eq:pert:d2v}.
\end{equation} 
This average is evaluated with a Monte Carlo integration on the stochastic configurations for each element of the matrix. This calculation costs $N_\text{conf}\times N^2$ and must be performed at each step. Luckily, this operation runs efficiently in parallel, as the average on the $N_\text{conf}$ configurations of the ensemble can be partitioned in separate subsets computed by independent processing units. Notably, within the Lanczos algorithm, there is no difference in the computational cost between computing the bubble only approximation of the self-energy (neglecting $\bDfour$ in \eqname~\ref{eq:self:energy:ansatz}) or the full expression. We report more details on how \eqname~\eqref{eq:pert:d2v} and the transposed of ${\mathcal L}^{\text{anh}}$ are computed, and how we account for crystal symmetries in \appendixname~\ref{app:lanczos}.

\section{Applications}
\label{sec:applications}

We now illustrate two applications of the TD-SCHA.
First, as a simple benchmark, we show how the TD-SCHA performs in a toy model constituted by a single one-dimensional particle in an external anharmonic potential, comparing the results with the exact (numerical) solution of the problem (\secname~\ref{sec:1d}).

Then, we apply the TD-SCHA in linear response regime to a real physical system, the phase III of solid high-pressure hydrogen, and simulate the IR and Raman spectroscopy. This system is challenging for calculating quantum nuclear time-correlation functions \emph{ab initio}: it is strongly anharmonic, quantum effects are important, and we have many atoms in the simulation cell ($N = 96$).

\subsection{1D particle in an external potential}
\label{sec:1d}
\begin{figure*}
    \centering
    \includegraphics[width=\textwidth]{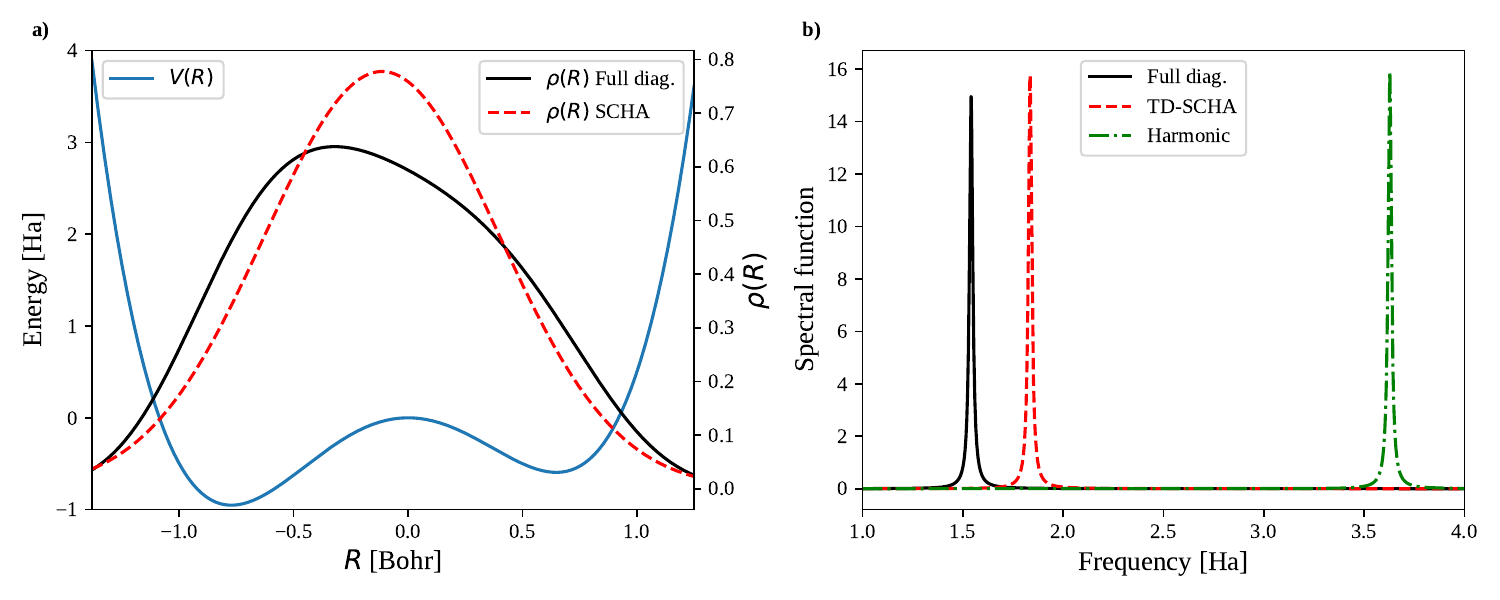}
    \caption{One dimensional anharmonic model. Panel \textbf{a}: The Born-Oppenheimer energy landscape $V(R)$ is plotted with the exact ground state $\psi(R)$ probability density (full diagonalization) and the SCHA equilibrium distribution $\rho(R)$. Panel \textbf{b}: the dynamical spectral function of the model. Comparison between the exact result (full diagonalization), the linear response from TD-SCHA, and the Harmonic approximation. The finite line-width arise from a smearing we introduced for presentation purposes.}
    \label{fig:bo:spect}
\end{figure*}

To benchmark the TD-SCHA, we compare it with the numerical solution of the Schroedinger equation.
This is possible only for small systems. In this section, we simulate a one-dimensional particle in a strongly anharmonic external potential.
The external potential $V(R)$ plays the role of Born-Oppenheimer energy landscape, its shape is reported in \figurename~\ref{fig:bo:spect}~(\textbf{a}). In Hartree atomic units it is:
\begin{equation}
    V(R) = 3 R^4  + \frac 12 R^3 - 3 R^2
\end{equation}
This example is performed at $T = \SI{0}{\kelvin}$, where the density matrix is a pure state. Our particle has the mass of an electron.
We compare the exact probability distribution of the ground state wave-funciton $\ket{\psi_{GS}}$ (obtained diagonalizing the real Hamiltonian) with the SCHA in \figurename~\ref{fig:bo:spect}(\textbf{a}).
The SCHA solution is Gaussian, while the exact ground state is a complex function.
In \figurename~\ref{fig:bo:spect}(\textbf{b}), we present the linear response to a dynamical perturbation. 
We plot the one-phonon spectral function multiplied by $\omega$, defined as:
\begin{equation}
    \Scal(\omega) = - \frac{\omega}{\pi}  \Im \mathcal G(\omega).
\end{equation}
In this way, $\Scal(\omega)$ satisfy the sum rule that its integral is proportional to the number of phonons\cite{Bianco2018} (the intensity of a peak does not depend on the phonon frequency).
We compare the results computed with the harmonic approximation and the TD-SCHA to the exact (numerical) solution.

The exact solution is calculated with the Lehmann representation of the Green function, where we performed the full diagonalization on the Hamiltonian $\hat H$ to obtain the excited sates $\ket{\psi_i}$:
\begin{align}
    \Gcal(\omega) &= \sum_{i = 1}^{\infty} \frac{\left|\braket{\psi_i|\sqrt m (\hat R - \Rcal)| \psi_{GS}}\right|^2}{\hbar\omega - E_i + E_{GS} + i 0^+} + \nonumber \\
    & + \sum_{i = 1}^{\infty}\frac{\left|\braket{\psi_i|\sqrt m (\hat R - \Rcal)| \psi_{GS}}\right|^2}{\hbar\omega + E_i - E_{GS} + i 0^+}
\end{align}
\begin{equation} 
\Rcal = \braket{\psi_{GS} | \hat R|\psi_{GS}}
\end{equation}

Since we have a 1D system, we can calculate the TD-SCHA response analytically. In particular, we can write the $\mathcal L$ matrix explicitly (\appendixname~\ref{app:full:system}):
\begin{equation}
    \mathcal L = \bpm 
    -\frac{\Dfour}{2\omega_s} - 2\omega_s^2 & 4\omega_s^2 & 4\omega_s \Dthree \\ 
    0 & 0 & 0 \\ 
    \frac{\Dthree}{8\omega_s^2} & 0 & - \omega_s^2\epm
\end{equation}
where $\omega_s$ is the frequency of the self-consistent harmonic Hamiltonian of the equilibrium SCHA. Since we are at $T = \SI{0}{\kelvin}$ the line of $\mathcal L$ that represent the evolution of $\Re A^{(1)}$ is zero, as the system is in a pure state. Therefore, $\Re A$ does not affect the dynamics and we can remove it:
\begin{equation}
    \mathcal L = \bpm 
    -\frac{\Dfour}{2\omega_s} - 2\omega_s^2  & 4\omega_s \Dthree \\
    \frac{\Dthree}{8\omega_s^2} & - \omega_s^2\epm
\end{equation}
The full anharmonic Green function is:
\begin{equation}
    \bm G(\omega) = - (\mathcal L + \omega^2)^{-1}
\end{equation}
The one-phonon Green function $\Gcal(\omega)$ is the element in the $\bRcal$ block (the last) of the full Green function $\bG(\omega)$. Performing the inversion analytically we can compute the self-energy:
\begin{equation}
    \Gcal(\omega)^{-1} = \omega^2 - \omega_s^2 - \Pi(\omega)
\end{equation}
\begin{equation}
    \Pi(\omega) =  - \frac{\Dthree^2 /(2\omega_s)}{\omega^2 - 4\omega_s^2- \Dfour/(2\omega_s)}
    \label{eq:self:1d}
\end{equation}

As we proved for the general case in \secname~\ref{sec:ansatz}, this self-energy is the same as the one obtained by exploiting the dynamical ansatz formulated in ref.\cite{Bianco2017}:
\begin{equation}
    \Pi(\omega) = \Dthree\left(-\frac 12 \chi(\omega)\right)\left[1 - \Dfour\left(-\frac 12 \chi(\omega)\right)\right]^{-1}\Dthree
\end{equation}
where in 1D at $T = \SI{0}{\kelvin}$ we have:
\begin{equation}
    \chi(\omega) =  \frac{1}{2\omega_s^2} \frac{1}{4\omega_s^2 - \omega^2}
\end{equation}

As shown in \figurename~\ref{fig:bo:spect}, the strong anharmonicity downshifts the energy of the peak by 50 \% of the harmonic result. Here, the TD-SCHA displays a relative error with the exact solution on the phonon energy of 5 \%; an important improvement from the 100 \% of the harmonic approximation. The TD-SCHA still slightly overestimates the vibrational energy. This is a general feature of the method, as the TD-SCHA wave-function is more rigid than the real one, as we constrained its Gaussian form. 

\begin{figure}
    \centering
    \includegraphics[width=\columnwidth]{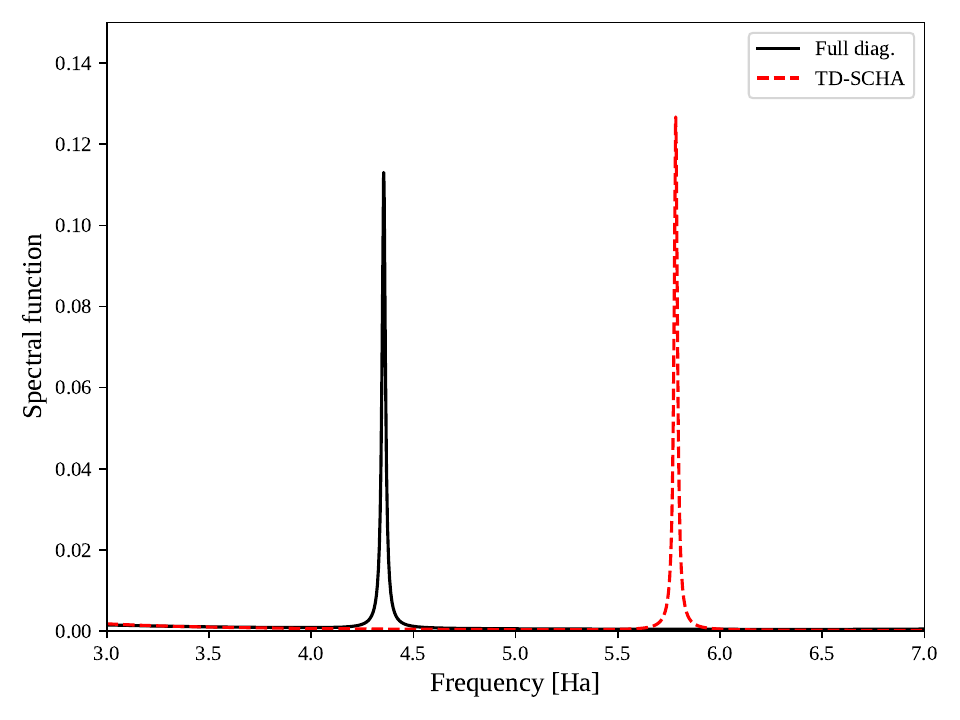}
    \caption{Zoom in the spectral function at high frequencies. We plot the comparison between the exact overtone and the TD-SCHA overtone.}
    \label{fig:overtone}
\end{figure}

Moreover, the TD-SCHA spectral function displays an overtone, originating from the anharmonic coupling between the oscillations of the average position and the fluctuations. This overtone arises from the pole of the self-energy in \eqname~\eqref{eq:self:1d}.
We plot, in \figurename~\ref{fig:overtone}, the spectral function zoomed in the energy region where the overtone appears. In this example, it is 100 times smaller than the principal peak. This is a consequence that the overtone is off-resonant: there are not one-phonon excitations in resonance. The TD-SCHA correctly reproduces the overtone intensity but with a relative error on its energy of 60\%.
Interestingly, the energy of the overtone is not twice the energy of the fundamental phonon mode, as there is a very strong anharmonicity.
The TD-SCHA is unable to simulate third or higher overtones, as we have only two poles from the Green function that are the zeros of the determinant of the inverse of $\bm G(\omega)$. 

\begin{figure*}
\centering
\includegraphics[width=\textwidth]{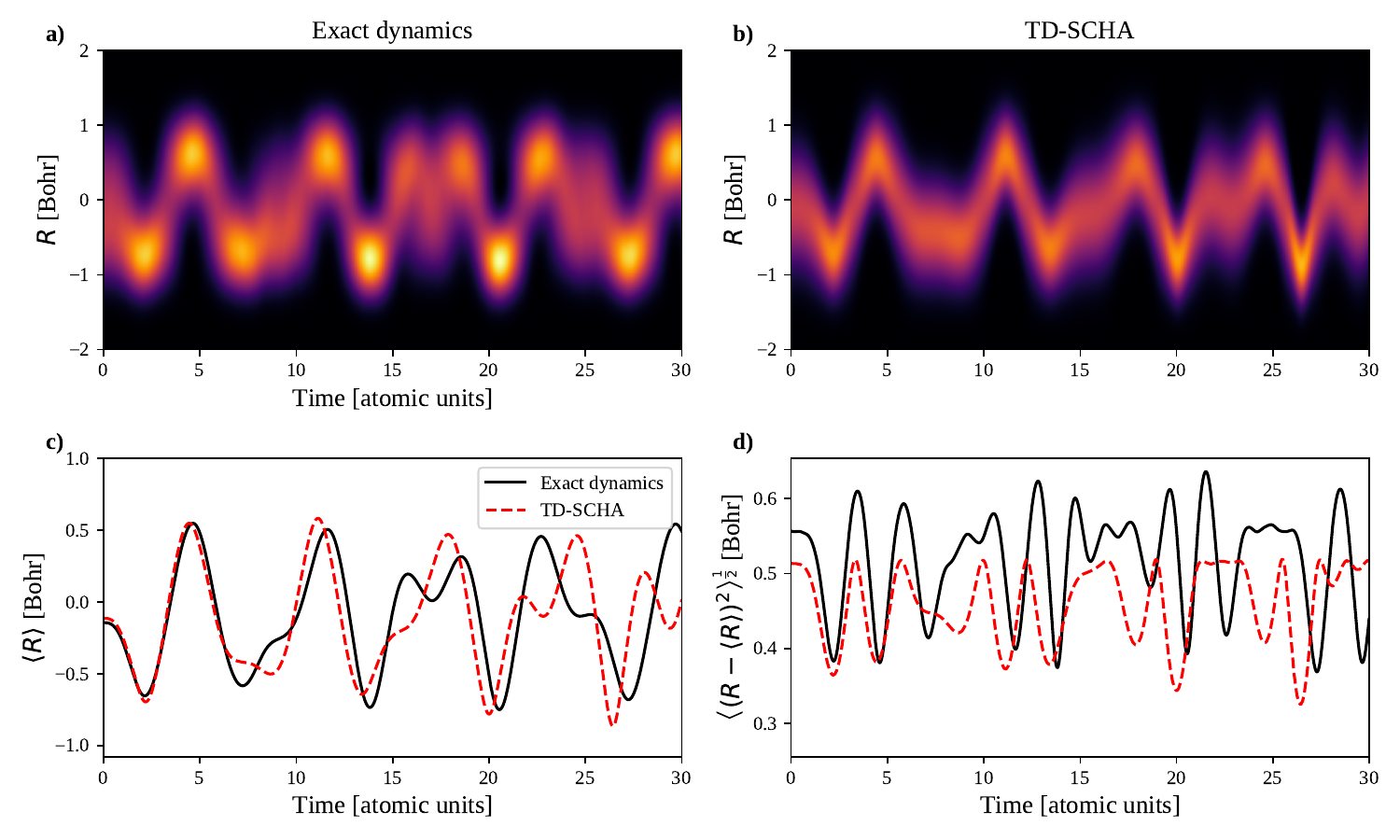}
\caption{Time evolution of the wave-function in a time-dependent external potential. The simulation starts from the equilibrium configuration of \figurename~\ref{fig:bo:spect}. Panel \textbf{a}: time evolution of the modulus square of the exact wave-function. Panel \textbf{b}: Time evolution of the TD-SCHA Gaussian distribution. Panel \textbf{c}: average position of the particle as a function of time. Panel \textbf{d}: quantum dispersion (mean square displacement).}
\label{fig:full:tdscha}
\end{figure*}

Finally, we can simulate the wave-function dynamics adding a time-dependent external potential in a nonlinear regime.
For this purpose, we integrate the TD-SCHA equation of motion and compare the result with the exact evolution.
We introduce, at $t_0 = 0$, a perturbation of the form:
\begin{equation}
    V^{(1)}(R, t) = E_0 R \sin(\omega_0 t)
\end{equation}
where $E_0 = 1\,\text{Ha/Bohr}$, and $\omega_0 = 1$~Ha.
The results are shown in \figurename~\ref{fig:full:tdscha}.

The TD-SCHA time evolution is very close to the exact dynamics in the first oscillations, where the wave-function is well localized. The two solutions deviate when the exact wave-function becomes delocalized at around 7 atomic units: The TD-SCHA probability distribution is more localized during the whole dynamics, as shown by the mean square displacement reported in \figurename~\ref{fig:full:tdscha}(\textbf{d}). However, even after that time, the accuracy of the TD-SCHA evolution is good. This shows how the TD-SCHA can reproduce well the nuclear dynamics even in the nonlinear regime.

\subsection{High-pressure hydrogen}
\label{sec:hydrogen}
In this section, we employ the linear response theory of the TD-SCHA to calculate the Raman and spectra of high-pressure hydrogen phase III.
The simulation of vibrational spectra in high-pressure phases of molecular hydrogen is very important to dissect the crystal geometry, as it is the only experimental signature directly related to the lattice. Both X-ray spectroscopy and neutron scattering are extremely challenging for the small samples of hydrogen available. Thus, the structure identification is possible only by comparing results from \emph{ab initio} simulations with experimental data. We already presented these results and deeply discussed their relevance and implications in ref.\cite{MonacelliNatPhys2020}. Here, instead, we focus on the details of the calculation of the Raman and IR response at \SI{260}{\giga\pascal} ($T = \SI{0}{\kelvin}$). 
Hydrogen phase III is a monoclinic crystal of symmetry group \emph{C2/c}, with 24 atoms in the primitive cell. 

We computed only the one-phonon contribution to the full response function: we approximate the Raman tensor as independent from the ionic displacement equal to the value on the SCHA equilibrium positions, as described in \secname~\ref{sec:ir:raman} and \ref{sec:two:phonons}. 
The simulations are performed on a 2x2x1 supercell to sample the Brillouin zone for phonons (96 atoms). The \emph{ab initio} energy landscape is simulated through DFT, with the BLYP\cite{BLYP} functional. 
For all the DFT calculations, we employed the Quantum ESPRESSO\cite{Giannozzi2009,Giannozzi2017} package, with a plane wave basis set and a norm-conserving pseudo-potential from the Pseudo Dojo library\cite{pseudodojo}.
The energy cutoff for the basis of the wave-functions was 60~Ry (240~Ry for the electronic density). The Brillouin zone for electrons is sampled on a 4x4x4 mesh in the phonon displaced supercell (2x2x1). The Raman tensor is obtained with linear response DFT as implemented in the PHonon package of quantum ESPRESSO within LDA\cite{Lazzeri_2003}.

\begin{figure}
    \centering
    \includegraphics[width=\columnwidth]{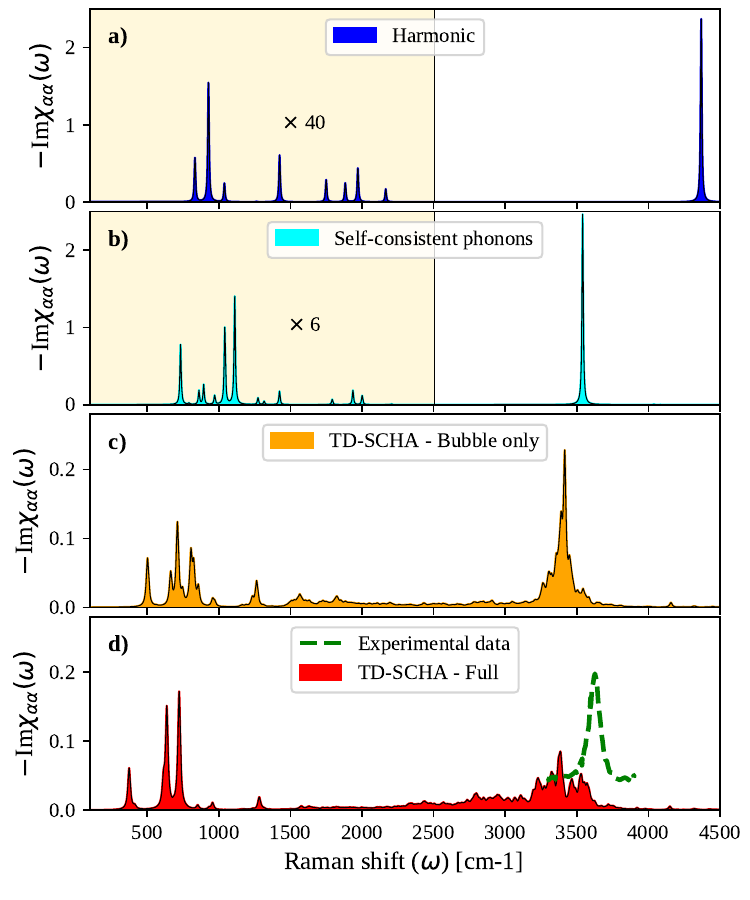}
    \caption{Raman spectra of high-pressure hydrogen phase III (\SI{260}{\giga\pascal}, \SI{0}{\kelvin}). \textbf{a}: Harmonic spectra. \textbf{b}: Spectrum obtained with from the equilibrium self-consistent phonons. These phonons solve the self-consistent harmonic Hamiltonian with the static density (the anharmonicity changes their energy, but they have infinite lifetimes). \textbf{c}: Anharmonic spectra within TD-SCHA within the bubble approximation (neglecting four phonon scattering tensor $\bDfour$ in the self-energy of \eqname~\ref{eq:self:energy:ansatz}). \textbf{d}: Full anharmonic spectra within TD-SCHA accounting for anharmonicity at any order. We performed 1000 iterations with a smearing of $\SI{10}{\per\centi\meter}$. The finite linewidth on panel \textbf{a} and \textbf{b} is for presentation porpoises, as those phonons have infinite lifetime. The spectrum is calculated with the incoming and out-coming polarizations along the direction aligned to the monoclinic plane. We employed the Lanczos algorithm described in \secname~\ref{sec:lanczos} with a smearing of $\SI{7}{\per\centi\meter}$ and 1000 steps for the simulations of panel \textbf{c} and \textbf{d}. Experimental data from ref.\cite{Goncharov_2001} are measured with unpolarized light, at $\SI{248}{\giga\pascal}$ and $\SI{140}{\kelvin}$.}
    \label{fig:V4}
\end{figure}

In \figurename~\ref{fig:V4} we compute the Raman spectra progressively switching on anharmonicity. 
In \figurename~\ref{fig:V4}(\textbf{a}), we show the harmonic Raman spectra, computed within perturbation theory.
We present in \figurename~\ref{fig:V4}(\textbf{b}) the Raman spectra of the SCHA equilibrium self-consistent harmonic Hamiltonian (\eqname~\ref{eq:H:scha}). Here, phonons have an infinite lifetime, as in the harmonic case. However, the peak positions and intensities are already strongly affected by anharmonicity through the equilibrium SCHA self-consistency. This response is the phonon spectra presented in the original derivation of the static SCHA\cite{Errea2014,Errea2015,ErreaRev2016}.
Indeed, phonons defined from the self-consistent harmonic Hamiltonian of the equilibrium density (the SSCHA auxiliary phonons) are not the correct dynamical response. To get the Raman spectrum, one has to calculate the dynamical response function, as introduced in \secname~\ref{sec:response}.
In \figurename~\ref{fig:V4}(\textbf{c},\textbf{d}), we report the anharmonic Raman spectrum calculated within TD-SCHA and the Lanczos algorithm introduced in \secname~\ref{sec:lanczos}. We compute the Raman signal within the bubble approximation in \figurename~\ref{fig:V4}(\textbf{c}), where we neglect four-phonon scattering processes in the expression of the self-energy (\eqname~\ref{eq:self:energy:ansatz}). This is equivalent to account only for the bubble dyagram of \figurename~\ref{fig:selfenergy}(\textbf{a}) replacing the interacting two-phonon propagator with the non interacting one. This is the standard way of computing dynamical properties starting from the SCHA and similar methods, as commonly done in TDEP\cite{TDEP}, and ALAMODE\cite{ALAMODE}. In \figurename~\ref{fig:V4}(\textbf{d}) we report the full dynamical response within TD-SCHA beyond the bubble approximation. Notably, the Raman signal strongly changes when we account for the full anharmonicity, both in the vibron (that acquires a higher line-width and non-Lorentzian shape) and low-energy phonons. Our Lanczos algorithm enables for the first time to systematically overcome the ``bubble'' approximation.

\begin{figure}
    \centering
    \includegraphics[width=\columnwidth]{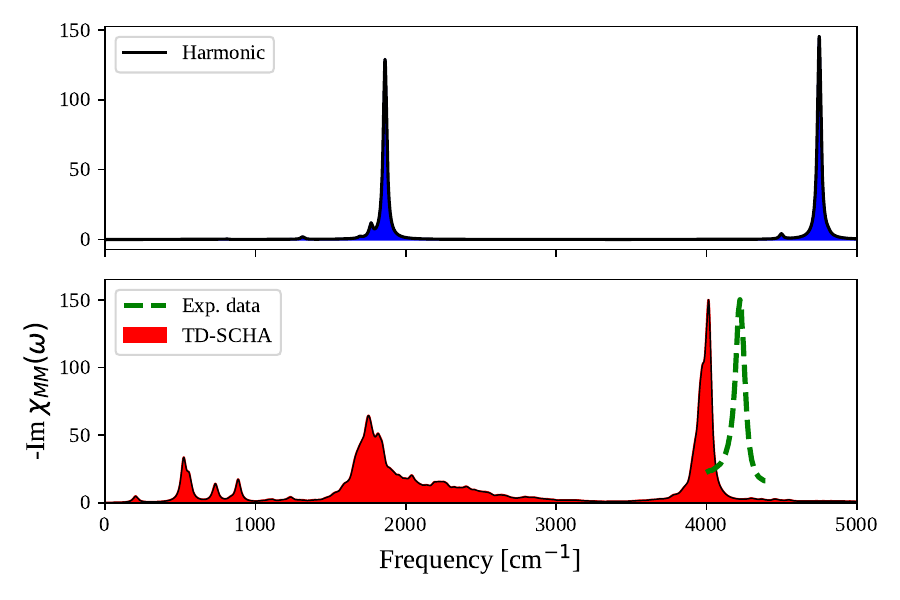}
    \caption{Simulation of the IR signal of high-pressure hydrogen at $\SI{260}{\giga\pascal}$ and $\SI{0}{\kelvin}$. We compare the Harmonic approximation (top panel) with the full anharmonic TD-SCHA spectrum beyond the bubble approximation (bottom panel). We also report the vibron peak measured at $\SI{248}{\giga\pascal}$ and $\SI{140}{\kelvin}$\cite{Goncharov_2001}. To compute the TD-SCHA spectrum with the Lanczos algorithm we employed a smearing of $\SI{20}{\per\centi\metre}$ and 1000 steps.} 
    \label{fig:ir}
\end{figure}

The downshift of the vibron peak (the strongest peak in the spectrum at high frequencies) from the Harmonic (\figurename~\ref{fig:V4}\textbf{a}) to the full anharmonic spectrum (\figurename~\ref{fig:V4}\textbf{d}) occurs already when considering equilibrium self-consistent harmonic phonons. The vibron acquires a finite lifetime when we calculate the response function with the Lanczos, deviating from the Lorentzian shape. We report the comparison with experimental data, obtained under similar conditions\cite{Goncharov_2001}. More details on the comparison with experiments have been extensively discussed elsewhere\cite{MonacelliNatPhys2020}. In particular, the bubble only approximation matches better the experimental data than the full TD-SCHA expression. However, this is an artifact of the DFT functional (BLYP) used to represent the nuclear energy landscape, which exasperates the quantum melting and dissociation of the \ch{H2} molecules under pressure\cite{MonacelliNatPhys2020,Drummond2015}. Thus the apparent good agreement of Bubble only is due to an error cancellation between higher order anharmonicity and the exchange correlation error in the DFT simulation. We report also the simulation of the IR and the comparison with experiments\cite{Goncharov_2001} in \figurename~\ref{fig:ir}.

\begin{figure}
    \centering
    \includegraphics[width=\columnwidth]{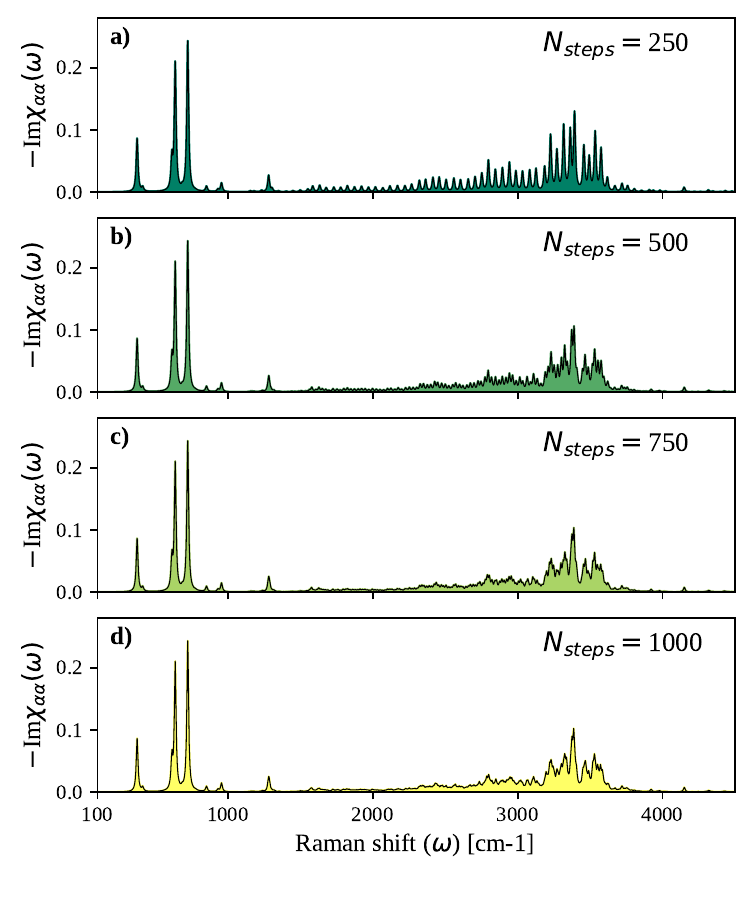}
    \caption{Convergence of the Raman signal with the number of steps in the Lanczos chain. Here we fixed the smearing $\eta = \SI{7}{\per\centi\meter}$. Lower values of smearing requires higher number of steps to converge. The exact solution of the Lanczos is recovered in the limit $\eta\rightarrow 0$ and $N_\text{steps} \rightarrow\infty$.}
    \label{fig:steps}
\end{figure}

In \figurename~\ref{fig:steps} we report how the Raman spectrum depends on the number of iterations in the Lanczos algorithm. As we increase the number of Lanczos iterations the spectrum gets refined.
To converge, we must do the limit $\eta \rightarrow 0$ and $N_\text{steps}\rightarrow\infty$; each step introduces a new pole in the Green function and a peak in the spectrum. When the distance between peaks is smaller than the smearing, the spectrum becomes smooth and does not change if we add more iterations. 

When the number of iterations is small, the use of a terminator (introduced in  \secname~\ref{sec:lanczos}) strongly improves the convergence with lower values of $\eta$. However, when the number of iterations is several hundred, it does not affect the spectral function.


Since the energy of the vibron is more than \SI{3000}{\per\centi\meter}, the temperature needed to populate excited states is above \SI{4000}{\kelvin}. There is no hope to simulate this lattice vibration with standard AIMD, that neglects quantum fluctuations. For this reason, previous studies on high-pressure hydrogen that neglected quantum effects reported a much more modest anharmonicity in the Raman signal\cite{Magdau_2013,Zhang_2018}.

These results indicate that the linear response of TD-SCHA can tackle complex open physical problems from first principles, providing an unprecedented precision on the evaluation of dynamical correlation functions that is sufficient to compare with experimental results\cite{MonacelliNatPhys2020}.

\section{Conclusions}

In this work, we introduced a new time-dependent theory for lattice dynamics based on the self-consistent harmonic approximation that can be applied to real systems with a first-principle treatment of the electrons. We discussed the linear response on the static SCHA equilibrium solution of the new equations and derived an efficient algorithm that computes the response function at any frequency with a single calculation. 
This algorithm, with the simple knowledge of two observables of ionic positions, can compute efficiently time-correlation function, with the inclusion of thermal and quantum effects. Notably, this result is not empirical but derived from a grounded basis: the least action principle.
We benchmark the theory both in a simple strongly anharmonic one-dimensional case and on phase III of solid hydrogen, a real complex molecular crystal, with \emph{ab initio} treatment of electrons.

The theory proved able to accurately reproduce complex phonon line-widths that arise from strong anharmonic coupling.
The TD-SCHA paves the way to predict from first-principles the outcome of most experiments that probe the lattice dynamics, like IR and Raman spectroscopy, Neutron, and X-Ray scattering. The theory can be employed in strongly anharmonic regimes, even when quantum fluctuations are dominant and where perturbative approach fails, as in cryogenic conditions, when light ions are present, or in systems close to a second order structural phase transition, as multiferroics, charge density waves, ferroelectrics and thermoelectric materials. Moreover, TD-SCHA describes also out-of-equilibrium dynamics, enabling the study of chemical reactions with light atoms, as the proton transfer in biomolecules, and the simulation of pump-probe spectroscopies.

\begin{appendices} 

\section{Equilibrium}
\label{app:equilibrium}

In this appendix we show that, among all possible stationary density matrices in \eqname~\eqref{eq:rho:scha}, the one that minimizes the functional:
$$
F = E - TS
$$
is the SCHA solution, i.e. the one where temperature is uniform on all the degrees of freedom. 

Since we are in a stationary solution, we can write everything in the basis of $\boldsymbol{e_\mu}$ vectors that diagonalizes the $\bUps$, $\bA$ and $\Avgclassic{\partial^2 V / \partial \tilde\bR ^2}$.

The $F$ functional is:
\begin{equation}
    F[\hat\rho] = \Avgquantum{\hat K + \hat V} - TS[\hat \rho]
\end{equation}
The average of the kinetic energy (\eqname~\ref{eq:kin:energy}) is:
\begin{equation}
    \Avgquantum{\hat K} = \frac{\hbar^2}{2} \sum_\mu \left(\frac{\tilde\Upsilon_\mu}{4} + \tilde A_\mu\right)
\end{equation}
Substituting \eqname~\eqref{eq:density:eq}, we get:
\begin{equation}
\Avgquantum{\hat K} = \sum_{\mu} \frac{\hbar \omega_\mu(2n_\mu + 1)}{4}
\end{equation}

The entropy $S[\hat \rho]$ can be obtained as the sum of the entropies of the Harmonic oscillators on the $\mu$ modes:
\begin{equation}
    S = \frac{k_b}{2}\sum_\mu \left[\frac{\beta_\mu \hbar\omega_\mu}{\tanh\frac{\beta_\mu\hbar\omega}{2}} - 2 \log \left(\sinh\frac{\beta_\mu\hbar\omega_\mu}{2}\right)\right]
\end{equation}

We must impose that
\begin{equation}
    \frac{\partial F}{\partial\beta_\mu} = 0
\end{equation}
\begin{equation}
  \frac{\partial \braket{\hat K}}{\partial \beta_\mu} + \frac{\partial \braket{\hat V}}{\partial \beta_\mu} - T \frac{\partial S}{\partial\beta_\mu} = 0
\end{equation}
\begin{equation}
    \frac{\partial \braket{\hat K}}{\partial\beta_\mu} = - \frac{\hbar^2 \omega_\mu^2}{8 \sinh^2\left(\frac{\beta_\mu\hbar\omega_\mu}{2}\right)}
\end{equation}

The derivative of the average of the potential can be computed exploiting the formalism introduced by Bianco et al.\cite{Bianco2017}. In particular, they showed that the derivative of any observable with respect to a the quantum fluctuations is:
\begin{equation}
    \frac{\partial \Avgclassic{V(\bR)}}{\partial \Upsilon_\mu} = \frac 12 \sum_{ab} \frac{\partial {{\tilde \Upsilon}^{-1}}_{ab}}{\partial \Upsilon_\mu} \Avgclassic{\frac{\partial^2 V}{\partial \tilde R_a \partial \tilde R_b}}
\end{equation}
We can write it in the polarization basis (exploiting \eqname~\eqref{eq:density:eq}:
\begin{equation}
    \frac{\partial \Avgclassic{V(\bR)}}{\partial \Upsilon_\mu} = - \frac 12 {\tilde \Upsilon}_\mu^{-2} \omega_\mu^2
\end{equation}
From which we get the average of the potential:
\begin{equation}
    \frac{\partial \Avgclassic{V(\bR)}}{\partial \beta_\mu} = 
    - \frac{\hbar^2 \omega_\mu^2}{8 \sinh^2\left(\frac{\beta_\mu \hbar\omega_\mu}{2}\right)}
\end{equation}
While the derivative of the entropy is:
\begin{equation}
    \frac{\partial S}{\partial \beta_\mu} = - \frac{\beta_\mu \hbar^2 \omega_\mu^2 k_b}{4 \sinh^2\left(\frac{\beta_\mu \hbar\omega_\mu}{2}\right)}
\end{equation}
Putting all together we get:
\begin{equation}
    \frac{2\hbar^2\omega_\mu^2 - k_b T 2\beta_\mu \hbar^2 \omega_\mu^2}{8 \sinh^2\left(\frac{\beta_\mu \hbar\omega_\mu}{2}\right)} = 0
\end{equation}
And we get the condition that $\beta_\mu$ must satisfy to minimize the free energy:
\begin{equation}
    \beta_\mu = \frac{1}{k_bT}
\end{equation}

\section{Least action}
\label{app:least:action}

Here, we prove that for a Gaussian pure state expressed in \eqname~\eqref{eq:psi:t0},  the action (\eqname~\ref{eq:action}) with the TD-SCHA dynamics is stationary.

First we break the action in three parts: kinetic, potential, and time:
\begin{equation}
    A_1 = \frac{1}{t_2 - t_1} \int_{t_1}^{t_2} dt\, \braket{\psi(t)| \hat K |\psi(t)}\label{eq:action:kin}
\end{equation}
\begin{equation}
    A_2 = \frac{1}{t_2 - t_1} \int_{t_1}^{t_2} dt\, \braket{\psi(t)| \hat \Vtot |\psi(t)}
    \label{eq:action:v}
\end{equation}
\begin{equation}
    A_3 = -\frac i \hbar \frac{1}{t_2 - t_1} \int_{t_1}^{t_2} dt\, \braket{\psi(t)| \frac{d}{dt} |\psi(t)}
    \label{eq:action:dt}
\end{equation}
by substituting \eqname~\eqref{eq:psi:t0} in \eqname~\eqref{eq:action:kin}, \eqref{eq:action:v} and \eqref{eq:action:dt}, we get:
\begin{equation}
    A_1 = \frac{\hbar^2}{t_2 - t_1} \int_{t_1}^{t_2} dt\left\{
    \sum_a \frac{Q_a^2}{2m_a} + 
    \Tr{\frac{\tilde\bUps}{8} + 2 {\tilde \bC}_{ab} {\tilde\bUps}^{-1}\tilde \bC}\right\}
\end{equation}
From which the derivatives of the action are:
\begin{equation}
    \frac{\delta A_1}{\delta \tilde\bC} = 2\hbar^2 {\tilde \bUps}^{-1} \tilde\bC + 2\hbar^2 \tilde\bC {\tilde \bUps}^{-1}
\end{equation}
\begin{equation}
    \frac{\delta A_1}{\delta Q_a} = \frac{\hbar^2 Q_a}{m_a}
    \qquad
    \frac{\delta A_1}{\delta \bRcal} = 0
\end{equation}
\begin{equation}
    \frac{\delta A_1}{\delta{\tilde\Upsilon}_{ab}} = - \frac{\hbar^2}{8}\sum_{hk} \tilde\Upsilon_{ch}\tilde\Upsilon_{ck}\frac{\partial {\tilde\Upsilon}^{-1}_{hk}}{\partial\tilde\Upsilon_{ab}} + 2\hbar^2
    \sum_{cde}\tilde C_{cd}\frac{\partial{\tilde\Upsilon}^{-1}_{de}}{\partial\tilde\Upsilon_{ab}} \tilde C_{ec}
\end{equation}
In the same way $A_2$:
\begin{equation}
    \frac{\delta A_2}{\delta\Upsilon_{ab}} = \frac 12 \sum_{cd}\frac{\partial \Upsilon^{-1}_{cd}}{\partial\Upsilon_{ab}} \Avgclassic{\frac{\partial^2 \Vtot}{\partial R_c\partial R_d}}
\end{equation}
\begin{equation}
    \frac{\delta A_2}{\delta \Rcal_a} = - \Avgclassic{f_a^{(tot)}}
\end{equation}
\begin{equation}
    \frac{\delta A_2}{\delta Q_a} =\frac{\delta A_2}{\delta C_{ab}} = 0
\end{equation}
And for $A_3$:
\begin{equation}
    \frac{\delta A_3}{\delta Q_a} = \hbar \frac{d\Rcal_a}{dt}
\end{equation}
\begin{equation}
    \frac{\delta A_3}{\delta \Rcal_a} = -\hbar \frac{dQ_a}{dt}
\end{equation}
\begin{equation}
    \frac{\delta A_3}{\delta C_{ab}} = -\hbar \sum_{cd} \frac{\partial \Upsilon^{-1}_{ab}}{\partial\Upsilon_{cd}} \frac{d\Upsilon_{cd}}{dt}
\end{equation}
\begin{equation}
    \frac{\delta A_3}{\delta \Upsilon_{ab}} = \hbar \sum_{cd}\frac{\partial\Upsilon^{-1}_{cd}}{\partial\Upsilon_{ab}} \frac{d C_{cd}}{dt}
\end{equation}

By imposing that:
$$
\delta A = \delta A_1 + \delta A_2 + \delta A_3 = 0
$$
we get the equation of motion.
$$
\frac{\delta A}{\delta Q} = 0
$$
\begin{equation}
    \frac{d\Rcal_{a}}{dt} =  \frac{\hbar Q_a}{m_a}
\end{equation}
$$
\frac{\delta A}{\delta \Rcal} = 0
$$
\begin{equation}
    \frac{dQ_a}{dt} = \frac 1 \hbar \Avgclassict{ f_a^{(tot)}}
    \label{eq:newton:t0}
\end{equation}
Indeed, \eqname~\eqref{eq:newton:t0} is equal to \eqname~\eqref{eq:newton}. We can proceed to get the equations for $\bC$ and $\bUps$.
To proceed, we need to remove the $\partial\bUps^{-1}/\partial \bUps$ tensor:
\begin{equation}
\frac{\partial\Upsilon^{-1}_{cd}}{\partial \Upsilon_{ab}} = -\frac 12  \Upsilon_{ac}^{-1}\Upsilon_{bd}^{-1} - \frac 12 \Upsilon_{cb}^{-1}\Upsilon_{ad}^{-1}
\end{equation}
$$
\frac{\delta A}{\delta C_{ab}} = 0
$$
\begin{equation}
    \frac{d\tilde\bUps}{dt} = 2\hbar\left({\tilde \bUps} \tilde\bC + \tilde\bC {\tilde \bUps}\right) 
    \label{eq:ups:t0}
\end{equation}
\eqname~\eqref{eq:ups:t0} is equal to \eqname~\eqref{eq:ups:dyn} if we set $\bA = 0$.

Lastly,
$$
\frac{\delta A}{\delta \Upsilon_{ab}} = 0
$$
\begin{equation}
\frac{d\tilde\bC}{dt} = \frac {1}{2\hbar} \Avgclassic{\frac{\partial^2 V}{\partial \tilde\bR\partial\tilde\bR}} -
\frac{\hbar}{8} \tilde\bUps\tilde\bUps + 2\hbar \tilde\bC \tilde\bC
\label{eq:C:t0}
\end{equation}
Indeed, also in this case, \eqname~\eqref{eq:C:t0} is equal to \eqname~\eqref{eq:C:dyn} if we set $\bA = 0$.

Therefore, we found that the dynamical equations for obtained with the least action principle coincides with those obtained from the TD-SCHA.

\section{Equations of motion}
\label{app:full:dynamical}

In this section, we report the details of the derivation of the dynamical equation of motion in \eqname~\eqref{eq:dynamics}.

We need to cast the density matrix definition \eqname~\eqref{eq:density:matrix} into the TD-SCHA equation (\eqname~\ref{eq:tdscha:ft}).

First, we compute the left-hand side of the dynamical equation of motion:
$$
i\hbar \frac{d}{dt} \hat \rho
$$
where $\hat \rho$ is the general Gaussian density matrix defined in \eqname~\eqref{eq:density:matrix}.
We write the density matrix in real space $\rho(\bR, \bR')$.

\begin{align}
  \frac{1}{\rho}\frac{d\rho}{dt} &= \sum_{ab}(R_a - \Rc_a)(R_b - \Rc_b)\left[ i \frac{d}{dt} C_{ab} - \frac 1 4\frac{d}{dt} \Theta_{ab}\right] +\nonumber\\
  & +  \sum_{ab}(R_a' - \Rc_a)(R_b' - \Rc_b)\left[ -i \frac{d}{dt} C_{ab} - \frac 1 4\frac{d}{dt} \Theta_{ab}\right] +\nonumber\\
  & + \sum_{ab}(R_a' - \Rc_a)(R_b - \Rc_b)\frac{d}{dt} A_{ab} + \nonumber\\
  & - \sum_{ab}(R - \Rc_a)\left[2i C_{ab} - \frac 12 \Theta_{ab} + A_{ab}\right]\frac{d}{dt}\Rc_b + \nonumber\\
  & - \sum_{ab}(R' - \Rc_a)\left[-2i C_{ab} - \frac 12 \Theta_{ab} + A_{ab}\right]\frac{d}{dt}\Rc_b + \nonumber\\
  & - i\sum_{a}(R_a - \Rc_a)\frac{dQ_a}{dt} + i \sum_a(R'_a - \Rc_a)\frac{dQ_a}{dt} + \frac{1}{N} \frac{dN}{dt}
  \label{eq:drhodt}
\end{align}
This is a polynomial in $(\bR - \bRcal)$ and $(\bR' - \bRcal)$. On the right-hand side of the TD-SCHA equation (\eqname~\ref{eq:tdscha:ft}) we have the Liouville operator:
\begin{equation}
\Hcal[\rho] \hat \rho - \hat\rho \Hcal[\rho],
\label{eq:liouville:comm}
\end{equation}
where the Hamiltonian is:
\begin{align}
\Hcal[\rho] &=  \frac 12 \sum_{ab}(R_a -\Rc_a)\Avgclassict{\frac{d^2 \Vtot}{d R_adR_b}} (R_b - \Rc_b) + \nonumber \\ 
& -
\sum_a\Avgclassict{f_a}(R_a - \Rc_a) + \sum_{a}\frac{p_a^2}{2m_a}. 
\end{align}

We compute first the commutator with the kinetic operator: 
$$
p_a^2 \rightarrow - \hbar^2 \frac{\partial^2}{\partial R_a^2}
$$

We start computing the first derivative of the $R_a$ variable:
\begin{align}
\frac{1}{\rho}\frac{d\rho}{dR_a}& = \sum_b(R_b - \Rc_b)\left[2i C_{ab} - \frac{\Theta_{ab}}{2}\right] + \nonumber \\
& + \sum_b(R_b' - \Rc_b)A_{ab} - i Q_a
\end{align}
Then, we perform the second derivative:
\begin{widetext}
\begin{align}
  \frac{1}{\rho}\frac{d^2\rho}{dR_a^2} &= \sum_{bc}(R_b - \Rc_b)(R_c - \Rc_c)\left[-4 C_{ab}C_{ac} + \frac{\Theta_{ab}\Theta_{ac}}{4} -iC_{ab}\Theta_{ac} -i C_{ac}\Theta_{ab}\right] + \nonumber\\
  &+\sum_{bc}(R_b' - \Rc_b)(R_c' - \Rc_c)A_{ab}A_{ac}+
  \sum_{bc}(R_b - \Rc_b)(R'_c - \Rc_c)\left(4iC_{ab}A_{ac} - \Theta_{ab}A_{ac}\right) + \nonumber\\
  &+\sum_b(R_b - \Rc_b)(4C_{ab}Q_a + iQ_a\Theta_{ab}) - \sum_b(R'_b - \Rc_b)2iA_{ab}Q_a - Q_a^2 + 2iC_{aa} - \frac 12 \Theta_{aa}
  \label{al:kin:mix}
\end{align}
\end{widetext}

If we change the derivative of $R$ with $R'$ (that is obtained from $\hat \rho\Hcal[\rho]$), we get the complex conjugate of \eqname~\eqref{al:kin:mix} where $R$ and $R'$ variables exchanged.

Therefore, we collect all the terms in \eqname~\eqref{eq:drhodt} and \eqname~\eqref{eq:liouville:comm} quadratic in $\bR$, i.e. the coefficients of $\sum_{ab}(R_a - \Rc_a)(R_b-\Rc_b)$).

\begin{align}
i\hbar \left(i \frac{dC_{ab}}{dt} - \frac14 \frac{d\Theta_{ab}}{dt}\right) & = -\frac 12 \Avgclassic{\frac{d^2\Vtot}{dR_adR_b}} + \nonumber \\
&- \sum_c\frac{\hbar^2}{2m_c} \bigg[4 C_{ac}C_{cb}  - \frac 14 \Theta_{ac}\Theta_{cb} + \nonumber \\
& +2i C_{ac}\Theta_{cb} + A_{ac} A_{bc}\bigg]
\end{align}
We can split the imaginary and real part to get the first two equation of motion for $\bC$ and $\bTheta$:
\begin{align}
\frac{dC_{ab}}{dt} =  \frac{1}{2\hbar}& \Avgclassic{\frac{d^2V}{dR_adR_b}} + \sum_c\frac{\hbar}{2m_c} \bigg[4 C_{ac}C_{cb} + \nonumber \\
& - \frac 14 \Theta_{ac}\Theta_{cb}  + \Re\left[A_{ac} A_{bc}\right]\bigg]
\end{align}
\begin{align}
\frac{d\Theta_{ab}}{dt} = \sum_c\frac{2\hbar}{m_c}\left(C_{ac}\Theta_{cb} + C_{bc}\Theta_{ca} + \Im \left[A_{ac} A_{bc}\right]\right)
\end{align}

We now consider the equality between \eqname~\eqref{eq:drhodt} and \eqref{eq:liouville:comm} of the coefficient that multiplies $(\bR - \bRcal)(\bR' - \bRcal)$.

\begin{align}
i\hbar \frac{dA_{ab}}{dt} = \sum_c\frac{\hbar^2}{2m_c} &\big[4 i C_{ac}A_{cb} + 4iC_{bc}A_{ac} +\nonumber \\
& - \Theta_{ac}A_{cb} + \Theta_{bc}A_{ac}\big] \label{eq:A:brutta}
\end{align}

We can split also this equation in real and imaginary part:
\begin{align}
\Re \frac{dA_{ab}}{dt} = \sum_c \frac{\hbar}{2m_c} &\big( 4 C_{ac}\Re A_{cb} + 4 C_{bc} \Re A_{ca} +\nonumber \\ 
&- \Theta_{ac}\Im A_{cb} + \Theta_{bc}\Im A_{ac}\big)
\end{align}

\begin{align}
 \Im \frac{dA_{ab}}{dt} =
\sum_{c} \frac{\hbar}{2m_c}&\big[4 C_{ac} \Im A_{cb} + 4 C_{bc}\Im A_{ac} + \nonumber \\ 
& +  \Theta_{ac}\Re A_{cb} - \Theta_{bc}\Re A_{ac}\big]
\end{align}

The last two equations for $\bQ$ and $\bRcal$ are obtained from the real and imaginary part of the terms linear in $\bR$

\begin{align}
-i\hbar&\sum_b\left(2iC_{ab} - \frac 12 \Theta_{ab} + A_{ab}\right)\frac{d\Rc_b}{dt} - \hbar \frac{dQ_a}{dt}  = \nonumber \\
& = \sum_c\frac{\hbar^2}{2m_c}\left(4C_{ca}Q_c +iQ_c \Theta_{ca} -2i A_{ca}Q_c\right) - \braket{f_a^{(tot)}}
\end{align}
Separating real and imaginary parts we get:
\begin{align}
\sum_b&\left(2 C_{ab} + \Im A_{ab}\right)\frac{d\Rcal_b}{dt} -  \frac{dQ_a}{dt} = \nonumber \\ & = \sum_b\frac{\hbar}{m_b}\left(2 C_{ab} + \Im A_{ab}\right)Q_b - \frac{\braket{f^{(tot)}_a}}{\hbar}
\end{align}
\begin{align}
\sum_b \hbar& \left(\frac 12 \Theta_{ab} -  \Re A_{ab}\right)\frac{d\Rc_b}{dt} = \nonumber \\
& = \sum_b \frac{\hbar^2}{2m_b}\left(\Theta_{ab} -2 \Re A_{ab}\right)Q_b  
\end{align}

Simplifying we get the last equations:
\begin{equation}
\frac{d\Rc_a}{dt} = \frac{\hbar Q_a}{m_a}
\end{equation}
\begin{equation}
\frac{dQ_a}{dt} =  \frac{\braket{f_a}}{\hbar}
\end{equation}

The last term of the equivalence between \eqname~\eqref{eq:drhodt} and \eqref{eq:liouville:comm} is the term that does depend in neither $(\bR - \bRcal)$ nor $(\bR' - \bRcal)$. This last equation expresses the time evolution for the norm $N(t)$, and it is just the conservation of the density matrix normalization.

The final equations of motion are obtained substituting the expression of $\bUps$ (\eqname~\ref{eq:upsilon}) and dividing for the masses to obtain the mass-rescaled matrices (\eqname~\ref{eq:mass:rescaled}).
The convenience of rescaling the masses is that we can express all the summations as standard rows-by-columns products.

\section{Energy conservation}
\label{app:energy:cons}

Here we show that the TD-SCHA equation of motion (\eqname~\ref{eq:tdscha:ft}) satisfies energy conservation when we switch off the time-dependent perturbation, as we expect from a closed system. This is not trivial: it does not happen in other methods for finite temperature dynamics of ions, as the finite temperature multi-configurational time-dependent Hartree  method\cite{Meyer2003}.

The total energy is computed as the average of the BO Hamiltonian on the time-dependent density matrix:
\begin{equation}
    E(t) = \Avgquantumt{\hat H} = \tr\left[ \hat \rho(t) \hat H\right].
\end{equation}

By dividing into kinetic and potential contribution, the total energy is:
\begin{equation}
    E(t) = \Avgquantumt{\hat K} + \Avgquantumt{\hat V}
\end{equation}

The kinetic energy can be evaluated analytically directly from the density matrix of \eqname~\eqref{eq:density:matrix}:
\begin{align}
    \Avgquantumt {\hat K} = 
    \frac{\hbar^2}{2}\bigg[&\tr\bigg(4\tilde \bC(\tilde \bUps)^{-1} \tilde \bC -  \Im\tilde \bA (\tilde \bUps)^{-1} \Im \tilde \bA\bigg) + \nonumber \\ 
    & - 2 \tr\bigg(\tilde \bC (\tilde \bUps)^{-1} \Im \tilde \bA\bigg)
    + \nonumber \\
    & + \frac 14 \tr \tilde \bUps +
    \tr \tilde \bA\bigg] + \sum_{i = 1}^n \frac{\hbar^2 Q_i^2}{2m_i}
    \label{eq:kin:energy}
\end{align}
Also here, we dropped the explicit time dependency of the parameters that represent $\hat\rho(t)$.

The last term is the kinetic energy of classical particles, the rest is the contribution of the quantum and thermal fluctuations to the kinetic energy.

To prove that $\frac{d}{dt} E(t) = 0$, we compute the derivative of the average kinetic energy:
\begin{align}
    \frac{d\Avgquantumt{\hat K}}{dt} &= \frac{\hbar^2}{2}\bigg[  \frac 14 \tr\frac{d\tilde \bUps}{dt}
    + \sum_{i = 1}^{3n} \frac{Q_i }{m_i}\frac{dQ_i}{dt} +\nonumber \\
   & + 2\tr\left(4\tilde \bC (\tilde \bUps)^{-1} \frac{d {\tilde \bC}}{dt} -  \tilde \bC(\tilde \Upsilon)^{-1} \frac{d\Im \tilde \bA}{dt} \right) +   \nonumber\\
   & + \tr \frac{d \Re \tilde \bA}{dt} -2\tr\left(\frac {d\tilde \bC}{dt}(\tilde \bUps)^{-1} \Im \tilde \bA\right) +\nonumber \\
    &- 2\tr\left(\Im \tilde \bA (\tilde\bUps)^{-1} \dot{\Im\tilde \bA} +  \tilde \bC \frac{d(\tilde\bUps)^{-1}}{dt} \Im \tilde \bA\right) + \nonumber \\
    & + \tr\left(4 \tilde \bC \frac{d(\tilde\bUps)^{-1}}{dt} \tilde \bC - \Im\tilde \bA \frac{d(\tilde\bUps)^{-1}}{dt} \Im \tilde \bA\right)\bigg].
\end{align}

By substituting the equation of motion (\eqname~\ref{eq:dynamics})  we get the final derivative of the kinetic energy:

\begin{align}
    \frac{d\Avgquantumt{\hat K}}{dt} &= \sum_i \frac{\hbar Q_i \Avgclassic{f_i}}{m_i} + \nonumber \\
    & + 2\hbar \tr \left[\Avgclassict{\frac{\partial^2V}{\partial\tilde \bR \partial\tilde \bR}} (\tilde\bUps)^{-1} \tilde \bC \right] + \nonumber \\
    & -  \hbar
    \tr \left[\Avgclassict{\frac{\partial ^2V}{\partial \tilde \bR \partial\tilde \bR}} (\tilde\bUps)^{-1} \Im\tilde \bA\right]
\end{align}

In the same way, we can compute the derivative of the average potential:
\begin{align}
    &\frac{d\Avgquantumt {\hat V}}{dt}  =
    - \sum_{ab}\Avgclassict{(R_a -\Rc_a)(R_b - \Rc_b) V}
    \frac 12 \frac{d\Upsilon_{ab}}{dt} + \nonumber \\
    & + 
    \sum_{ab}\Avgclassict{(R_a - \Rc_a)V}\Upsilon_{ab} \frac{d\Rc_b}{dt} + 
    \Avgclassict{V} \frac {1}{\norm} \frac{d\norm}{dt}
    \label{eq:dV:dt}
\end{align}

By substituting the equation of motion and integrating by parts the averages, it is possible to show that:
\begin{equation}
    \frac{d}{dt}\Avgquantumt{V} = - \frac{d}{dt} \Avgquantumt{K}
\end{equation}

To finally prove this equation, we calculate the time derivative of the average potential (\eqname~\ref{eq:dV:dt}), showing that it balances the time derivative of the average kinetic energy, proving that the total energy is conserved by the equation of motions.

For this purpose, we use formalism introduced by Bianco et al\cite{Bianco2017} (appendix C, \eqname~C1): given an observable $O(\bR)$, the average of its derivative can be written as:
\begin{equation}
    \Avgclassic{\frac{dO}{dR_a}} = \sum_{b} \Upsilon_{ab}\Avgclassic{(R_b - \Rcal_b) O}
    \label{eq:bianco}
\end{equation}

For this reason, we can write:
\begin{equation}
    \sum_{ab} \frac{d\Rcal_b}{dt} \Upsilon_{ab} \Avgclassic{(R_a - \Rcal_a)V} = -\sum_b \frac{d\Rcal_b}{dt}\Avgclassic{f_a(\bR)}
    \label{eq:finalav1}
\end{equation}
For simplicity, we define $\bu = (\bR - \bRcal)$. The other term of \eqname~\eqref{eq:dV:dt} is
\begin{align}
    \sum_{ab}\Avgclassic{u_au_b V} \frac 12 \frac{d\Upsilon_{ab}}{dt}& = \sum_{abcd}\Avgclassic{u_cu_b V} \Upsilon_{cd}\Upsilon^{-1}_{da}\frac 12 \frac{d\Upsilon_{ab}}{dt} \nonumber \\
    & = \sum_{abd} \Avgclassic{\frac{d(u_b V)}{dR_d}} \Upsilon^{-1}_{da}\frac 12 \frac{d\Upsilon_{ab}}{dt} 
\end{align}
The derivative gives:
\begin{align}
    \sum_{ab}\Avgclassic{u_au_b V} \frac 12 \frac{d\Upsilon_{ab}}{dt} & =  \sum_{ab} \Avgclassic{V} \Upsilon^{-1}_{ba}\frac 12 \frac{d\Upsilon_{ab}}{dt}  + \nonumber \\
    & + \sum_{abd} \Avgclassic{u_b \frac{dV}{dR_d}} \Upsilon^{-1}_{da}\frac 12 \frac{d\Upsilon_{ab}}{dt} 
\end{align}

The first term is zero, because:
\begin{equation}
    \sum_{ab}\Upsilon^{-1}_{ba} \frac{d\Upsilon_{ab}}{dt} = \frac 12\frac{d}{dt}\Tr{\bUps^{-1} \bUps} = 0
\end{equation}

While for the second term we can proceed again:
\begin{equation}
     \sum_{ab}\Avgclassic{u_au_b V} \frac 12 \frac{d\Upsilon_{ab}}{dt} = \sum_{abcde} \Avgclassic{u_e \frac{dV}{dR_d}}\Upsilon_{ec}\Upsilon^{-1}_{cb} \Upsilon^{-1}_{da}\frac 12 \frac{d\Upsilon_{ab}}{dt} 
\end{equation}

\begin{equation}
     \sum_{ab}\Avgclassic{u_au_b V} \frac 12 \frac{d\Upsilon_{ab}}{dt} = \sum_{abcd} \Avgclassic{ \frac{d^2V}{dR_cdR_d}}\Upsilon^{-1}_{cb} \Upsilon^{-1}_{da}\frac 12 \frac{d\Upsilon_{ab}}{dt} 
\end{equation}
Introducing the mass re-scaled for $\bUps^{-1}$:
\begin{equation}
    {\tilde\Upsilon}^{-1}_{ab}=  \sqrt{m_a m_b} \Upsilon^{-1}_{ab},
\end{equation}
we get the final equivalence:

\begin{equation}
    \frac 12 \Tr{\Avgclassic{\bu\otimes \bu V} \frac{d\bUps}{dt}} = \frac 12\Tr{ \Avgclassic{\frac{d^2 V}{d\tilde\bR d\tilde\bR}} {\tilde \bUps}^{-1} {\tilde\bUps^{-1}} \frac{d\tilde\bUps}{dt}}.\label{eq:finalav2}
\end{equation}

Substituting \eqname~\eqref{eq:finalav1} and \eqref{eq:finalav2} into \eqname~\eqref{eq:dV:dt}, and substituting the equation of motion, it is easy to show that
$$
\frac{d}{dt} \Avgclassict{\hat K} = - \frac{d}{dt}\Avgclassict{\hat V}
$$

\section{Entropy conservation}
\label{app:entropy:cons}

The entropy defined on the many-body density matrix is:
\begin{equation}
    S[\hat\rho(t)] = - k_b \tr \left[\hat\rho(t)\log\hat\rho(t)\right]
\end{equation}

for simplicity, we drop the explicit $t$ dependence of $\hat\rho(t)$.

\begin{equation}
     B(\hat \rho) = \hat \rho\log\hat\rho
\end{equation}
\begin{equation}
    \frac{dS}{dt} = -k_b \tr \left[\frac{d\hat\rho}{dt}\frac{dB(\hat \rho)}{d\hat\rho}\right]
\end{equation}
\begin{equation}
    \frac{dS}{dt} = \frac{ik_b}{\hbar}\left[\tr\left(\Hcal[\rho] \hat\rho\frac{d B(\hat\rho)}{d\hat\rho}\right) - \tr\left(\hat\rho\Hcal[\rho]  \frac{d B(\hat\rho)}{d\hat\rho}\right)\right]
    \label{eq:ds:dt}
\end{equation}
By exploiting the cyclic permutation of the trace and the commutation between $d B(\hat \rho)/d\hat\rho$ and $\hat\rho$, \eqname~\eqref{eq:ds:dt} is zero.

\section{Steady state solutions}
\label{app:steady:state}
Substituting \eqname~\eqref{eq:stationary} inside \eqname~\eqref{eq:dynamics} we get the following conditions:
\begin{subequations}
\begin{equation}
\Avgclassic{f_a} = 0\label{eq:static:wyckoff}
\end{equation}
\begin{equation}
    \frac{1}{2\hbar} \Avgclassic{\frac{\partial^2 V}{\partial \tilde R_a \partial \tilde R_b}} = \frac\hbar 2 \left( \frac 1 4 \tilde\Theta_{ab}^2 - \Re {\tilde A}^2_{ab}\right)
\end{equation}
\end{subequations}
\eqname~\eqref{eq:static:wyckoff} is a necessary condition for the equilibrium SCHA: the average of the BO forces is the gradient of the SCHA free energy versus the centroids positions $\bRcal$\cite{Errea2014}. 
Moreover, from \eqname~\eqref{eq:dia:dt} we have that $\tilde\bTheta$ and $\Re\tilde \bA$ commute in the static solution, from which also $\tilde \bUps$ and $\Re\tilde \bA$ commute (that is a direct consequence of the commutation between $\hat\rho$ and $\Hcal[\rho]$).
Since it is better to express our quantity as a function of the total dispersion $\tilde \bUps$ (\eqname~\ref{eq:upsilon}) we have:

\begin{equation}
    \frac{\tilde\bUps^2}{4} + \tilde\bUps \Re \tilde \bA  = \frac{1}{\hbar^2} \Avgclassic{\frac{\partial^2 V}{\partial \tilde \bR\partial \tilde \bR}}.\label{eq:static1}
\end{equation}
If we express \eqname~\eqref{eq:static1} in the basis that diagonalizes both $\tilde \bUps$ and $\Re\Tilde \bA$, the same basis must diagonalize also $\braket{\frac {\partial^2 V}{\partial\tilde \bR\partial\tilde \bR}}$. 
Let us define $\boldsymbol{e_\mu}$ the eigenvector of $\braket{\frac {\partial^2 V}{\partial\tilde \bR\partial\tilde \bR}}$, $\tilde\bUps$ and $\Re\tilde\bA$ and $\omega_\mu^2$, $\tilde\Upsilon_\mu$ and $\Re\tilde A_\mu$ the corresponding eigenvalues:
\begin{equation}
    \sum_{b} \Avgclassic{\frac{\partial^2 V}{\partial \tilde R_a \partial\tilde R_b}} e_\mu^b = \omega_\mu^2 e_\mu^a,
    \label{eq:scha:freqs}
\end{equation}
\eqname~\eqref{eq:static1} becomes:
\begin{equation}
    \frac{\tilde \Upsilon^2_\mu}{4} + \tilde\Upsilon_\mu \Re \tilde A_\mu = \frac {\omega_\mu^2}{ \hbar^2}. \label{eq:scha_mu}
\end{equation}

Without loss of generality, if we change variable introducing a new parameter $n_\mu$ and define:
\begin{subequations}
\begin{equation}
    \tilde\Upsilon_\mu = \frac{2\omega_\mu}{\hbar(2n_\mu + 1)},
\end{equation}
from \eqname~\eqref{eq:scha_mu}, we get:
\begin{equation}
    \Re\tilde A_\mu = \frac{2\omega_\mu n_\mu(n_\mu +1)}{\hbar(2n_\mu + 1)}.
\end{equation}
\label{eq:density:eq}
\end{subequations}
In these equations, $n_\mu$ is a free parameter for each frequency: the TD-SCHA equations are stationary for any choice of $n_\mu$.
In particular, we can write $n_\mu$ as the Bose-Einstein occupation number with a temperature that depends on $\mu$:
\begin{equation}
    n_\mu = \frac{1}{e^{\beta_\mu\hbar\omega_\mu} - 1}\label{eq:bose:einstein}
\end{equation}

The stationary density matrix identified by \eqname~\eqref{eq:density:eq} and \eqref{eq:stationary} is the product of equilibrium densities matrices of harmonic oscillators of frequencies $\omega_\mu$ and temperatures $\beta_\mu$.

\section{Derivation of the linear response system}
\label{app:linear:response}
Here, we apply perturbation theory on the TD-SCHA equations for small perturbations. Starting from this paragraph through the rest of the paper, we drop the $^{(0)}$ index and all the quantities without $^{(1)}$ or the explicit time dependency refer to equilibrium quantities.
Since we are expanding around equilibrium solutions, we have:
\begin{equation}
\bC = 0 \qquad
\bQ = 0 \qquad
\Im \bA = 0
\end{equation}
When $t \ge t_0$, we add an external time dependent perturbation to the non interacting potential:
\begin{equation}
V(\bR, t) = V^{(0)}(\bR) +  V^{(1)}(\bR, t)
\end{equation}
And we want to study the dynamics of the system close to equilibrium. This perturbation affects the dynamics (\eqname~\ref{eq:dynamics}) only in the two averages:
\begin{equation}
\Avgclassic{f_a} \qquad \Avgclassic{\frac{d^2V}{dR_adR_b}}
\end{equation}

Since the perturbation changes the parameters of the density matrix, they also change the ionic probability distribution:
\begin{equation}
    \rho(\bR, t) = \rho(\bR) + \rho^{(1)}(\bR, t)
\end{equation}

This affect also the generic average over an observable:
\begin{equation}
    \Avgclassict{O(\bR)} = \Avgclassiceq{O(\bR)} + \Avgclassicpert{O^{(0)}(\bR)}
\end{equation}

By expanding the probability distribution $\rho^{(1)}(\bR, t)$ as a function of the perturbed centroid position $\bRcal^{(1)}$ and the perturbed fluctuations $\bUps^{(1)}$,
we get:
\begin{align}
\Avgclassicpert{O} = \frac 1 2 \sum_{ab}\bigg[&\Upsilon_{ab}\left(\Avgclassiceq{Ou_a}\Rc_b^{(1)} +  
 \Rc_a^{(1)}\Avgclassiceq{Ou_b}\right) + \nonumber \\
 & - \Upsilon_{ab}^{(1)}\left(\Avgclassiceq{u_au_b O} - \Upsilon^{-1}_{ba} \Avgclassiceq{O}\right)\bigg]
\end{align}
where $u$ is the displacement with respect to the equilibrium centroid.
In particular, we are interested in the averages of forces and second derivative of the BO potential:
\begin{align}
\Avgclassicpert{\frac{dV}{dR_a}} &= \frac 12 \sum_{hk}\bigg[ \Upsilon_{hk}\Avgclassiceq{\frac{dV}{dR_a}u_h}\Rc^{(1)}_k +  \nonumber \\ 
& + \Upsilon_{hk} \Rc_h^{(1)}\Avgclassiceq{\frac{dV}{dR_a} u_k} 
- \Upsilon_{hk}^{(1)}\Avgclassiceq{u_hu_k\frac{dV}{dR_a}}\bigg]
\end{align}

Using the definition of the 3 and 4 phonon scattering tensor $\bPhithree$ and $\bPhifour$ (ref.\cite{Bianco2017}), we have:
\begin{equation}
    \Phi_{ab} = \sum_p \Upsilon_{ap} \Avgclassiceq{u_p \frac{dV}{dR_b}}
\end{equation}
\begin{equation}
\Phithree_{abc} =  \sum_{pq} \Upsilon_{ap}\Upsilon_{bq}\Avgclassiceq{u_pu_q \frac{dV}{dR_c}}
\end{equation}
\begin{equation}
\Phifour_{abcd} =  \sum_{pqr} \Upsilon_{ap}\Upsilon_{bq}\Upsilon_{cr}\Avgclassiceq{u_pu_qu_r \left(\frac{dV}{dR_d} + \sum_k\Phi_{dk}u_k\right)}
\end{equation}
Thus we have:
\begin{align}
\Avgclassicpert{\frac{dV}{dR_a}} =  -\frac 12&
\sum_{hkpqrs} \Upsilon^{(1)}_{hk}\Upsilon^{-1}_{kp}\Upsilon_{ps}\Upsilon^{-1}_{hq}\Upsilon_{qr} \Avgclassiceq{u_ru_s\frac{dV}{dR_a}} + \nonumber \\
& + \sum_{h} \Phi_{ah}\Rc_h^{(1)}
\end{align}

\begin{equation}
\Avgclassicpert{\frac{dV}{dR_a}} = \sum_{h} \Phi_{ah}\Rc_h^{(1)} -\frac12
\sum_{hkpq} \Upsilon^{(1)}_{hk}\Upsilon^{-1}_{kp}\Upsilon^{-1}_{hq}\Phithree_{qpa}
\end{equation}

To simplify notation, in the rest of this section we use the convention:
$$
\braket{\cdot} = \Avgclassiceq{\cdot}
$$

in an analogous way we can get the other term of the perturbation:
\begin{align}
\Avgclassicpert{\frac{d^2V}{dR_adR_b}} = \frac 12 \sum_{hk}&\bigg[ \Upsilon_{hk}\braket{\frac{d^2V}{dR_adR_b}u_h}\Rc^{(1)}_k + \nonumber \\
& + \Upsilon_{hk} \Rc_h^{(1)}\braket{\frac{d^2V}{dR_adR_b} u_k} +\nonumber \\
& -\Upsilon_{hk}^{(1)}\braket{u_hu_k\frac{d^2V}{dR_adR_b}} +\nonumber \\
&+  \Upsilon_{hk}^{(1)}\Upsilon^{-1}_{kh}\braket{\frac{d^2V}{dR_adR_b}} \bigg]
\end{align}

Using the Bianco formalism introduced in \eqname~\eqref{eq:bianco}, and integrating by parts, we get:

\begin{align}
  \Avgclassicpert{\frac{d^2V}{dR_adR_b}} =   - \frac 12& \sum_{hkpq} \Upsilon^{-1}_{kp}\Upsilon^{-1}_{hq}\stackrel{(4)}{\Phi}_{abpq} \Upsilon^{(1)}_{hk} + \nonumber \\
  & + \sum_{h} \stackrel{(3)}{\Phi}_{abh}\Rc_h^{(1)}
\end{align}

We can derive now the explicit expression of all the perturbed equation of motion:
\begin{align}
m_a\ddot{\Rc_a^{(1)}} &= \braket{f_a^{(1)}} -\sum_{h} \Phi_{ah}\Rc_h^{(1)} + \nonumber \\ 
&+  \frac12 \sum_{hkpq} \Upsilon^{(1)}_{hk}\Upsilon^{-1}_{kp}\Upsilon^{-1}_{hq}\stackrel{(3)}{\Phi_{qpa}}\label{eq:pert:R}
\end{align}

\begin{align}
\dot{\Upsilon^{(1)}}_{ab} &= \sum_c \frac{\hbar}{m_c}\bigg[
\Upsilon_{cb}(2C_{ac}^{(1)} - \Im A_{ac}^{(1)}) + \Upsilon_{ac}(2 C_{cb}^{(1)} +
\nonumber \\
& + \Im A_{cb}^{(1)}) + \Im A_{cb}^{(1)}\Re A_{ac} - \Im A_{ac}^{(1)}\Re A_{cb}\bigg]
\end{align}

\begin{align}
\Im \dot {A_{ab}^{(1)}} &= \sum_c \frac{\hbar}{2m_c} \Big(\Theta_{ac}^{(1)} \Re A_{cb} + \nonumber \\ 
& -\Theta_{bc}^{(1)}\Re A_{ac} + \Theta_{ac}\Re A_{cb}^{(1)} - \Theta_{bc}\Re A_{ac}^{(1)}\Big)
\end{align}

\begin{align}
\Re \dot{A_{ab}^{(1)}} &= \sum_c \frac{\hbar}{2m_c}\Big(4C_{ac}^{(1)}\Re A_{cb} + 4C_{bc}^{(1)}\Re A_{ca} +\nonumber \\ 
& - \Theta_{ac}\Im A_{cb}^{(1)} + \Theta_{bc} \Im A_{ac}^{(1)}\Big)
\end{align}

\begin{align}
  \dot{C_{ab}^{(1)}} &= \frac{1}{2\hbar}\braket{\frac{d^2 V^{(1)}}{dR_adR_b}} +  \frac {1}{2\hbar} \sum_{h} \stackrel{(3)}{\Phi}_{abh}\Rc_h^{(1)} + \nonumber \\ 
  &- \frac {1}{4\hbar} \sum_{hkpq} \Upsilon^{-1}_{kp}\Upsilon^{-1}_{hq}\stackrel{(4)}{\Phi}_{abpq} \Upsilon^{(1)}_{hk} + \nonumber\\
  & + \sum_c \frac{\hbar}{2m_c}\bigg(-\frac{1}{4}\Theta_{ac}^{(1)}\Theta_{cb} - \frac{1}{4}\Theta_{ac}\Theta_{cb}^{(1)} + \nonumber \\
  &+\Re A_{ac} \Re A_{bc}^{(1)} + \Re A_{ac}^{(1)}\Re A_{bc}\bigg)
\end{align}

\begin{equation}
\Theta^{(1)}_{ab}= \Upsilon_{ab}^{(1)} + 2\Re A_{ab}^{(1)}
\end{equation}

We can remove the theta dependence on ${\Im \dot \bA}$ and $\dot \bC$:
\begin{align}
\Im \dot{A_{ab}^{(1)}} &=   \sum_c \frac{\hbar}{2m_c}\Big(\Re A_{cb}\Upsilon^{(1)}_{ac} - \Re A_{ac}\Upsilon^{(1)}_{bc} +\nonumber \\ 
& +\Re A_{cb}^{(1)}\Upsilon_{ac} - \Re A_{ac}^{(1)}\Upsilon_{bc}\Big)
\end{align}
\begin{align}
\dot C^{(1)}_{ab} &= \frac{1}{2\hbar}\braket{\frac{d^2 V^{(1)}}{dR_adR_b}} +  \frac {1}{2\hbar} \sum_{h} \stackrel{(3)}{\Phi}_{abh}\Rc_h^{(1)} + \nonumber \\
&- \frac {1}{4\hbar} \sum_{hkpq} \Upsilon^{-1}_{kp}\Upsilon^{-1}_{hq}\stackrel{(4)}{\Phi}_{abpq} \Upsilon^{(1)}_{hk} + \nonumber\\
  & + \sum_c \frac{\hbar}{8m_c}\Big(2\Re A_{ac}\Upsilon_{cb}^{(1)} + 2\Re A_{cb}\Upsilon_{ac}^{(1)} + \nonumber \\
  &+2 \Re A_{ac}^{(1)} \Upsilon_{cb} + 2\Re A_{cb}^{(1)}\Upsilon_{ac} - \Upsilon_{ac}\Upsilon^{(1)}_{cb} - \Upsilon_{cb}\Upsilon^{(1)}_{ac}\Big)
\end{align}

By further deriving $\dot{\bUps}^{(1)}$ and $\Re \dot {\bA^{(1)}}$ we delete two variables from the equations ($\bC$ and $\Im \bA$).

\begin{subequations}
\begin{equation}
    \frac{d^2 {\tilde \bUps}^{(1)}}{dt^2} =
    {\bar {\bm X}} {\tilde\bUps}^{(1)} + {\bar{\bm Y}}{\Re \tilde \bA}^{(1)} + {\bar{\bm Z}} {\tilde \bRcal}^{(1)} + \bf^{(1)}_\Upsilon,\label{eq:Y:pert}
\end{equation}
\begin{equation}
    \frac{d^2 {\Re \tilde \bA}^{(1)}}{dt^2} =
     {\bar {\bm X}}' {\tilde\bUps}^{(1)} + {\bar {\bm Y}}'{\Re \tilde \bA}^{(1)} + {\bar{\bm Z}}' {\tilde \bRcal}^{(1)} + \bf^{(1)}_{\Re A},
    \label{eq:RA:pert}
\end{equation}
\begin{equation}
    \frac{d^2 {\tilde \bRcal}^{(1)}}{dt^2} =
    {\bar{\bm X}}'' {\tilde\bUps}^{(1)} + {\bar {\bm Z}}'' {\tilde \bRcal}^{(1)} + \bf_\Rc^{(1)}.
    \label{eq:r:perturb}
\end{equation}
\label{eq:perturb:my}
\end{subequations}
Even if $\Re\tilde\bA(t)$ does not affect the average of the observable directly, we need to keep it, as $\bUps(t)$ depends explicitly on its dynamics.
In \eqname~\eqref{eq:perturb:my}, the bar $\bar{\cdot}$ over a symbol indicates a tensor.
$\bar{\bm X}$, $\bar{\bm Y}$, $\bar{\bm X'}$, and $\bar{\bm Y'}$ are 4-rank tensors, $\bar{\bm Z}$, $\bar{\bm Z'}$, and $\bar{\bm X''}$ are 3-rank tensors, while $\bar{\bm Z''}$ is a 2-rank tensor.
We recall that $\tilde\bUps^{(1)}$ and $\Re\tilde\bA^{(1)}$ are 2-rank tensor, as well as $\bf_{\Upsilon}^{(1)}$ and $\bf_{\Re A}^{(1)}$, while $\bRcal^{(1)}$ and $\bf_{\Rcal}^{(1)}$ are vectors (1-rank).

The product between tensors is the defined by the operator on the left.
4-rank tensors $\bar{\bm X}$, $\bar{\bm Y}$, $\bar{\bm X'}$, and $\bar{\bm Y'}$ are contracted on the last two indices:
\begin{subequations}
\begin{equation}
    \left(\bar {\bm X} {\tilde\bUps}^{(1)}\right)_{ab} = \sum_{cd}{\bar X}_{abcd} {\tilde\Upsilon}^{(1)}_{cd},
\end{equation}
also 3-rank tensors $\bar{\bm Z}$, $\bar{\bm Z'}$ are contracted on the last two indices:
\begin{equation}
    \left(\bar{\bm Z} {\tilde\bUps}^{(1)}\right)_a = \sum_{bc} \bar Z_{abc}{\tilde\Upsilon}^{(1)}_{bc}.
\end{equation}
The 3-rank tensor $\bar{\bm X''}$ is contracted only on the last index:
\begin{equation}
    \left(\bar{\bm X''} {\tilde\bRcal}^{(1)}\right)_{ab} = \sum_{bc} \bar X''_{abc}{\tilde\Rcal}^{(1)}_{c}.
\end{equation}
\label{eq:tensor:conv}
\end{subequations}
The 2-rank tensor $\bar{\bm Z''}$ is contracted on the last index, as the standard matrix rows-by-columns product.

The explicit expression of the coefficients in $\bar{\bm X}$, $\bar{\bm Y}$, $\bar{\bm Z}$... is reported in \appendixname~\ref{app:full:system}.

The tensors introduced in \eqname~\eqref{eq:perturb:my} account for the free time-evolution of the system with the full anharmonic interaction: they are defined by the static unperturbed Hamiltonian $\Hcal[\rho]$. In particular, they depend on phonon scattering vertexes: the 4-phonon scattering tensor $\bDfour$ and 3-phonon scattering tensor $\bDthree$, and the free evolution $\bDtwo$.
 Due to the Gaussian constrain on the density matrix, the TD-SCHA does not account directly for higher-order phonon scattering processes. However, $\bDthree$, $\bDfour$, and $\bDtwo$ depend ``self-consistently'' on higher-order anharmonicities, as they are averaged on the equilibrium distribution. In fact, they are temperature dependent.

In particular, $\bar {\bm X}$ and $\bar {\bm X'}$ depend on $\bDfour$, while $\bar{\bm Z}$, $\bar{\bm Z'}$ and $\bar{\bm X''}$ depend on $\bDthree$. The $\bar {\bm X}$, $\bar {\bm X'}$, $\bar{\bm Y}$, $\bar{\bm Y'}$ and $\bar{\bm Z''}$ contain terms of the free evolution, that are non zero even if the system is a perfect harmonic crystal. \tablename~\ref{tab:d3d4} summarizes these dependencies.
\begin{table}
\centering
\newcommand{\itsok}{$\boldsymbol{\circ}$}
\setlength{\tabcolsep}{18pt}
{\setlength{\extrarowheight}{3pt}
\begin{tabular}{c|ccc}
\hline 
\hline
 & $\bDtwo$ & $\bDthree$ & $\bDfour$ \\
 \hline
$\bar {\bm X}$ & \itsok & & \itsok \\
$\bar {\bm Y}$ & \itsok & &  \\
$\bar {\bm Z}$ & & \itsok & \\
$\bar {\bm X'}$ & \itsok & & \itsok \\
$\bar {\bm Y'}$ & \itsok & &  \\
$\bar {\bm Z'}$ & & \itsok & \\
$\bar {\bm X''}$ & & \itsok & \\
$\bar {\bm Z''}$ &\itsok &  & \\
\hline
\hline
\end{tabular}}
\caption{Dependency of the coefficients of the linear response system (\eqname~\ref{eq:perturb:cart}) on the free evolution ($\bDtwo$), on anharmonic coupling $\bDthree$ (\eqname~\ref{eq:D3}) and $\bDfour$ (\eqname~\ref{eq:D4}). A \itsok~in the grid indicates that the tensor on the first column depends on the corresponding term in the first row. This table helps to see which terms can be set to zero if we neglect $\bDthree$ or $\bDfour$, and to visualize how anharmonicity couples different degrees of freedom}
\label{tab:d3d4}
\end{table}

The $\bDthree$ and $\bDfour$ terms are obtained from the $\bPhithree$ and $\bPhifour$ when transforming in the mass-rescaled variables.
\begin{equation}
    \Dthree_{abc} = \frac{\Phithree_{abc}}{\sqrt{m_am_bm_c}}
\end{equation}
\begin{equation}
    \Dfour_{abcd} = \frac{\Phifour_{abcd}}{\sqrt{m_am_bm_cm_d}}
\end{equation}

\section{Full expression of the Linear Perturbation system}
\label{app:full:system}
Here we report the full expressions of the 2,3,4-rank tensors that define the linear response system of the TD-SCHA. These are obtained writing the full expression of the system as derived in appendix~\ref{app:linear:response}.
These are expressed in the polarization basis, i.e. the basis that diagonalizes the SCHA dynamical matrix at equilibrium:

\begin{widetext}
\begin{align}
{\bar X}_{\mu\nu\eta\lambda} &= 
- \frac{\hbar (2n_\eta+1)(2n_\lambda + 1)(2\omega_\mu n_\nu + 2\omega_\nu n_\mu + \omega_\mu + \omega_\nu)}{4(2n_\mu + 1)(2n_\nu + 1)\omega_\eta\omega_\lambda} {\Dfour}_{\mu\nu\eta\lambda} - \frac{\delta_{\mu\eta}\delta_{\nu\lambda} + \delta_{\mu\lambda}\delta_{\nu\eta}}{2}\left(\omega_\mu^2 + \omega_\nu^2 + \frac{2\omega_\mu\omega_\nu}{(2n_\mu + 1)(2n_\nu + 1)}\right)
\label{eq:Xmu}
\end{align}
\begin{align}
    {\bar X'}_{\mu\nu\eta\lambda}
    &= - \frac{\hbar \left[\omega_\mu n_\mu(n_\mu + 1)(2n_\nu + 1) + \omega_\nu n_\nu(n_\nu + 1)(2n_\mu + 1)\right](2n_\eta + 1)(2n_\lambda + 1)}{4(2n_\mu + 1)(2n_\nu + 1) \omega_\eta\omega_\lambda} \Dfour_{\mu\nu\eta\lambda}  + \nonumber \\
    & - \left(\frac{\delta_{\mu\eta}\delta_{\nu\lambda} + \delta_{\mu\lambda}\delta_{\nu\eta}}{2}\right) \frac{(2n_\mu n_\nu + n_\mu + n_\nu)(2n_\mu n_\nu + n_\mu + n_\nu + 1)2\omega_\mu \omega_\nu}{(2n_\mu + 1)(2 n_\nu + 1)}
\label{eq:X1mu}
\end{align}
\end{widetext}
\begin{equation}
    {\bar Y}_{\mu\nu\eta\lambda} = - \frac{4(\delta_{\mu\eta} \delta_{\nu\lambda} + \delta_{\mu\lambda}\delta_{\nu\eta})\omega_\mu\omega_\nu}{(2n_\mu + 1)(2n_\nu + 1)}
\label{eq:Ymu}
\end{equation}
\begin{align}
    {\bar Y'}_{\mu\nu\eta\lambda}& = \frac{\delta_{\mu\eta}\delta_{\nu\lambda} + \delta_{\mu\lambda}\delta_{\nu\eta}}{2}\cdot \nonumber\\ &\cdot \left(\frac{2\omega_\mu\omega_\nu}{(2n_\mu + 1)(2n_\nu + 1)} - \omega_\mu^2 - \omega_\nu^2\right)
\label{eq:Y1:mu}
\end{align}
\begin{equation}
    {\bar Z}_{\mu\nu\eta} = 
    \frac{2[(2n_\mu + 1)\omega_\nu + (2n_\nu + 1)\omega_\mu]}{\hbar(2n_\mu + 1)(2n_\nu +1)}\Dthree_{\mu\nu\eta}
    \label{eq:Zmu}
\end{equation}
\begin{align}
    {\bar Z'}_{\mu\nu\eta} &=
    \frac{2\omega_\mu n_\mu(n_\mu + 1)(2n_\nu + 1)}{\hbar (2n_\mu + 1)(2n_\nu + 1)} \Dthree_{\mu\nu\eta} + \nonumber \\
    & + \frac{2\omega_\nu n_\nu(n_\nu + 1)(2n_\mu + 1)}{\hbar (2n_\mu + 1)(2n_\nu + 1)}\Dthree_{\mu\nu\eta} 
    \label{eq:Z1mu}
\end{align}
\begin{equation}
    {\bar X''}_{\mu\nu\eta} = \frac{\hbar^2 (2n_\nu + 1)(2n_\eta + 1)}{8\omega_\nu \omega_\eta} \Dthree_{\mu\nu\eta}
    \label{eq:X2mu}
\end{equation}
\begin{equation}
    {\bar Z''}_{\mu\nu} = -\delta_{\mu\nu} \omega_\mu^2
    \label{eq:Z2mu}
\end{equation}

\section{Proof of the dynamical ansatz}
\label{app:ansatz:proof}
Here we compute the TD-SCHA self energy:
$$
\bGcal^{-1}(\omega) = ({\bm Z}'' + \omega^2) - \bPi(\omega)
$$

By looking at \eqname~\eqref{eq:green:one:phonon}, $\bPi(\omega)$ is obtained by how the last row interact with the rest of the big $\bm G(\omega)$ matrix through $\bar{\bm X''}$. 

The complete one-phonon green function is given by \eqname~\eqref{eq:one:phonon:full}.

\begin{equation}
\arraycolsep=0.8pt
\Gcal_{\mu\nu}(\omega) = \bpm 0&0 & \boldsymbol{\delta_\mu}\epm
{
\arraycolsep=-2pt\def\arraystretch{1}
\bpm \bar{\bm X} + \omega^2  & 
\bar {\bm Y} & \bar{\bm Z} \\
\bar{\bm X'} & \bar{\bm Y'} + \omega^2 & 
\bar {\bm Z'} \\
\bar{\bm X''} & 0 & \bar{\bm Z''} + \omega^2\epm^{-1}
}
\bpm 0 \\ 0 \\ \boldsymbol{\delta_\nu}\epm.
\label{eq:green:one:phonon}
\end{equation}

Thus, the self-energy is
\begin{equation}
    \bPi(\omega) = {\bar{\bm X}}'' \frac{\partial {\tilde \bUps}^{(1)}}{\partial {\tilde\bRcal}^{(1)}}(\omega)
\end{equation}
Here, products between tensors follows the same convention as \eqname~\eqref{eq:tensor:conv}, where the number of indices to be contracted is determined by the tensor on the left.
The inversion of a 4-rank tensor is equivalent of inverting a matrix where we group the first two and last two indices:
\begin{equation}
    \bar X_{abcd} = \bar X_{(ab)(cd)}.
\end{equation}
and $\bar{\bm X''}$ is contracted with the indices of ${\tilde \bUps}^{(1)}$.
In particular, this term indicates how a perturbation in the quantum fluctuations ${\tilde\bUps}^{(1)}$ affects the average positions ${\tilde\bRcal}^{(1)}$.
If $\bDthree \neq 0$, the dynamics in ${\tilde\bRcal}^{(1)}$ affects ${\tilde\bUps}^{(1)}$ and $\Re{\tilde\bA}^{(1)}$ through $\bar{\bm Z}$ and $\bar{\bm Z'}$ (\eqname~\ref{eq:Y:pert} and \ref{eq:RA:pert}). They evolve freely (harmonic propagation) and interacting anharmonically through $\bDfour$. Finally, ${\tilde\bUps}^{(1)}$ affects back $\tilde{\bRcal}^{(1)}$ through $\bar{\bm X''}$ (\eqname~\eqref{eq:r:perturb}).

To calculate $\bPi(\omega)$, we need to get an explicit expression from the dependency of ${\tilde\bUps}^{(1)}$ and ${\tilde\bRcal}^{(1)}$. It is convenient to remove the $\Re{\tilde\bA}^{(1)}(\omega)$ firstly (\eqname~\ref{eq:RA:pert}):
\begin{equation}
    \bar{\bm X'}{\tilde \bUps}^{(1)} + 
    (\bar{\bm Y'} + \omega^2)\Re{\tilde\bA}^{(1)} + \bar{\bm Z'} {\tilde \rschatrial}^{(1)} = 0
\end{equation}

\begin{equation}
\Re{\tilde\bA}^{(1)} = -(\bar{\bm Y'} + \omega^2)^{-1} \left(\bar{\bm X'} {\tilde\bUps}^{(1)} + \bar{\bm Z'} {\tilde\rschatrial}^{(1)}\right).
\end{equation}
We substitute it in the ${\tilde\bUps}^{(1)}(\omega)$ equation (\eqname~\ref{eq:Y:pert}):
\begin{align}
    &\left[(\bar{\bm X} + \omega^2) - \bar{\bm Y}(\bar{\bm Y'} + \omega^2)^{-1}\bar{\bm X'}\right]{\tilde\bUps}^{(1)} +\nonumber \\
    &+ \left[\bar {\bm Z} - \bar{\bm Y}(\bar{\bm Y'} + \omega^2)^{-1}\bar{\bm Z'}\right]{\tilde\rschatrial}^{(1)} = 0
\end{align}
From which we get the relationship between ${\tilde \bUps}^{(1)}(\omega)$ and ${\tilde\rschatrial}^{(1)}(\omega)$.

The final result we get is:
\begin{align}
    \bPi(\omega) = -  \bar{\bm X''} &\left[ (\bar {\bm X} + \omega^2) - \bar {\bm Y}(\bar {\bm Y'} + \omega^2)^{-1}\bar{\bm X'}\right]^{-1}\cdot \nonumber \\
    &\cdot \left[ \bar{\bm Z} - \bar{\bm Y}(\bar {\bm Y'} + \omega^2)^{-1}\bar{\bm Z'}\right]
\end{align}

Now, we prove that the previous equation correspond to the dynamical \emph{ansatz} proposed by Bianco et al.\cite{Bianco2017}, reported in \eqname~\eqref{eq:self:energy:ansatz}.

To simplify the expression, we define the tensor $\bm U$, $\bm P$ and $\bm T$,  that in the polarization basis are:
\begin{equation}
    \bar {X''}_{\alpha\beta\gamma} = U_{\alpha\beta}\Dthree_{\alpha\beta\gamma}
\end{equation}
\begin{equation}
    \bm P = (\bar {\bm X} + \omega^2) - \bar {\bm Y}(\bar {\bm Y'} + \omega^2)^{-1}\bar{\bm X'}
\end{equation}
\begin{equation}
    T_{\alpha\beta} = \frac{\bar Z_{\alpha\beta\gamma}}{\Dthree_{\alpha\beta\gamma}} - {\bar Y}_{\alpha\beta}\left(\bar{Y'}_{\alpha\beta} + \omega^2\right)^{-1} \frac{Z'_{\alpha\beta\gamma}}{\Dthree_{\alpha\beta\gamma}}\label{eq:Tdef}
\end{equation}

Both $\bm U$ and $\bm T$ are 2-rank tensors, while $\bm P$ is in general a 4-rank tensor. However, if $\bDfour = 0$ also $\bm P$ becomes a 2-rank tensor. For this reason it is easier to split $\bm P$ in two contribution:
\begin{equation}
    \bm P = {\bm P^{(0)}} + {\bm P^{(1)}}
\end{equation}
where $
{\bm P^{(0)}} = \bm P$ if $\bDfour = 0$.

With this expression, the self-energy becomes:
\begin{equation}
\label{eq:sigmanew}
    \bPi(\omega) = \bDthree {\bm U} {\bm P}^{-1} \bm T \bDthree
\end{equation}

Here, the inversion of the 4-rank $\bm P$ tensor is equal to the inversion of a big rank 2 tensor in which the first two and last two indices are grouped together. Moreover, $\bm T$ has only two indices even if in the right-hand expression of \eqname~\eqref{eq:Tdef} three indices appears, as both $ \bar Z_{\alpha\beta\gamma}$ and $\bar { Z'}_{\alpha\beta\gamma}$ loose the dependence on the $\gamma$ index if divided by $\Dthree_{\alpha\beta\gamma}$.

First of all, lets consider the simple case for which $\bDfour = 0$ but $\bDthree  \neq 0$. In this case the Bianco self-energy reduces to the bubble diagram:
\begin{equation}
    \bPi(\omega) = \bDthree \bLambda(\omega) \bDthree
\end{equation}
If we set $\bDfour = 0$ in our \eqname~\eqref{eq:sigmanew} we get:
\begin{equation}
    \bPi(\omega) = \bDthree {\bm U} {\bm P^{(0)}}^{-1} {\bm T} \bDthree
\end{equation}

The two equations are equal if:
\begin{equation}
    \bLambda(\omega) = {\bm U} {\bm P^{(0)}}^{-1} {\bm T}
    \label{eq:lambda:1}
\end{equation}

This equality can be proved by a simple algebric calculations. In fact, $\bm P^{(0)}$ is a diagonal 4-rank tensor in the polarization basis, and can be inverted by inverting its elements. 

The explicit expression in the polarization basis of the $\bm T$, $\bm P^{(0)}$ and $\bm U$ are:
\begin{equation}
    U_{\alpha\beta} = - \hbar^2 \frac{(2n_\alpha + 1)(2n_\beta + 1)}{8\omega_\alpha\omega_\beta}
\end{equation}

\begin{widetext}
\begin{align}
    T_{\gamma\delta} &= 
    \bigg[\frac{(4n_\gamma + 2)\omega_\delta + (4n_\delta + 2)\omega_\gamma}{\hbar (2n_\gamma + 1)(2n_\delta + 1)} +
    \frac{8\omega_\gamma\omega_\delta}{(2n_\gamma+1)(2n_\delta + 1)}\frac{(2n_\gamma + 1)(2n_\delta + 1)}{(\omega^2 - \omega_\gamma^2 - \omega_\delta^2)(2n_\gamma + 1)(2n_\delta + 1) + 2\omega_\gamma\omega_\delta}\cdot \nonumber\\ 
    & \cdot \left(\frac{2\omega_\gamma n_\gamma(n_\gamma + 1)}{\hbar (2n_\gamma + 1)} + \frac{2\omega_\delta n_\delta(n_\delta + 1)}{\hbar(2n_\delta + 1)}\right)\bigg]
\end{align}
\begin{align}
    T_{\gamma\delta} &= 
    \bigg[\frac{(4n_\gamma + 2)\omega_\delta + (4n_\delta + 2)\omega_\gamma}{\hbar (2n_\gamma + 1)(2n_\delta + 1)} +
    \frac{8\omega_\gamma\omega_\delta}{(\omega^2 - \omega_\gamma^2 - \omega_\delta^2)(2n_\gamma + 1)(2n_\delta + 1) + 2\omega_\gamma\omega_\delta} \left(\frac{2\omega_\gamma n_\gamma(n_\gamma + 1)}{\hbar (2n_\gamma + 1)} + \frac{2\omega_\delta n_\delta(n_\delta + 1)}{\hbar(2n_\delta + 1)}\right)\bigg]
\end{align}
\begin{align}
    P_{\alpha\beta}^{(0)} &= \omega^2 - \omega_\alpha^2 - \omega_\beta^2  - \frac{2\omega_\alpha\omega_\beta}{(2n_\alpha + 1)(2n_\beta + 1)} - \frac{8\omega_\alpha\omega_\beta}{(\omega^2 - \omega_\alpha^2 - \omega_\beta^2)(2n_\alpha + 1)(2n_\beta + 1) + 2\omega_\alpha\omega_\beta}\cdot \nonumber \\ 
    &\cdot\frac{2\omega_\alpha\omega_\beta(2n_\alpha n_\beta + n_\alpha + n_\beta)(2n_\alpha n_\beta + n_\alpha + n_\beta + 1)}{(2n_\alpha + 1)(2n_\beta + 1)}
\end{align}
\begin{align}
    P_{\alpha\beta}^{(0)} &= \omega^2 - \omega_\alpha^2 - \omega_\beta^2  - \frac{2\omega_\alpha\omega_\beta}{(2n_\alpha + 1)(2n_\beta + 1)}\left[1 + \frac{8\omega_\alpha\omega_\beta(2n_\alpha n_\beta + n_\alpha + n_\beta)(2n_\alpha n_\beta + n_\alpha + n_\beta + 1)}{(\omega^2 - \omega_\alpha^2 - \omega_\beta^2)(2n_\alpha + 1)(2n_\beta + 1) + 2\omega_\alpha\omega_\beta}\right]
    \label{eq:P0}
\end{align}
\end{widetext}

With some algebric manipulation, it is straightfoward to show that
\begin{align}
    \frac{U_{\alpha\beta} T_{\alpha\beta}}{P^{(0)}_{\alpha\beta}} = - \frac{\hbar^2}{4\omega_\alpha\omega_\beta} \bigg[&\frac{(\omega_\alpha + \omega_\beta)(n_\alpha + n_\beta + 1)}{(\omega_\alpha + \omega_\beta)^2 - \omega^2} + \nonumber \\ 
    & -\frac{(\omega_\alpha - \omega_\beta)(n_\alpha - n_\beta)}{(\omega_\alpha - \omega_\beta)^2 - \omega^2}\bigg]
\end{align}
That is exactly the expression in the polarization basis of the $\bLambda(\omega)$ tensor. Therefore we proved \eqname~\eqref{eq:lambda:1}, and the dynamical ansatz of Bianco in the case $\bDfour = 0$. 
To proceed with the case $\bDfour \neq 0$ we must add the $\bm P^{(1)}$. 

\begin{equation}
    \bPi(\omega) = \bDthree \bU \left[ {\bm P^{(0)}} + \bar { \bm P^{(1)}}\right]^{-1} {\bm T} \bDthree
\end{equation}
\begin{equation}
    \bPi(\omega) = \bDthree \bU  \left[ \mathbbm{1} +  {\bm P^{(0)}}^{-1}\bar { \bm P^{(1)}}\right]^{-1} {\bm P^{(0)}}^{-1}{\bm T} \bDthree
    \label{eq:sigma:prefinal}
\end{equation}

We already proved that
\begin{equation}
    \bLambda(\omega) = \bU {\bm P^{(0)}}^{-1} {\bm T}
\end{equation}

Therefore, it is trivial to show that \eqname~\eqref{eq:sigma:prefinal} is the Bianco self-energy (\eqname~\ref{eq:self:energy:ansatz}) if we prove that:
\begin{equation}
    {\bm P^{(0)}}^{-1}\bar { \bm P^{(1)}} = {\bm P^{(0)}}^{-1}{\bm T}\bDfour \bU .\label{eq:final:proof}
\end{equation}
In fact, if we substitute \eqname~\eqref{eq:final:proof} into \eqname~\eqref{eq:sigma:prefinal}, and we perform the Taylor expansion for small $\bDfour$, we get:
\begin{equation}
\bPi(\omega) = \bDthree \bU {\bm P^{(0)}}^{-1}{\bm T} \bDthree + 
\bDthree \bU {\bm P^{(0)}}^{-1}{\bm T}\bDfour \bU {\bm P^{(0)}}^{-1}
{\bm T} \bDthree + \cdots\nonumber
\end{equation}
\begin{equation}
\bPi(\omega) = \bDthree \bLambda(\omega) \bDthree + 
\bDthree \bLambda(\omega)\bDfour \bLambda(\omega) \bDthree + \cdots
\end{equation}
That is the correct diagrammatic expansion of the self-energy, as defined in ref.\cite{Bianco2017}.

We give here the explicit expression of the $\bm P^{(1)}$ 4-rank tensor.
\begin{widetext}
\begin{align}
    P^{(1)}_{\alpha\beta\gamma\delta} &=
    - \frac{\hbar(2n_\gamma + 1)(2 n_\delta + 1)(2 \omega_\alpha n_\beta + 2\omega_\beta n_\alpha + \omega_\alpha + \omega_\beta)}{4(2n_\alpha + 1)(2n_\beta + 1)\omega_\gamma\omega_\delta} \Dfour_{\alpha\beta\gamma\delta} -
    \frac{8\omega_\alpha\omega_\beta}{(\omega^2 - \omega_\alpha^2 - \omega_\beta^2)(2n_\alpha + 1)(2n_\beta + 1) + 2\omega_\alpha\omega_\beta}\cdot\nonumber \\
    &\cdot  
    \frac{\hbar\left[\omega_\alpha n_\alpha(n_\alpha + 1)(2n_\beta + 1) + \omega_\beta n_\beta(n_\beta + 1)(2n_\alpha + 1)\right](2n_\gamma + 1)(2n_\delta + 1)}{4 (2n_\alpha + 1)(2n_\beta + 1)\omega_\gamma\omega_\delta} \Dfour_{\alpha\beta\gamma\delta}
\end{align}
\begin{equation}
    P^{(1)}_{\alpha\beta\gamma\delta} =
    - \frac{\hbar(2n_\gamma + 1)(2 n_\delta + 1)\Dfour_{\alpha\beta\gamma\delta}}{4(2n_\alpha + 1)(2n_\beta + 1)\omega_\gamma\omega_\delta} 
    \left[
    2 \omega_\alpha n_\beta + 2\omega_\beta n_\alpha + \omega_\alpha + \omega_\beta + 
    \frac{8\omega_\alpha\omega_\beta [\omega_\alpha n_\alpha(n_\alpha + 1)(2n_\beta + 1) + \omega_\beta n_\beta(n_\beta + 1)(2n_\alpha + 1)]}{(\omega^2 - \omega_\alpha^2 - \omega_\beta^2)(2n_\alpha + 1)(2n_\beta + 1) + 2\omega_\alpha\omega_\beta}
    \right]
\end{equation}
\end{widetext}

Again, with straightfoward algebra we can prove that:
\begin{equation}
    \frac{P^{(1)}_{\alpha\beta\gamma\delta}}{P^{(0)}_{\alpha\beta}} = \frac{T_{\alpha\beta}\Dfour_{\alpha\beta\gamma\delta}U_{\gamma\delta}}{P^{(0)}_{\alpha\beta}}
\end{equation}
This concludes the proof that the one-phonon self-energy proposed as \emph{ansatz} by Bianco et al.\cite{Bianco2017} can be formally derived in a full dynamical contex within the TD-SCHA.

\section{Harmonic two-phonon propagator}
\label{app:twophonons}
In this section we derive the Harmonic two-phonon propagator from the response system of \eqname~\eqref{eq:perturb:cart}.

The harmonic limit is obtained as $\bDthree = 0$ and $\bDfour = 0$. In this limit, the one-phonon and two-phonon Green functions are decoupled. The two-phonon propagation is described by the variables $\Re\bA$ and $\bUps$, so we isolate these contributions in the green function (\eqname~\eqref{eq:full:green:function}).

The two phonon green function is obtained inverting the matrix:
\begin{equation}
    G^{(2ph)}(\omega) = \bpm -\bar{\bm X} - \omega^2 & - \bar{\bm Y} \\
    - {\bar {\bm X'}} & - \bar {\bm Y'} -\omega^2\epm^{-1}
    \label{eq:G2:pre}
\end{equation}

In the polarization basis, the tensor $\bar{\bm X}$, $\bar{\bm Y}$, $\bar{\bm X'}$ and $\bar{\bm Y'}$ are diagonal.
Therefore, once we identify a couple of modes $\mu\nu$, we only need to invert a 2x2 matrix in \eqname~\eqref{eq:G2:pre}.

Since we are interested in the fluctuations-fluctuations correlation function, we only need the response in the $\bUps$ block (the fluctuations are the convariance matrix $\bUps^{-1}$).
This term is:
\begin{equation}
    G^{(\Upsilon)}_{\mu\nu}(\omega) = \frac{-\bar{ Y'}_{\mu\nu\mu\nu} - \omega^2}{\bar{X'}_{\mu\nu\mu\nu} \bar{Y}_{\mu\nu\mu\nu} + (\bar X_{\mu\nu\mu\nu} + \omega^2)(\bar {Y'}_{\mu\nu\mu\nu} + \omega^2)}
\end{equation}

Subsittuting the expression defined in \ref{app:full:system} (taking care of the symmetry exchanging $\mu\leftrightarrow\nu$) we get:

\begin{align}
     G^{(\Upsilon)}_{ab}(\omega) &= \frac{2}{\hbar \left(4 {n}_{a} {n}_{b} + 2 {n}_{a} + 2 {n}_{b} + 1\right)}\cdot \nonumber \\
     &\cdot \bigg[ \frac{- 2 \omega^{2} {\omega}_{a} {n}_{b} - \omega^{2} {\omega}_{a} - 2 \omega^{2} {\omega}_{b} {n}_{a} - \omega^{2} {\omega}_{b}}{[(\omega_a - \omega_b)^2 - \omega^2][(\omega_a + \omega_b)^2 - \omega^2]} + \nonumber \\
     &+\frac{2 {\omega}_{a}^{3} {n}_{b} + {\omega}_{a}^{3} - 2 {\omega}_{a}^{2} {\omega}_{b} {n}_{a} - {\omega}_{a}^{2} {\omega}_{b}}{[(\omega_a - \omega_b)^2 - \omega^2][(\omega_a + \omega_b)^2 - \omega^2]} + \nonumber \\ 
     &+  \frac{- 2 {\omega}_{a} {\omega}_{b}^{2} {n}_{b} - {\omega}_{a} {\omega}_{b}^{2} + 2 {\omega}_{b}^{3} {n}_{a} + {\omega}_{b}^{3}}{[(\omega_a - \omega_b)^2 - \omega^2][(\omega_a + \omega_b)^2 - \omega^2]}\bigg]
     \label{eq:G:ups}
\end{align}

This green function has the poles in the correct position, when:
$$
\omega^2 = (\omega_a \pm \omega_b)^2
$$

Since we want the fluctuation-fluctuation correlation function, we need to change variable for the Green function from $\bUps$ to $\bUps^{-1}$.
This is achieved knowing how the perturbation in $(\bUps^{-1})^{(1)}$ depend on those on $\bUps^{(1)}$:

\begin{equation}
    (\bUps^{-1})^{(1)}_{ab} = - \sum_{cd} \Upsilon^{-1}_{ac} \Upsilon^{-1}_{bd}\Upsilon^{(1)}_{cd}
\end{equation}

From this expression we can compute the fluctuation-fluctuation response function:
\begin{equation}
    \chi_{abcd}(\omega) = G^{(\Upsilon^{-1})}_{abcd}(\omega) = 
    -\sum_{ef} \Upsilon^{-1}_{ae} \Upsilon^{-1}_{bf} G^{(\Upsilon)}_{efcd}(\omega)
\end{equation}
If we substitute the expressions in the polarization basis, we obtain:
\begin{align}
    \chi_{\mu\nu}(\omega)& = 
    - \frac{\hbar \left({\omega}_{\mu} - {\omega}_{\nu}\right) \left({n}_{\mu} - {n}_{\nu}\right)}{2 \left(\omega - {\omega}_{\mu} + {\omega}_{\nu}\right) \left(\omega + {\omega}_{\mu} - {\omega}_{\nu}\right) {\omega}_{\mu} {\omega}_{\nu}} + \nonumber \\ 
    &+\frac{\hbar \left({\omega}_{\mu} + {\omega}_{\nu}\right) \left({n}_{\mu} + {n}_{\nu} + 1\right)}{2 \left(\omega^{2} - \left({\omega}_{\mu} + {\omega}_{\nu}\right)^{2}\right) {\omega}_{\mu} {\omega}_{\nu}}
\end{align}

That is the standard two-phonon propagator.

\section{Additional details on the Lanczos algorithm}
\label{app:lanczos}

Here, we prove \eqname~\eqref{eq:apply:L}, and discuss the stochastic calculation of \eqname~\eqref{eq:pert:d2v} as well as how we implemented symmetries in the Lanczos algorithm. 
If we have a perturbation vector
\begin{equation}
    \rho^{(1)} \rightarrow \bpm {\tilde\bUps}^{(1)} \\\
    \Re{\tilde \bA}^{(1)} \\ 
    {\tilde\bRcal}^{(1)}\epm,
\end{equation}
the averages on the perturbed ensemble have been computed in \appendixname~\ref{app:linear:response}:
\begin{equation}
  \Avg{\frac{d\mathbb{V}}{dR_a}}{\rho^{(1)}} =- \frac 12 \sum_{hkqp}{\tilde\Upsilon}^{(1)}_{hk} {\tilde\Upsilon}^{-1}_{kp}{\tilde\Upsilon}^{-1}_{hq} \Dthree_{qpa}
\end{equation}
\begin{equation}
  \Avg{\frac{d^2\mathbb{V}}{d\tilde R_a d\tilde R_b}}{\rho^{(1)}}  = 
  \sum_{c} \Dthree_{abc}{\tilde\Rcal}^{(1)}_c -\frac12 \sum_{hkpq} \tilde\Upsilon^{(1)}_{hk} {\tilde\Upsilon}^{-1}_{kp}{\tilde\Upsilon}^{-1}_{hq} \Dfour_{qpab}
\label{eq:d2Vbb}
\end{equation}
Thus, substituting this expression inside the equation of the motion, it is straightforward to show that:
\begin{equation}
     {\mathcal L}^{\text{anh}}\bpm {\tilde \bUps}^{(1)} \\ {\tilde\Re\bA}^{(1)} \\ {\tilde\bRcal}^{(1)} \epm = \bpm \tilde\bUps \Avgclassicpert{\frac{d^2\mathbb{V}}{d\tilde\bR d\tilde\bR}} + \Avgclassicpert{\frac{d^2\mathbb{V}}{d\tilde\bR d\tilde\bR}} \tilde\bUps \\ 
     \Re \tilde\bA \Avgclassicpert{\frac{d^2\mathbb{V}}{d\tilde\bR d\tilde\bR}} + \Avgclassicpert{\frac{d^2\mathbb{V}}{d\tilde\bR d\tilde\bR}} \Re\tilde\bA \\ 
     - \Avgclassicpert{\frac{d{\mathbb V}}{d\tilde\bR}}\epm,
\end{equation}

However, computing \eqname~\eqref{eq:d2Vbb} by explicitly calculating both $\bDthree$ and $\bDfour$ is a terrible choice. The calculation of $\bDfour$ itself is challenging, as it must be computed with a stochastic average for each of the $N^4$ elements, with a computational cost scaling as $N_\text{conf} N^4$, prohibitive for systems with hundreds of atoms in the simulation cell.

Instead, we use the importance sampling to calculate these averages:
\begin{equation}
    \Avgclassicpert{\frac{d\bbV}{d\tilde R_a}} = - \Avgclassiceq{\bbf_a \frac{\rho^{(1)}(\bR)}{\rho(\bR)}}\label{eq:d1v:partial}
\end{equation}
\begin{align}
    \Avgclassicpert{\frac{d^2\bbV}{d\tilde R_ad\tilde R_b}} &= - \sum_c \tilde\Upsilon_{ac}  \Avgclassiceq{\tilde u_c \tilde \bbf_b \frac{\rho^{(1)}(\bR)}{\rho(\bR)}} \label{eq:d2v:partial}
\end{align}
where $\bbf$ is the difference between the BO force $\bf$ and the force of the equilibrium SCHA auxiliary Hamltonian
\begin{equation}
    \bbf_a(\bR) = f_a(\bR) + \sum_b \Avgclassiceq{\frac{d^2V}{dR_adR_b}} R_b,
\end{equation}
and $\bu = \bR - \bRcal$ is the displacement from the average centroid position.
We got \eqname~\eqref{eq:d2v:partial}  by integrating by parts, as done in ref.\cite{Bianco2017}.
Thus, the averages are computed with new weights $w_i^{(1)}$ on each ionic configuration $\bR_{\{i\}}$:
\begin{equation}
    w_i^{(1)} = \frac{\rho^{(1)}(\bR_{\{i\}})}{\rho(\bR_{\{i\}})}
\end{equation}
By Taylor expanding the density matrx, we get:
\begin{equation}
    w_i^{(1)} = -\frac 12 \tilde\bu_{\{i\}} {\tilde\bUps}^{(1)} \tilde\bu_{\{i\}}  + \tilde\bu_{\{i\}} \tilde\bUps {\tilde\bRcal}^{(1)} + \frac 12 \Tr{{\tilde\bUps}^{-1} {\tilde\bUps}^{(1)}}
    \label{eq:full:weight}.
\end{equation}

Since the last term is a constant factor that does not depend on the specific configuration, does not contribute to the averages; we remember that the averages of \eqname~\eqref{eq:d1v:partial} and \eqname~\eqref{eq:d2v:partial} are on the SCHA gradient for centroid positions and auxiliary force constants, thus any term of $w^{(1)}_i$ not depending on the configuration gives zero at equilibrium.

We compute \eqname~\eqref{eq:d1v:partial} and \eqname~\eqref{eq:d2v:partial} with a standard weighted average:

\begin{equation}
    \Avgclassicpert{\frac{d\bbV}{d\tilde R_a}} = - \frac{1}{N_\text{conf}}\sum_{i = 1}^{N_\text{conf}} \bbf_a(\bR_{\{i\}}) w_i^{(1)}\label{eq:d1v:final}
\end{equation}

\begin{equation}
    \Avgclassicpert{\frac{d^2\bbV}{d\tilde R_a d\tilde R_b}} = - \frac{1}{N_\text{conf}}\sum_{i = 1}^{N_\text{conf}} \left(\sum_c \tilde\Upsilon_{ac} \tilde u_c\right) \bbf_b(\bR_{\{i\}}) w_i^{(1)}\label{eq:d2v:final}
\end{equation}
\eqname~\eqref{eq:d1v:final} and \eqref{eq:d2v:final}, as well as the weights \eqname~\eqref{eq:full:weight}, require at most $N^2$ operations for each configuration, granting an overall computational cost of $N_\text{conf}N^2$.

The point group symmetries of a crystal are accounted for by unwrapping the stochastic ensemble. For each symmetry and configuration in the ensemble, we generate all the equivalent configurations according to the symmetry. The equivalent configuration is obtained by:
\begin{equation}
    u_{s(a)}^\alpha = \sum_{\beta = 1}^3 S^{\alpha\beta} u_a^\beta
\end{equation}
where $s(a)$ is where the atom equivalent to $a$ by the $S$ symmetry and $S_{\alpha\beta}$ is the symmetry matrix. In the same way, we obtain the BO force of the new configuration. Here, we make explicit the atomic lower index with Latin letters and Cartesian upper indices with the ancient Greek alphabet. In this way, the stochastic average is computed over a number of effective configurations equal to $N_\text{conf}N_\text{sym}N_T$, where $N_\text{sym}$ is the number of symmetries in the point-group and $N_T$ is the number of translations inside the super-cell. Notably, symmetries cannot be applied on the final result itself of \eqname~\eqref{eq:d1v:final} and \eqref{eq:d2v:final}, as the perturbed vector $\rho^{(1)}(\bR)$ violates point group symmetries.

The other symmetry not explicitly accounted for in \eqname~\eqref{eq:d1v:final} and \eqref{eq:d2v:final} is that  $\bDthree$ and $\bDfour$ are invariant under indices permutation. To enforce this symmetry we need to sum separately each part of the weights (\eqname~\ref{eq:full:weight}) into the perturbed averages, and swapping the index of forces $\bbf$ with the one of the displacements $\tilde\bUps \tilde\bu$.

\end{appendices}

\bibliography{biblio}

\end{document}